\documentclass[onecolumn]{IEEEtran}
\pdfoutput=1
\usepackage[cmex10]{amsmath}
\usepackage{amssymb,amsthm,mathrsfs}
\usepackage{graphicx}
\usepackage{tikz}
 \usetikzlibrary{decorations}
 \usetikzlibrary{decorations.pathreplacing}
 \usetikzlibrary{arrows}
\usepackage{subfigure}
\usepackage{multirow}
\usepackage{hyperref}
\usepackage[sort,compress]{cite}
\usepackage[nomarkers,nolists]{endfloat}
 
\newtheorem{theorem}{Theorem}
\newtheorem{lemma}[theorem]{Lemma}

\newtheorem{corollary}[theorem]{Corollary}
\theoremstyle{definition}
 \newtheorem{property}{Property}

\DeclareMathOperator*{\lcm}{LCM}

\begin{document}

\title{Two-User Interference Channels with Local Views:\\ On Capacity Regions of TDM-Dominating Policies}
\author{\authorblockN{David T.-H. Kao and Ashutosh Sabharwal}
 \thanks{This work has been supported in part by NSF grant CNS-1012921.}\\
 \authorblockA{Department of Electrical \& Computer Engineering \\
  Rice University -- Houston, TX 77005 \\
  \texttt{\{davidkao,ashu\}@rice.edu}
 }
}
\maketitle

\begin{abstract}
We study the capacity regions of two-user interference channels where transmitters base their transmission schemes on local views of the channel state. Under the local view model, each transmitter knows only a subset of the four channel gains, which may be mismatched from the other transmitter.

We consider a set of seven local views, and find that for five out of the seven local views, TDM is sufficient to achieve the qualified notion of capacity region for the linear deterministic interference channel which approximates the Gaussian interference channel. For these five local views, the qualified capacity result implies that no policy can achieve a rate point outside the TDM region without inducing a corner case of sub-TDM performance in another channel state. 
The common trait shared by the two remaining local views --- those with the potential to outperform TDM --- is transmitter knowledge of the outgoing interference link accompanied by some common knowledge of state, emphasizing their importance in creating opportunities to coordinate usage of more advanced schemes.

Our conclusions are extended to bounded gap characterizations of the capacity region for the Gaussian interference channel.
\end{abstract}


\pdfoutput=1
\section{Introduction}

Interference occurs in wireless networks when multiple transmitters simultaneously attempt to convey information to distinct receivers using a shared medium. This phenomenon is a direct result of superposition of electromagnetic waveforms in wireless communications, and becomes more prevalent as the demand for wireless devices increases. The interference channel (IC) is a mathematical model of this phenomenon wherein for each transmitter there is one unique receiver, and salient effects of the medium are captured in a probabilistic input-output relationship. The capacity region of the IC in general remains unknown, including that of the Gaussian IC, a canonical model for wireless networks. 

The best known achievable region for the two-user IC results from the Han-Kobayashi (HK) coding scheme, introduced in the 1980s~\cite{HK:81}. HK codes enable higher rates than any other known scheme through exploitation of structure in coded transmission: in a HK code, separate codebooks are created for public and private components of each source's message, and the component codes are structured with the characteristics of the channel so as to provide ``space'' within an unintended receiver's channel output for its corresponding transmitter to convey information. Exploiting the structure of the channel with well-designed codes facilitates higher rate for both users, and it was shown in~\cite{ETW:08} that a simple type of HK code is near optimal for the two-user Gaussian interference channel; i.e., the scheme is within one-bit per user of the Gaussian interference channel capacity region. 

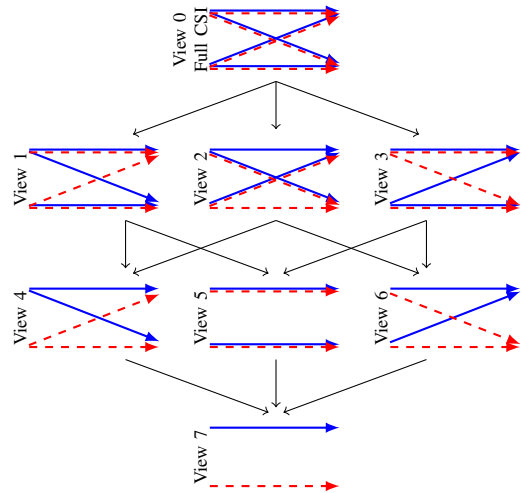
\begin{figure}[h]
	\centering\begin{tikzpicture}[xscale=0.4,yscale=0.37]
	\draw[blue,thick,-latex,shorten >=3pt,shorten <=2pt](0.125,6.05) -- (4.875,6.05);
	\draw[blue,thick,-latex,shorten >=3pt,shorten <=2pt](0.125,4.05) -- (4.875,6.05);
	\draw[blue,thick,-latex,shorten >=3pt,shorten <=2pt](0.125,4.05) -- (4.875,4.05);
	\draw[blue,thick,-latex,shorten >=3pt,shorten <=2pt](0.125,6.05) -- (4.875,4.05);
	\draw[red,dashed,thick,-latex,shorten >=3pt,shorten <=2pt](0.125,3.95) -- (4.875,3.95);
	\draw[red,dashed,thick,-latex,shorten >=3pt,shorten <=2pt](0.125,5.95) -- (4.875,3.95);
	\draw[red,dashed,thick,-latex,shorten >=3pt,shorten <=2pt](0.125,5.95) -- (4.875,5.95);
	\draw[red,dashed,thick,-latex,shorten >=3pt,shorten <=2pt](0.125,3.95) -- (4.875,5.95);
	\draw[] (-0.35,5) node[rotate=90]{\parbox{5em}{\centering\scriptsize View~0\\Full CSI}};
	\draw[blue,thick,-latex,shorten >=3pt,shorten <=2pt](-5.875,1.05) -- (-1.125,1.05);
	\draw[blue,thick,-latex,shorten >=3pt,shorten <=2pt](-5.875,1.05) -- (-1.125,-0.95);
	\draw[blue,thick,-latex,shorten >=3pt,shorten <=2pt](-5.875,-0.95) -- (-1.125,-0.95);
	\draw[red,dashed,thick,-latex,shorten >=3pt,shorten <=2pt](-5.875,-1.05) -- (-1.125,-1.05);
	\draw[red,dashed,thick,-latex,shorten >=3pt,shorten <=2pt](-5.875,-1.05) -- (-1.125,0.95);
	\draw[red,dashed,thick,-latex,shorten >=3pt,shorten <=2pt](-5.875,0.95) -- (-1.125,0.95);
	\draw[] (-6,0) node[rotate=90]{\parbox{5em}{\centering\scriptsize View~1}};
	\draw[blue,thick,-latex,shorten >=3pt,shorten <=2pt](0.125,1.05) -- (4.875,1.05);
	\draw[blue,thick,-latex,shorten >=3pt,shorten <=2pt](0.125,1.05) -- (4.875,-0.95);
	\draw[blue,thick,-latex,shorten >=3pt,shorten <=2pt](0.125,-0.95) -- (4.875,1.05);
	\draw[red,dashed,thick,-latex,shorten >=3pt,shorten <=2pt](0.125,-1.05) -- (4.875,-1.05);
	\draw[red,dashed,thick,-latex,shorten >=3pt,shorten <=2pt](0.125,-1.05) -- (4.875,0.95);
	\draw[red,dashed,thick,-latex,shorten >=3pt,shorten <=2pt](0.125,0.95) -- (4.875,-1.05);
	\draw[] (0,0) node[rotate=90]{\parbox{5em}{\centering\scriptsize View~2}};
	\draw[blue,thick,-latex,shorten >=3pt,shorten <=2pt](6.125,1.05) -- (10.875,1.05);
	\draw[blue,thick,-latex,shorten >=3pt,shorten <=2pt](6.125,-0.95) -- (10.875,1.05);
	\draw[blue,thick,-latex,shorten >=3pt,shorten <=2pt](6.125,-0.95) -- (10.875,-0.95);
	\draw[red,dashed,thick,-latex,shorten >=3pt,shorten <=2pt](6.125,-1.05) -- (10.875,-1.05);
	\draw[red,dashed,thick,-latex,shorten >=3pt,shorten <=2pt](6.125,0.95) -- (10.875,-1.05);
	\draw[red,dashed,thick,-latex,shorten >=3pt,shorten <=2pt](6.125,0.95) -- (10.875,0.95);
	\draw[] (6,0) node[rotate=90]{\parbox{5em}{\centering\scriptsize View~3}};
	\draw[blue,thick,-latex,shorten >=3pt,shorten <=2pt](-5.875,-3.95) -- (-1.125,-3.95);
	\draw[blue,thick,-latex,shorten >=3pt,shorten <=2pt](-5.875,-3.95) -- (-1.125,-5.95);
	\draw[red,dashed,thick,-latex,shorten >=3pt,shorten <=2pt](-5.875,-6.05) -- (-1.125,-6.05);
	\draw[red,dashed,thick,-latex,shorten >=3pt,shorten <=2pt](-5.875,-6.05) -- (-1.125,-4.05);
	\draw[] (-6,-5) node[rotate=90]{\parbox{5em}{\centering\scriptsize View~4}};
	\draw[blue,thick,-latex,shorten >=3pt,shorten <=2pt](0.125,-3.95) -- (4.875,-3.95);
	\draw[blue,thick,-latex,shorten >=3pt,shorten <=2pt](0.125,-5.95) -- (4.875,-5.95);
	\draw[red,dashed,thick,-latex,shorten >=3pt,shorten <=2pt](0.125,-6.05) -- (4.875,-6.05);
	\draw[red,dashed,thick,-latex,shorten >=3pt,shorten <=2pt](0.125,-4.05) -- (4.875,-4.05);
	\draw[] (0,-5) node[rotate=90]{\parbox{5em}{\centering\scriptsize View~5}};
	\draw[blue,thick,-latex,shorten >=3pt,shorten <=2pt](6.125,-3.95) -- (10.875,-3.95);
	\draw[blue,thick,-latex,shorten >=3pt,shorten <=2pt](6.125,-5.95) -- (10.875,-3.95);
	\draw[red,dashed,thick,-latex,shorten >=3pt,shorten <=2pt](6.125,-6.05) -- (10.875,-6.05);
	\draw[red,dashed,thick,-latex,shorten >=3pt,shorten <=2pt](6.125,-4.05) -- (10.875,-6.05);
	\draw[] (6,-5) node[rotate=90]{\parbox{5em}{\centering\scriptsize View~6}};
	\draw[blue,thick,-latex,shorten >=3pt,shorten <=2pt](0.125,-8.95) -- (4.875,-8.95);
	\draw[red,dashed,thick,-latex,shorten >=3pt,shorten <=2pt](0.125,-11.05) -- (4.875,-11.05);
	\draw[] (0,-10) node[rotate=90]{\parbox{5em}{\centering\scriptsize View~7}};
\draw[->,shorten >=3pt] (2.5,3.5) -- (2.5,1.5);
\draw[->,shorten >=3pt] (2.5,3.5) -- (-2.5,1.5);
\draw[->,shorten >=3pt] (2.5,3.5) -- (7.5,1.5);

\draw[->,shorten >=3pt] (-2.5,-1.5) -- (-2.5,-3.5);
\draw[->,shorten >=3pt] (-2.5,-1.5) -- (2.5,-3.5);

\draw[->,shorten >=3pt] (2.5,-1.5) -- (-2.5,-3.5);
\draw[->,shorten >=3pt] (2.5,-1.5) -- (7.5,-3.5);

\draw[->,shorten >=3pt] (7.5,-1.5) -- (2.5,-3.5);
\draw[->,shorten >=3pt] (7.5,-1.5) -- (7.5,-3.5);

\draw[->,shorten >=3pt] (-2.5,-6.5) -- (2.5,-8.5);
\draw[->,shorten >=3pt] (2.5,-6.5) -- (2.5,-8.5);
\draw[->,shorten >=3pt] (7.5,-6.5) -- (2.5,-8.5);
\end{tikzpicture}
\caption[Local Views]{Local Views: solid blue edges are known to Transmitter~$a$ and dashed, red edges are known to Transmitter~$b$. Digraph also depicts relationships between preference between local views. An arrow from one view to another signifies that the first has a strictly more complete understanding of channel state, and its TDM-dominating capacity region contains the capacity region of the other.}
\label{fig:VIEWS}
\end{figure}

Despite these results, there exists a gap between what is theoretically near-optimal, and what is used in practice. Modern communication networks do not use HK codes when dealing with interference, and almost all commercial approaches to dealing with interference can be categorized as one of two methods: treating interference as noise and orthogonalization, both sub-optimal approaches. For both, there exist channel states where the gap between the rate achieved and capacity of the channel is unbounded.
Rationale for the use of suboptimal schemes often cite the challenge or high cost of gathering high-quality estimates of channel state~\cite{LHLGRA:08} and difficulty in coordinating numerous independent transmitter-receiver pairs~\cite{GKGO:07}. We consider precisely these limitations in this work, and are interested in the capacity region of the Gaussian interference channel when protocols are jointly designed to operate in a distributed manner: the exact transmission scheme used is based on mismatched knowledge of the channel. Conversely, we would like to better understand what information is necessary to enable efficient distributed usage of the wireless medium.
Thus, we study an interference channel with \emph{local views}, defined by the following three properties:
\begin{itemize}
 \item Nodes' knowledge of network state is \emph{incomplete}: 
	Each node only has knowledge of a subset of the four channels.
 \item Nodes' knowledge of network state is \emph{mismatched}: 
	The subset of channels known to a node may be different from the subset known to another node.
 \item Nodes' transmit decisions are \emph{distributed}: 
	Nodes must make independent decisions on how to communicate, each basing its decision on its incomplete and mismatched local view.
\end{itemize}
Each pair of local views (one for each transmitter node) considered specifies perfect, non-noisy knowledge of a subset of four link gains. Channel gains of links included in the local view are known without error and only the support of channel gains is known for links not in the local view. Each local view considered includes knowledge of the direct link, and we focus on cases where the local views of nodes are relatively symmetric: from each transmitter's point of view, the relative location of known channel gains is identical. For the two-user IC, this amounts to the seven pairs of local views depicted in Figure~\ref{fig:VIEWS}. 

In analyzing each view, our study focuses on transmission schemes, and the effect of incomplete information on design of distributed protocols. Although how a node arrives at a local view is not considered in this paper, each local view studied can be envisioned as the \emph{result} of some channel state learning mechanism. Symmetry of the views is likely in networks where each node uses a similar learning process. 
We also note that the class of local views we consider for the two-user IC is more flexible than the hop-based quantification of locality in~\cite{ALS:arx09, AAS:arx10}, which presupposes some message passing algorithm to learn channel state.

We propose a new definition of a local view capacity region, which naturally captures the impact of mismatched knowledge and design principles for distributed protocols. In the proposed definition, we limit transmission strategies to those which can universally dominate a time-division multiplexing (TDM) allocation for all possible 4-tuples of channel states. 
The reason for requiring TDM as the base-line measure of performance is that TDM can be achieved with the minimal local view of just knowing the direct channel gains at each transmitter.\footnote{We have implicitly assumed that the whole network is temporally synchronized irrespective of the amount of knowledge at any node.} 
With the requirement of performance universally ``at least as good as TDM,'' we are able to infer how views with more knowledge facilitate opportunistic performance gains. 

We next derive the exact local view capacity regions for a two-user linear deterministic interference channel, a class of channel models which provides the basis for approximate capacity region characterizations for two-user Gaussian IC. The seven symmetric local views can be classified into two categories: first where opportunistic Han-Kobayashi codes can exceed TDM region and second where TDM is an optimal approach.

The first category includes two local views out of seven possible symmetric views, and both require knowledge of three link gains \emph{including the outgoing interference link}.
Knowledge of the outgoing links permits each transmitter to opportunistically utilize HK-type codes, resulting in rates better than those achievable through TDM. These two particular views \emph{reveal opportunities} for multiuser codes, and confirms intuition that knowledge of the outgoing interference link is critical to employ such more complex schemes.

The second category includes the remaining five views, where there exists no policy that can universally exceed the performance of a TDM scheme. This includes even one case where, of four channel gains, only the outgoing interference gain is unknown to each transmitter. That a nearly complete view of the channel results in performance that can be achieved with less knowledge of the channel implies that knowledge acquisition mechanisms resulting in this view may be wasting resources in gathering information. 

Finally, we use the similarities between the linear deterministic and Gaussian channel models to analyze the gap between the capacity region of the linear deterministic interference channel and the capacity region of the Gaussian IC. For four local views, the gap is constant irrespective of the channel state. For the remaining three local views, the gap is dependent only on the relative channel gain values, which allows us to characterize the generalized degrees of freedom regions of the Gaussian IC with local views. 

Our results provide a number of key intuitions into the development of wireless networks. First, Although we consider a number of local views without assuming a learning mechanism (e.g., training, feedback, message passing, etc.), our results can provide intuition into what aspects of the network nodes should endeavor to learn. For instance, extra channel training may not be worth the cost if performance better than TDM is unattainable. On the other hand, if we desire spectral efficiency exceeding that which can be provided by TDM, our results may provide a guiding principle for determining what aspects of the network must be learned. Second, for currently deployed networks where nodes have very little knowledge about the network state, our results provide rationale for the orthogonalization-based approaches commonly used in practice. To some extent, our results validate approaches such as random access, high frequency reuse factors, and CDMA which all seek to orthogonalize transmission in distributed 
networks.

The remainder of this paper is structured as follows. 
Section~\ref{sec:litrev} details related work, and discusses how our contributions relate to previous findings. 
In Section~\ref{sec:prelim}, we formalize our analysis: we review Gaussian and linear deterministic interference channel models, describe the local view model of distributed channel state knowledge, discuss the relative importance of the local views studied, and provide mathematical preliminaries for the statement of main results. 
Within this section, we also clarify our notion of capacity regions for distributed networks and through analysis of a local view multiple-access channel, demonstrate some of the challenges posed by a local view model and how our definition of capacity regions is applied.
Section~\ref{sec:bounding}, describes the techniques used in defining inner and outer bounds for both channel models, and in Section~\ref{sec:lindet_results} we present our main results: capacity regions for the linear deterministic IC for each of the seven views studied. In Section~\ref{sec:gauss}, we examine the local view capacity regions of the Gaussian IC, and extend intuitions drawn in the linear deterministic interference channel to the Gaussian domain. 

\section{Related Work}
\label{sec:litrev}

Information theoretic study of interference began in the 1960s~\cite{Shannon:61}. In particular, the issue of the capacity region of the Gaussian interference channel was the impetus for development of many new achievable schemes and outer bounding techniques \cite{Carleial:75, Sato:78, HK:81, Sato:81, Costa:85, Kramer:04, Sason:04, BreslerT:08, ETW:08, MK:09, XKB:09, AV:09, BPT:10, AZ:arx10}. 

Among these, \cite{BreslerT:08} was instrumental in studying the linear deterministic interference channel model, a sub-class of the channels studied in \cite{EC:82}. The linear deterministic signal model, first introduced in \cite{ADT:11} is used within this work, and as shown in \cite{ADT:11,BreslerT:08} is an excellent approximation of the two-user Gaussian interference channel. Two desirable consequences of using the linear deterministic model will be discussed in the context of our paper: 
\begin{enumerate}
 \item Approximately-optimal layered approaches like the Han-Kobayashi code of \cite{ETW:08} and the lattice approach of \cite{BPT:10} are revealed naturally in the linear deterministic model.
 \item The relationship between the capacity regions of the linear deterministic channel and Gaussian channel and near-equivalence in the high-SNR regime (generalized degrees of freedom), provides intuition into solving real systems.
\end{enumerate}

In~\cite{CJ:08}, the authors introduced interference alignment (IA), a transmission scheme that is degrees of freedom optimal. While recent IA work also considers limited knowledge conditions, e.g., no-CSI~\cite{Jafar:arx09} and ``blind''~\cite{WGJ:arx10} cases. We consider a case without knowledge of channel statistics and our metric (capacity and approximate capacity \emph{regions}) reveal more about the performance capabilities of the network than degrees of freedom which is a single value that is not specific to a channel state.

Another way to model channel uncertainty is the approach taken in compound channels~\cite{BBT:59} which specify a set of possible channel states, and the objective is to define a scheme that will maximize rate regardless of the actual channel state. Within this domain, the work of \cite{RPV:09} is closest to our own. In their work, the authors define an achievable scheme such that the gap between the scheme and the capacity of the Gaussian IC is bounded by a constant. However, the uncertainty set of possible channel states in the compound IC is synchronized among the nodes. In contrast, our contribution emphasizes transmitters having different uncertainty sets although for one of the seven views considered (View~5) the results can be derived from the results of \cite{RPV:09} because the views are identical and the responses to the system may be coordinated by common knowledge.

Local views are a newer field of study~\cite{ALS:arx09, AAS:arx10}, and so far the emphasis has been on notions of sum-rate. We instead consider a full capacity region, thereby taking a first step towards considering alternate notions of optimal rate allocation, such as fairness metrics which require knowledge of the full feasible set of rates~\cite{LKCS:10}. Additionally, the uncertainty model in~\cite{ALS:arx09, AAS:arx10} quantify the locality of the view by counting the number of ``hops'' of information. Our approach flexible in the sense that we do not assume a mechanism for acquiring the view, and instead examine a comprehensive subset of views, each of which may result from a different knowledge acquisition mechanism. 

Commodity WiFi network architecture provides a common example of how network design results in the most limited of local views we consider: only the direct link between associated mobiles and access is measured, and therefore mobiles have no knowledge of how much interference they may inflict on a flow associated with a neighboring access point. Most 3G cellular networks are similar in this respect, since no spectral resources are dedicated to training for channel gains characterizing inter-cell interference. On the other hand, protocols using inverse power tones or echoes~\cite{WTSRLLJ:10:Allerton,MBH:01:Infocom} have also been proposed, which provide transmitters with an estimate of the amount of interference they inflict on neighboring nodes but not the quality of the other user's direct link. 

\pdfoutput=1
\section{Preliminaries}
\label{sec:prelim}

This section defines aspects that make up our problem formulation. In Section~\ref{sec:chmod} two interference channel models are defined: the two-user Gaussian interference channel used in modeling wireless networks, and the two-user linear deterministic interference channel which is a discrete abstraction of the Gaussian model. We define the \emph{local view} formalization of distributed knowledge in Section~\ref{sec:locview}. Section \ref{sec:prelim_infoth} reviews theoretical preliminaries relevant to our study, and in Sections~\ref{sec:policies} and ~\ref{sec:TDM}, we define notions of policies, achievability, and capacity specific to the local view uncertainty model. Finally, in \ref{sec:MACexample} we provide an example --- a local view multiple-access channel --- which clarifies our model and reinforces the importance of each aspect of our problem formulation.

\subsection{Channel Models}
\label{sec:chmod}
We consider both Gaussian and linear deterministic two-user interference channel models:
\subsubsection{Gaussian IC}
\begin{figure}[ht]
	\begin{center}
	\begin{tikzpicture}[font=\scriptsize,yscale=0.7]
		\node (times aa) at (1,3) [circle, draw, thick, inner sep=1pt] {$\times$};
		\node (times ba) at (3,2) [circle, draw, thick, inner sep=1pt] {$\times$};
		\node (times ab) at (3,1) [circle, draw, thick, inner sep=1pt] {$\times$};
		\node (times bb) at (1,0) [circle, draw, thick, inner sep=1pt] {$\times$};

		\node (plus a) at (3,3) [circle, draw, thick, inner sep=1pt] {$+$};
		\node (plus b) at (3,0) [circle, draw, thick, inner sep=1pt] {$+$};
		
		\node (xa) at (0,3) [left]{$X_a$};
		\node (xb) at (0,0) [left]{$X_b$};
		
		\node (ya) at (4,3) [right]{$Y_a$};
		\node (yb) at (4,0) [right]{$Y_b$};
		
		\draw[thick,-latex] (xa) -- (times aa);
		\draw[thick,-latex] (times aa) -- (plus a);
		\draw[thick,-latex] (plus a) -- (ya);
		\draw[thick,-latex] (1,3.5) node[above] {$h_{aa}$} -- (times aa);
		\draw[thick,-latex] (3.5,2) node[right] {$h_{ba}$} -- (times ba);
		\draw[thick,-latex] (3,3.5) node[above] {$Z_a$} -- (plus a);

		\draw[thick,-latex] (xb) -- (times bb);
		\draw[thick,-latex] (times bb) -- (plus b);
		\draw[thick,-latex] (plus b) -- (yb);
		\draw[thick,-latex] (1,-0.5) node[below] {$h_{bb}$} -- (times bb);
		\draw[thick,-latex] (3.5,1) node[right] {$h_{ab}$} -- (times ab);
		\draw[thick,-latex] (3,-0.5) node[below] {$Z_b$} -- (plus b);
		
		\draw[thick,-latex] (xa) -- (0.25,3) -- (2.25,1) -- (times ab);
		\draw[thick,-latex] (xb) -- (0.25,0) -- (2.25,2) -- (times ba);

		\draw[thick,-latex] (times ba) -- (plus a);
		\draw[thick,-latex] (times ab) -- (plus b);
	\end{tikzpicture}
	\end{center}
	\caption{Gaussian Interference Channel}
	\label{fig:mod_G}
\end{figure}
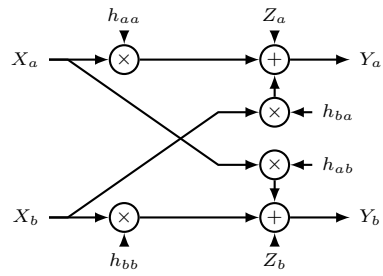

The Gaussian interference channel (GIC, shown in Figure~\ref{fig:mod_G}) consists of two transmitter-receiver pairs, labeled $a$ and $b$, and four point-to-point \emph{links} --- two direct and two interfering. The signal strengths of each point-to-point link are represented by the four complex gain values $h_{aa}$, $h_{ab}$, $h_{ba}$, and $h_{bb}$, collectively referred to as $H$.

The relationship between complex channel inputs $X_a$ and $X_b$ and the received channel outputs $Y_a$ and $Y_b$ is given by
\begin{align}
 Y_a ={}& h_{aa}X_{a} + h_{ba}X_{b} + Z_a,\\
 Y_b ={}& h_{bb}X_{b} + h_{ab}X_{a} + Z_b.
\end{align}
where we have a power constraint $\frac{1}{n}\sum_{t=1}^{n}|X_i|^2\leq 1$ for $i\in \{a,b\}$, and $Z_a$ and $Z_b$ are single samples of a zero-mean, unit-variance white Gaussian random process.

\subsubsection{Linear Deterministic IC}

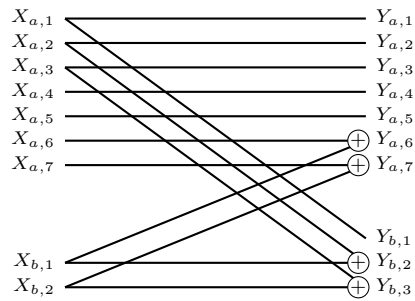
\begin{figure}[ht]
	\begin{center}
	\begin{tikzpicture}[font=\scriptsize,yscale=0.65]
\draw[white] (0,-0.25) -- (0,1);
		\draw[thick](0,6) -- (4,1.5);
		\draw[thick](0,5.5) -- (4,1);
		\draw[thick](0,5) -- (4,0.5);

		\draw[thick](0,1) -- (4,3.5);
		\draw[thick](0,0.5) -- (4,3);

		\draw[thick] (4,1.5) node[right] {$Y_{b,1}$};
		\draw[thick](0,1) node[left] {$X_{b,1}$} -- (4,1) node[right] {$Y_{b,2}$};
		\draw[thick](0,0.5) node[left] {$X_{b,2}$} -- (4,0.5) node[right] {$Y_{b,3}$};

		\draw[thick](0,6) node[left] {$X_{a,1}$} -- (4,6) node[right] {$Y_{a,1}$};
		\draw[thick](0,5.5) node[left] {$X_{a,2}$} -- (4,5.5) node[right] {$Y_{a,2}$};
		\draw[thick](0,5) node[left] {$X_{a,3}$} -- (4,5) node[right] {$Y_{a,3}$};
		\draw[thick](0,4.5) node[left] {$X_{a,4}$} -- (4,4.5) node[right] {$Y_{a,4}$};
		\draw[thick](0,4) node[left] {$X_{a,5}$} -- (4,4) node[right] {$Y_{a,5}$};
		\draw[thick](0,3.5) node[left] {$X_{a,6}$} -- (4,3.5) node[right] {$Y_{a,6}$};
		\draw[thick](0,3) node[left] {$X_{a,7}$} -- (4,3) node[right] {$Y_{a,7}$};
		
		\draw (3.9,3) node[fill=white,draw,circle, inner sep=0pt]{$+$};
		\draw (3.9,3.5) node[fill=white,draw,circle, inner sep=0pt]{$+$};
		\draw (3.9,1) node[fill=white,draw,circle, inner sep=0pt]{$+$};
		\draw (3.9,0.5) node[fill=white,draw,circle, inner sep=0pt]{$+$};
\end{tikzpicture}
	\end{center}
	\caption{A linear deterministic interference channel where $g_{aa}=7$, $g_{ab}=3$, $g_{ba}=2$, and $g_{bb}=2$.}
	\label{fig:lindet_IC}
\end{figure}

The linear deterministic signal model~\cite{ADT:11} captures the broadcast and superposition aspects of the wireless channel while abstracting the receiver noise into a signal level ``floor'' at each receiver. In doing so, the effects of noise become a constant effect, facilitating the analysis of the impact of interference. Figure~\ref{fig:lindet_IC} depicts an example of a linear deterministic interference channel (LDIC), that is used repeatedly within this document to demonstrate new concepts.

When considering the LDIC we refer to single-channel-use input (output) tuples of Transmitter~$i$ (Receiver~$j$) collectively as $\mathbf{X}_{i}$ ($\mathbf{Y}_{j}$), and to length-$n$ vectors of channel inputs (outputs) --- e.g., when considering $n$ channel uses --- as $\mathbf{X}_{i}^n$ ($\mathbf{Y}_{j}^n$) respectively. We also refer to the collection of all four channel gains as the \emph{channel state} $G = (g_{aa},g_{ab},g_{ba},g_{bb})$, and express the input-output relationship of the LDIC in the form of shift matrix operations and element-wise modulo-2 addition. Notice that we differentiate LDIC variables from those of the GIC with bold weight typeface for channel input and output variables and use of $G$ instead of $H$ for channel gains.
\begin{align}
 \mathbf{Y}_a ={}& S_{q_a}^{q_a-g_{aa}} \mathbf{X}_{a} \oplus S_{q_a}^{q_a-g_{ba}}\mathbf{X}_{b},\\
 \mathbf{Y}_b ={}& S_{q_b}^{q_b-g_{bb}} \mathbf{X}_{b} \oplus S_{q_b}^{q_b-g_{ab}}\mathbf{X}_{a},
\end{align}
where $q_a = \max\{g_{aa},g_{ba}\}$, $q_b = \max\{g_{bb},g_{ab}\}$, and $S_q$ is the $q\times q$ shift matrix 
\begin{equation*}
 S_q = 
 \begin{bmatrix}
 0 & 0 & 0 & \ldots & 0\\
 1 & 0 & 0 & \ldots & 0\\
 0 & 1 & 0 & \ldots & 0\\
 \vdots & \ddots & \ddots & \ddots & \vdots\\
 0 & \ldots & 0 & 1 & 0
 \end{bmatrix}.
\end{equation*}
By properties of matrix exponentiation, $S_q^0$ is a $q\times q$ identity matrix. 

Each set of GIC gains, $H$, is associated with an analogous set of LDIC gains, $G$, by the following relationship: 
\begin{align}
	g_{aa} ={}& \lfloor\log\left(|h_{aa}|^2\right)\rfloor^+ ,\label{eq:GtoLD1}\\
	g_{ab} ={}& \lfloor\log\left(|h_{ab}|^2\right)\rfloor^+ ,\\
	g_{ba} ={}& \lfloor\log\left(|h_{ba}|^2\right)\rfloor^+ ,\\
	g_{bb} ={}& \lfloor\log\left(|h_{bb}|^2\right)\rfloor^+.\label{eq:GtoLD4}
\end{align}
This linear deterministic approximation was proposed in~\cite{ADT:11} and it's usefulness for analyzing the GIC was demonstrated in~\cite{BreslerT:08} where it was shown that the capacity of a GIC is within a constant number of bits of the associated LDIC. Moreover, the intuition provided by segmentation of signal space into discrete levels provided an intuition on construction of nearly-optimal codes.

\subsection{Local Views: A Model for Distributed Knowledge}
\label{sec:locview}

In our model, Transmitter~$a$'s knowledge is composed of only a subset of all the gains in the network. When a link gain is known, it is known without error, and when it is unknown, Transmitter~$a$ has no knowledge of its value besides its support, which is the complex field for the GIC and all non-negative integers for the LDIC. We call the subset of channel gains known to Transmitter~$a$ its \emph{local view}, and denote it for the GIC as $\widehat{H}_a$. When specifying the many possible views, the symbol $\varnothing$ is used for unknown gains. 
As an example, if Transmitter~$a$ knew all of the gains besides that of the direct link between Transmitter~$b$ and Receiver~$b$, we would denote the relationship as
\begin{equation*}
\widehat{H}_a = (h_{aa},h_{ab},h_{ba},\varnothing).
\end{equation*}
The same conventions hold for Transmitter~$b$, as well as for the LDIC except we denote channels and views with $G$ instead of $H$ to distinguish between gains of the two different models. It is important to clarify that both transmitters are aware of the size of the network (two links in isolation), as well as the structure of the view of the other transmitter (which link gains are known and unknown). 

The receivers are assumed to have adequate knowledge to accommodate transmitter decisions and decode messages coherently. Ostensibly, it is sufficient to assume that receivers have a full view of all link gains, however it is feasible that receivers only know their incoming link gains and the transmitter decisions on codebook and rate are embedded in the transmission header. 

%
This study is restricted to cases where structures of the two views are \emph{symmetric} relative to perspective of the viewer: e.g., if gain $h_{ij}$ is (un)known to Transmitter~$a$, then $h_{ji}$ is (un)known to Transmitter~$b$. Symmetric views are only a subset in the full set of cases that may be considered, but for the two-user IC, symmetric views may describe cases where the same method is used by each transmitter to learn about the network. All eight symmetric views are depicted in Figure~\ref{fig:VIEWS}, and notice that while the structure of local views exhibit relative symmetry, the exact sets of channel gains known to each node for all but View~0 and View~5 are mismatched.

\pdfoutput=1
\subsection{Mathematical Preliminaries}
\label{sec:prelim_infoth}
Here we provide mathematical foundations defining theoretic analysis of reliable communication. In order to communicate, each transmitter uses an encoding function $c_{i,n}$ to encode a message $m_i$ drawn independently from the set $M_i = \{1, \ldots , 2^{nR_i} \}$ into a codeword of $n$ symbols, $\mathbf{X}_i^n = (\mathbf{X}_{i}[1] , \ldots, \mathbf{X}_{i}[n] )$, and subject to a unit power constraint $\frac{1}{n}\sum_{t=1}^{n}|X_i|^2\leq 1$ for the GIC.

Each receiver observes its channel outputs $(\mathbf{Y}_i[1],\ldots, \mathbf{Y}_i[n])$ and uses a decoding function $f_{i,n}$  to arrive at an estimate $\widehat{m}_i\in M_i$ of the encoded message $m_i$. An error occurs whenever $\widehat{m}_i \neq m_i$. The average probability of error for User~$i$ is given by
\begin{equation*}
 \epsilon_{i,n} = E[\Pr(\widehat{m}_i \neq m_i )],
\end{equation*}
where the expectation is taken with respect to the random choice of the transmitted messages $m_a$ and $m_b$.

A rate pair $(r_a,r_b)$ is achievable if there exists a family of codebook pairs $\left\{c_{a,n} , c_{b,n} \right\}_{n\in \mathbb{N}}$ indexed by the block length $n$ with codewords satisfying input constraints, and decoding functions $\left\{f_{a,n} (\cdot), f_{b,n} (\cdot)\right\}_{n\in \mathbb{N}}$, such that the average decoding error probabilities $\epsilon_{a,n}$ and $\epsilon_{b,n}$ vanish as block length $n$ goes to infinity. 
By applying Shannon's coding theorem for the point-to-point channel, the set of achievable rate points can be determined by the following. 
\begin{lemma} \cite[Lemma~1]{BreslerT:08}
 The rate point $(r_a,r_b)$ is achievable if and only if for every $\epsilon>0$ there exists a block length $n$ and distributions $p(\mathbf{X}_a^n)$ and $p(\mathbf{X}_b^n)$ such that
 \begin{align}
  nr_a - \epsilon \leq{}& I(\mathbf{X}_a^n;\mathbf{Y}_a^n)\label{eq:mutualinfo_a},\\
  nr_b - \epsilon \leq{}& I(\mathbf{X}_b^n;\mathbf{Y}_b^n)\label{eq:mutualinfo_b}.
 \end{align}\label{lem:1}
\end{lemma}
The capacity region $\mathcal{C}$ of the interference channel is the closure of the set of all achievable rate pairs.

\subsection{Distributed Policies}
\label{sec:policies}
In a centralized network, encoding functions may be designed jointly to adapt to the channel state. On the other hand, in our model each transmitter selects its encoding function based on its respective local view. Specifically, the encoding function used, $c_{i,n}(m_i;\widehat{G}_i)$, is dependent on the view, which implies that the resulting rates ($r_a(\widehat{G}_a)$, $r_b(\widehat{G}_b)$) and distributions of channel inputs ($p(\mathbf{X}_a;\widehat{G}_a)$, $p(\mathbf{X}_b;\widehat{G}_b)$) are also dependent on view. 

We assume that the mapping from view, $\widehat{G}_i$, to encoding function, $c_{i,n}(m_i;\widehat{G}_i)$, is both \emph{deterministic} and \emph{globally known}; i.e., although Transmitter~$a$ may not know Transmitter~$b$'s exact choice of codebook and rate due to mismatched views of $a$ and $b$, Transmitter~$a$ knows how $b$ would respond to a particular channel state.

The globally known deterministic mappings emulate predetermined protocols or \emph{policies} agreed upon by the two users or specified by the network architect. Therefore, throughout the paper we refer collectively to the mappings $c_{i,n}(m_i;\widehat{G}_i)$, $r_i(\widehat{G}_i)$, and $p(\mathbf{X}_i;\widehat{G}_i)$ as the policy of Transmitter~$i$ for $i\in\{a,b\}$.

A policy couples the uncertainty of the interferers' choice of encoding function to the uncertainty in channel state. Accordingly, the ability to coordinate selection of encoding functions and the resulting performance is still dependent on what knowledge will be available to each transmitter. In our formulation, we define achievability of a pair of policy-defined rates, $r_a(\widehat{G}_a)$ and $r_b(\widehat{G}_b)$, by extending Lemma~\ref{lem:1} to apply to view-dependent encoding functions, and requiring achievability of the respective encoding functions for all channel states consistent with the local view considered. Mathematically, achievability requires the existence of view-dependent input distributions $p(\mathbf{X}_a^n;\widehat{G}_a)$ and $p(\mathbf{X}_b^n;\widehat{G}_b)$ such that the following two expressions hold for all $G$.
\begin{align}
  nr_a(\widehat{G}_a) - \epsilon \leq{}& I(\mathbf{X}_a^n;\mathbf{Y}_a^n)\label{eq:pol_ach_constr_a},\\
  nr_b(\widehat{G}_b) - \epsilon \leq{}& I(\mathbf{X}_b^n;\mathbf{Y}_b^n)\label{eq:pol_ach_constr_b}.
\end{align}

\subsection{TDM-Dominating Capacity Region}
\label{sec:TDM}

The inequalities (\ref{eq:pol_ach_constr_a}) and (\ref{eq:pol_ach_constr_b}) provide necessary conditions for existence of coding schemes that achieve the target rates dictated by a pair of policies. However, (\ref{eq:pol_ach_constr_a}) and (\ref{eq:pol_ach_constr_b}) do not capture how policies may prescribe an aggressive scheme and maximize capacity for one channel state at the cost of low performance in another channel state. 

Policies based on time-division multiplexing (TDM) are unique in this regard because using a fixed time-division results in a rate pair that is the same for all channels with the same direct link gain. This feature of performance that is balanced, in addition to the characteristic that TDM is an achievable policy agnostic to which local view from Figure~\ref{fig:VIEWS} is being considered, provides a natural baseline for comparison of policies. Therefore, we consider only policies that guarantee performance equal to or better than a TDM policy regardless of channel state, thereby formalizing a notion of capacity qualified by a TDM minimum performance criterion.

Let $\overline{\mathcal{R}}^\mathsf{TDM}$ be defined as the closure of the non-zero-rate boundary of the rate region achieved using TDM. Specifically,
\begin{align}
\overline{\mathcal{R}}^\mathsf{TDM} = \left\{(r_a,r_b): r_a = (1-\tau)C_a, r_b = \tau C_b, \tau\in[0,1]\right\}, 
\end{align}
where $C_a ={} g_{aa}$, $C_b ={} g_{bb}$ for the LDIC, and
$C_a ={} \log\left(1+\left|h_{aa}\right|^2\right)$, $C_b ={} \log\left(1+\left|h_{bb}\right|^2\right)$ for the GIC. We refer to a pair of policies as \emph{TDM-dominating} if there exists some $(r_a^\mathsf{TDM},r_b^\mathsf{TDM})\in\overline{\mathcal{R}}^\mathsf{TDM}$ such that for every channel state $G$ and views $\widehat{G}_a = V_a(G)$ and $\widehat{G}_b = V_b(G)$, 
\begin{align}
	r_a(\widehat{G}_a)\geq{}& r_a^\mathsf{TDM},\label{eq:tdm-dom1}\\
	r_b(\widehat{G}_b)\geq{}& r_b^\mathsf{TDM}.\label{eq:tdm-dom2}
\end{align}
We say that the policy strictly dominates TDM if (\ref{eq:tdm-dom1}) and (\ref{eq:tdm-dom2}) are satisfied for all channels and for at least one channel state, either (\ref{eq:tdm-dom1}) or (\ref{eq:tdm-dom2}) is strict.
The analogous definition of TDM-dominating policies applies to the GIC.

\vspace{0.25em}\noindent\textbf{TDM Minimum Performance Criterion:}
Our notion of capacity region considers only rate pairs achievable by TDM-dominating policies, which implies that policies considered must satisfy for every possible channel, $G$, the criterion
\begin{equation}
 \frac{r_a(\widehat{G}_a)}{C_{a}} + \frac{r_b(\widehat{G}_b)}{C_{b}} \geq 1.\label{eq:TDM}
\end{equation}
\ 

Under the TDM minimum performance criterion, if a local view is shown to have a strictly larger capacity region, then there exists a policy which strictly dominates TDM, or any other orthogonalized scheme, across all channel realizations.
Without any criterion, the standard notion of a capacity region for channel state $G$ is misleading in the context of distributed protocol design. For a given network state, a policy that prescribes a full view capacity achieving scheme \emph{always} exists, but, as we demonstrate in the next section, use of such a policy may limit what can be achieved in other channel states. By considering TDM-dominating policies and the resulting TDM-dominating capacity region, we present rate pairs that are achievable for one channel state without inducing a corner-case channel state with performance poorer than TDM.

We note that our problem formulation is an analysis of worst-case or robust communication within a network with distributed uncertainty. When available and valid, knowledge of statistics of the channel can provide a method of improving the average performance of the network. However, we contest that our formulation is readily applicable to scenarios where the wireless fading statistics are poorly understood or difficult to model. Additionally, when the measure of system performance cannot be measured ergodically averaging over many channel fades, e.g., systems sensitive to delay like streaming voice and video, first responder systems, and system critical control, our results can provide more insight into what guarantees can be made regarding a networks ability to communicate.
%
%
\subsection{Example: A Local View Multiple-Access Channel}
\label{sec:MACexample}
Consider the following two-user local view linear deterministic multiple-access channel (LV-MAC). The full-view capacity region for any particular channel state is~\cite{ADT:11} the set of rates $(r_a,r_b)$ satisfying
\begin{align}
 r_a \leq{}& g_a\label{eq:MAC1},\\
 r_b \leq{}& g_b\label{eq:MAC2},\\
 r_a+r_b \leq{}& \max\{g_a,g_b\}.\label{eq:MAC3}
\end{align}

The local views are such that each user only knows its direct link gain, i.e., $\widehat{G}_a = (g_a,\varnothing)$ and $\widehat{G}_b = (\varnothing,g_b)$. As in the local view IC, transmitters must select a codebook and rate given incomplete, mismatched knowledge of the channel. Consider a policy that when $g_a = 2$ and $g_b=1$ achieves, using random codebooks\footnote{Consider block codes of length $n$ where each entry of each $g_i\times n$ matrix codeword are drawn from a Bernoulli distribution with $p=\frac{1}{2}$.} and joint decoding, the rate point $(1,1)$ which is a corner point on the full knowledge capacity region, i.e., 
\begin{align}
 r_a( \widehat{G}_a )|_{g_a=2} = 1,\label{eq:MAC_ex_ra}\\
 r_b( \widehat{G}_b )|_{g_b=1} = 1.\label{eq:MAC_ex_rb}
\end{align}
In order to satisfy (\ref{eq:MAC3}) when $G = (1,1)$ or $G = (2,2)$, and assuming the policy chosen is such that (\ref{eq:MAC_ex_ra}) and (\ref{eq:MAC_ex_rb}) hold, we find the following constraints on policy responses
\begin{align}
 r_a( \widehat{G}_a )|_{g_a=1} \leq{}& \max\{1,1\} - r_b( \widehat{G}_b )|_{g_b=1}
 = 0,\\
 r_b( \widehat{G}_b )|_{g_b=2} \leq{}& \max\{2,2\} - r_a( \widehat{G}_a )|_{g_a=2}
 = 1.
\end{align}
For the channel states $G = (1,1)$, $G = (2,1)$, and $G = (2,2)$, the policy results in rate points on the boundary of the respective ideal capacity regions. However for the case where $G = (1,2)$, the resulting extremal rate point, $r = (0,1)$, is not only an interior point, but also less efficient than TDM. Therefore, although the policy designed thus far strictly dominates TDM for some channel states, the performance gain comes at the expense of what may occur in other states. 

Furthermore, we can show that no policy strictly dominating TDM exists. Consider the two channel states $G = (1,1)$ and $G = (2,2)$. In each, TDM is capacity achieving even with a full view. Let the time-division parameters in state $G = (1,1)$ be defined as $\tau_a(1)$ and $\tau_b(1)$ where $\tau_a(1)+\tau_b(1) = 1$. Similarly, we define $\tau_a(2)$ and $\tau_b(2)$ where $\tau_a(2)+\tau_b(2) = 1$. The rates resulting from this policy are $r_a(\widehat{G}_a)|_{g_a=s} = s\tau_a(s)$ and $r_b(\widehat{G}_b)|_{g_b=t} = t\tau_b(t)$.

Assume that $\tau_a(1)\leq\tau_a(2)$, which implies $\tau_b(2)\leq\tau_b(1)$. We now consider the channel state $G = (1,2)$ and notice
\begin{equation}
 \frac{r_a(\widehat{G}_a)}{g_a} + \frac{r_b(\widehat{G}_b)}{g_b} = \tau_a(1) + \tau_b(2) \leq 1,
\end{equation}
with equality if and only if $\tau_a(1)=\tau_a(2)$ and $\tau_b(2)=\tau_b(1)$, i.e., if the rate achieved in each channel state dominates TDM, then not only are the capacity regions of all four possible channel states the TDM region, but also all four states are tied to the same operating point (time-division allocation).

In fact, the confinement to a single time-division regardless of channel state holds true for a more general case as well: the capacity of the $K$-user linear deterministic multiple-access channel where each transmitter only knows the gain of his direct link cannot strictly dominate TDM. Moreover, all channel states are bound to the same time divisions. The proof can be found in Appendix~\ref{append:LV-MAC}.
\begin{theorem}[$K$-User LV-MAC TDM-Dominating Capacity Region] 
	Let a LV-MAC be defined as a $K$-user multiple-access channel where for each transmitter, $\widehat{G}_k = (\varnothing,\ldots,\varnothing,g_k,\varnothing,\ldots)$. 
	The TDM-dominating capacity region, $\mathcal{C}_{MAC}$, is equal to the set of rate tuples achievable by TDM, $\mathcal{R}_{MAC}^\mathsf{TDM}$, i.e.,
	\begin{equation}
	 \mathcal{C}_{MAC} = \mathcal{R}_{MAC}^\mathsf{TDM}.
	\end{equation}
%
\end{theorem}

\pdfoutput=1
\section{Bounding Techniques}
\label{sec:bounding}
We characterize capacity regions through analysis of inner and outer bounds. Though the analysis for each local view varies, the basic techniques employed are summarized in this section.

\subsection{Inner Bounds}
\label{sec:inner}
In this work, we reference only two types of achievable schemes: Time-Division Multiplexing (TDM) and the simple Han-Kobayashi scheme (HK) from \cite{ETW:08} whose achievable region was shown (in the same reference) to be within one bit per user of the full view two-user GIC capacity region. 

\subsubsection{Time-Division Multiplexing}
\label{sec:in_TDM}
The rate pairs achievable though TDM (the boundary of which was specified in Section~\ref{sec:TDM}) are achieved through time orthogonalization. 
%
We note that in the case of the GIC our TDM-based rate region assumes no power scaling, however the gap between the two regions can be shown to be less than two bits per user.

\subsubsection{Simple Han-Kobayashi Codes}
\label{sec:HK}
In general HK schemes, each transmitter splits the contents of its message into a common message and a private message. The simple HK codes of~\cite{ETW:08} use random Gaussian codebooks for both the common and private encoding functions, with a division in power between the two chosen such that the private component of the message is received at the unintended receiver ``in the noise floor''; i.e.,  the private and common codebooks are drawn from independent zero-mean Gaussian distributions with variances $P_{i,p} = \min\left\{\frac{1}{|h_{ij}|^2},1\right\}$ ($i\neq j$) and $P_{i,c} = 1-P_{i,p}$ respectively. 

At the receiver the private message of the undesired signal is treated as noise, thereby at most doubling the power of the interference-noise floor. The receiver jointly decodes both common messages and the desired private message, forming a virtual three user multiple-access channel. The resulting rate region $\mathcal{R}_G^\mathsf{HK}$ was shown to be approximately capacity achieving
and is given by all rate pairs $(r_a, r_b)$ such that $r_a \leq r_{a,p}+r_{a,c}$ and $r_b \leq r_{b,p}+r_{b,c}$ satisfying
 \begin{align}
 r_{a,p} \geq{}& 0\\
 r_{a,c} \geq{}& 0\\
 r_{b,p} \geq{}& 0\\
 r_{b,c} \geq{}& 0\\
 r_{a,p} \leq{}& \log\left( 1 + \frac{ \min\left\{\frac{|h_{aa}|^2}{|h_{ab}|^2},|h_{aa}|^2\right\}}{1+\min\left\{|h_{ba}|^2,1\right\}}\right)\\
 r_{a,c} \leq{}& \min\left\{
	 \log\left( 1 + \frac{|h_{aa}|^2-\min\left\{\frac{|h_{aa}|^2}{|h_{ab}|^2},|h_{aa}|^2\right\}}{1+\min\left\{|h_{ba}|^2,1\right\}}\right),
	 \log\left( 1 + \frac{|h_{ab}|^2-\min\left\{1,|h_{ab}|^2\right\}}{1+\min\left\{|h_{ab}|^2,1\right\}}\right)\right\}\\
 r_{a,p} + r_{a,c} \leq{}& 
	\log\left( 1 + \frac{|h_{aa}|^2}{1+\min\left\{|h_{ba}|^2,1\right\}}\right)\\
 r_{a,p} + r_{b,c} \leq{}& 
	\log\left( 1 + \frac{\min\left\{\frac{|h_{aa}|^2}{|h_{ab}|^2},|h_{aa}|^2\right\} + |h_{ba}|^2-\min\left\{1,|h_{ba}|^2\right\}}{1+\min\left\{|h_{ba}|^2,1\right\}}\right)\\
 r_{a,p} + r_{a,c} + r_{b,c} 
	\leq{}& \log\left( 1 + \frac{|h_{aa}|^2 + |h_{ba}|^2-\min\left\{1,|h_{ba}|^2\right\}}{1+\min\left\{|h_{ba}|^2,1\right\}}\right)
\end{align}
\begin{align}
 r_{b,p} \leq{}& \log\left( 1 + \frac{ \min\left\{\frac{|h_{bb}|^2}{|h_{ba}|^2},|h_{bb}|^2\right\}}{1+\min\left\{|h_{ab}|^2,1\right\}}\right)\\
 r_{b,c} \leq{}& \min\left\{
	 \log\left( 1 + \frac{|h_{bb}|^2-\min\left\{\frac{|h_{bb}|^2}{|h_{ba}|^2},|h_{bb}|^2\right\}}{1+\min\left\{|h_{ab}|^2,1\right\}}\right),
	 \log\left( 1 + \frac{|h_{ba}|^2-\min\left\{1,|h_{ba}|^2\right\}}{1+\min\left\{|h_{ba}|^2,1\right\}}\right)\right\}\\
 r_{b,p} + r_{b,c} \leq{}& 
	\log\left( 1 + \frac{|h_{bb}|^2}{1+\min\left\{|h_{ab}|^2,1\right\}}\right)\\
 r_{b,p} + r_{a,c} \leq{}& 
	\log\left( 1 + \frac{\min\left\{\frac{|h_{bb}|^2}{|h_{ba}|^2},|h_{bb}|^2\right\} + |h_{ab}|^2-\min\left\{1,|h_{ab}|^2\right\}}{1+\min\left\{|h_{ab}|^2,1\right\}}\right)\\
 r_{b,p} + r_{b,c} + r_{a,c} 
	\leq{}& \log\left( 1 + \frac{|h_{bb}|^2 + |h_{ab}|^2-\min\left\{1,|h_{ab}|^2\right\}}{1+\min\left\{|h_{ab}|^2,1\right\}}\right)\\
 r_{a,c} + r_{b,c} \leq{}& \min\left\{
	\log\left( 1 + \frac{|h_{aa}|^2-\min\left\{\frac{|h_{aa}|^2}{|h_{ab}|^2},|h_{aa}|^2\right\} + |h_{ba}|^2-\min\left\{1,|h_{ba}|^2\right\}}{1+\min\left\{|h_{ba}|^2,1\right\}}\right),\right.\nonumber\\
	&{}\left.\log\left( 1 + \frac{|h_{bb}|^2-\min\left\{\frac{|h_{bb}|^2}{|h_{ba}|^2},|h_{bb}|^2\right\} + |h_{ab}|^2-\min\left\{1,|h_{ab}|^2\right\}}{1+\min\left\{|h_{ab}|^2,1\right\}}\right)\right\}.
\end{align}

\begin{figure}[ht]
	\begin{center}
	\begin{tikzpicture}[font=\scriptsize,yscale=0.65]
		\draw[thick](0,6) -- (4,1.5);
		\draw[thick](0,5.5) -- (4,1);
		\draw[thick](0,5) -- (4,0.5);

		\draw[thick](0,1) -- (4,3.5);
		\draw[thick](0,0.5) -- (4,3);

		\draw[thick] (4,1.5) node[right] {$Y_{b,1}$};
		\draw[thick](0,1) node[left] {$X_{b,1}$} -- (4,1) node[right] {$Y_{b,2}$};
		\draw[thick](0,0.5) node[left] {$X_{b,2}$} -- (4,0.5) node[right] {$Y_{b,3}$};

		\draw[thick](0,6) node[left] {$X_{a,1}$} -- (4,6) node[right] {$Y_{a,1}$};
		\draw[thick](0,5.5) node[left] {$X_{a,2}$} -- (4,5.5) node[right] {$Y_{a,2}$};
		\draw[thick](0,5) node[left] {$X_{a,3}$} -- (4,5) node[right] {$Y_{a,3}$};
		\draw[thick](0,4.5) node[left] {$X_{a,4}$} -- (4,4.5) node[right] {$Y_{a,4}$};
		\draw[thick](0,4) node[left] {$X_{a,5}$} -- (4,4) node[right] {$Y_{a,5}$};
		\draw[thick](0,3.5) node[left] {$X_{a,6}$} -- (4,3.5) node[right] {$Y_{a,6}$};
		\draw[thick](0,3) node[left] {$X_{a,7}$} -- (4,3) node[right] {$Y_{a,7}$};
		
		\draw (3.9,3) node[fill=white,draw,circle, inner sep=0pt]{$+$};
		\draw (3.9,3.5) node[fill=white,draw,circle, inner sep=0pt]{$+$};
		\draw (3.9,1) node[fill=white,draw,circle, inner sep=0pt]{$+$};
		\draw (3.9,0.5) node[fill=white,draw,circle, inner sep=0pt]{$+$};
		
		\draw [thick,decorate,decoration={brace}] (-0.75,4.8) -- node[sloped,above]{Common} (-0.75,6.2);
		\draw [thick,decorate,decoration={brace}] (-0.75,2.8) -- node[sloped,above]{Private} (-0.75,4.7);

		\draw [thick,decorate,decoration={brace}] (-0.75,0.3) -- node[sloped,above]{Common} (-0.75,1.2);
	\end{tikzpicture}
	\end{center}
	\caption{Separation of Usable Linear Deterministic Channel Levels into Common and Private Components}
		\label{fig:lindet_HK}
\end{figure}
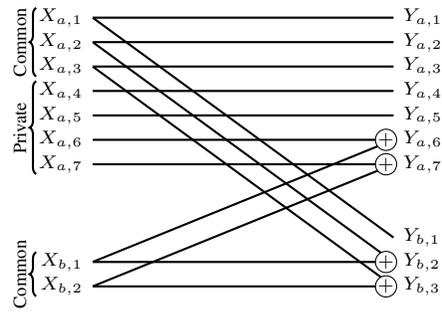

The analogous approach in the LDIC is to similarly split each user's message into common and private parts, where the private message is carried by the signal levels that are not seen at the unintended receiver (shown in Figure~\ref{fig:lindet_HK}). 
As with simple HK codes for the GIC, receivers decode the two common messages and desired private message jointly, resulting in the capacity-achieving rate region, $\mathcal{R}_D^\mathsf{HK}$, given by all rate pairs $(r_a, r_b)$ such that $r_a \leq r_{a,p}+r_{a,c}$ and $r_b \leq r_{b,p}+r_{b,c}$ satisfying
\begin{align}
 r_{a,p} \geq{}& 0 \label{eq:rates_lindet_HK_1},\\
 r_{a,c} \geq{}& 0 \label{eq:rates_lindet_HK_2},\\
 r_{b,p} \geq{}& 0 \label{eq:rates_lindet_HK_3},\\
 r_{b,c} \geq{}& 0 \label{eq:rates_lindet_HK_4},\\
 r_{a,p} \leq{}& (g_{aa}-g_{ab})^+ \label{eq:rates_lindet_HK_5},\\
 r_{a,c} \leq{}& \min\{g_{aa},g_{ab}\} \label{eq:rates_lindet_HK_6},\\
 r_{a,p} + r_{a,c} \leq{}& g_{aa}\\
 r_{a,p} + r_{b,c} \leq{}& \max\{g_{aa}-g_{ab},g_{ba}\} \label{eq:rates_lindet_HK_7},\\
 r_{a,p} + r_{a,c} + r_{b,c} \leq{}& \max\{g_{aa},g_{ba}\} \label{eq:rates_lindet_HK_8},\\
 r_{b,p} \leq{}& (g_{bb}-g_{ba})^+ \label{eq:rates_lindet_HK_9},\\
 r_{b,c} \leq{}& \min\{g_{bb},g_{ba}\} \label{eq:rates_lindet_HK_10},\\
 r_{b,p} + r_{b,c} \leq{}& g_{bb}\\
 r_{b,p} + r_{a,c} \leq{}& \max\{g_{bb}-g_{ba},g_{ab}\} \label{eq:rates_lindet_HK_11},\\
 r_{b,p} + r_{b,c} + r_{a,c} \leq{}& \max\{g_{bb},g_{ab}\} \label{eq:rates_lindet_HK_12}\\
 r_{a,c} + r_{b,c} \leq{}& \min\{\max\{g_{aa},g_{ba}\},\max\{g_{bb},g_{ab}\}\} \label{eq:rates_lindet_HK_x}.
\end{align}
These expressions can also be adapted from~\cite{BreslerT:08} by removing redundant constraints, but are presented in a form that emphasizes similarities to the virtual 3-user MAC seen at each receiver in the GIC and facilitates comparison between achievable component rates. By maximizing the difference between the right hand side of each pair of analogous expressions, we have
\begin{lemma}
 Let a GIC channel, $H$, and analogous LDIC channel, $G$, defined by (\ref{eq:GtoLD1})--(\ref{eq:GtoLD4}) be given. 
 If the tuple ($r_a^p,r_a^c,r_b^p,r_b^c$) achievable by the Gaussian simple HK scheme, the tuple ($r_a^p-2,r_a^c-2,r_b^p-2,r_b^c-2$) is achievable in the associated linear deterministic scheme.
 Conversely, if the tuple ($r_a^p,r_a^c,r_b^p,r_b^c$) achievable by the linear deterministic HK scheme, the tuple ($r_a^p-2,r_a^c-2,r_b^p-2,r_b^c-2$) is achievable in the associated Gaussian scheme.

 Consequently, if ($r_a,r_b$) is achievable for either channel model using the described HK scheme, then ($r_a-4,r_b-4$) is achievable in the alternate model.
  \label{lem:codegap}
\end{lemma}

Although a more concise set of inequalities from~\cite{BreslerT:08} and more generally from~\cite{EC:82} is shown in (\ref{eq:reg_view0_1})--(\ref{eq:reg_view0_7}) below for the LDIC (the same region approximates the GIC), the component-separated expressions (\ref{eq:rates_lindet_HK_1})--(\ref{eq:rates_lindet_HK_12}) reveals where opportunities for increased rate over orthogonalized schemes exist. We will show that transmitters may only capitalize on an available opportunity if that opportunity is revealed by a local view.

\begin{align}
r_a \leq{}& g_{aa}\label{eq:reg_view0_1}\\
r_b \leq{}& g_{aa}\label{eq:reg_view0_2}\\
r_a + r_b \leq{}& (g_{aa} - g_{ba})^+ +\max\{g_{bb},g_{ba}\} \label{eq:reg_view0_3}\\
r_a + r_b \leq{}& (g_{bb} - g_{ab})^+ +\max\{g_{aa},g_{ab}\} \label{eq:reg_view0_4}\\
r_a + r_b \leq{}& \max\{g_{ab},(g_{aa} - g_{ba})^+\} 
	+ \max\{g_{ba},(g_{bb} - g_{ab})^+\} \label{eq:reg_view0_5}\\
2r_a + r_b \leq{}& \max\{g_{aa},g_{ab}\} + (g_{aa} - g_{ba})^+ 
	+ \max\{g_{ba},(g_{bb} - g_{ab})^+\} \label{eq:reg_view0_6}\\
r_a + 2r_b \leq{}& \max\{g_{bb},g_{ba}\} + (g_{bb} - g_{ab})^+ 
	+ \max\{g_{ab},(g_{aa} - g_{ba})^+\}. \label{eq:reg_view0_7}
\end{align}

\subsection{Outer Bounds}
We build our outer bounds on the capacity region of local view policies on two key techniques.
\subsubsection{Expanded Virtual Z-Channel }
\label{sec:virtualZ}
In an IC, construction of each transmitter's encoding scheme is faced with two objectives:
\begin{itemize}
 \item On the direct link, a transmitter seeks to adapt its signal to increase its rate (increase entropy) in the presence of an interference signal.
 \item On the out-going interference link, a transmitter seeks to minimize its impact (reduce entropy).
\end{itemize}
A special case of IC, known as the Z-channel, often provides clarity regarding these competing objectives by considering the effect of only one of the two interference links that must be considered. The relationship between the Z-channel and IC has been noted previously, e.g., in derivation of outer bounds~\cite{Kramer:04}. 

\begin{figure}[ht]
	\begin{center}
\begin{tikzpicture}[font=\scriptsize,yscale=0.65]
		\draw[thick](0,6) -- (4,1.5);
		\draw[thick](0,5.5) -- (4,1);
		\draw[thick](0,5) -- (4,0.5);

		\draw[thick](0,8.5) -- (4,3.5);
		\draw[thick](0,8) -- (4,3);

		\draw[thick](0,8.5) node[left] {$X_{b^\prime,1}$} -- (4,8.5) node[right] {$Y_{b^\prime,1}$};
		\draw[thick](0,8) node[left] {$X_{b^\prime,2}$} -- (4,8) node[right] {$Y_{b^\prime,2}$};

		\draw[thick](0,1) node[left] {$X_{b,1}$} -- (4,1) node[right] {$Y_{b,2}$};
		\draw[thick](0,0.5) node[left] {$X_{b,2}$} -- (4,0.5) node[right] {$Y_{b,3}$};

		\draw[thick](0,6) node[left] {$X_{a,1}$} -- (4,6) node[right] {$Y_{a,1}$};
		\draw[thick](0,5.5) node[left] {$X_{a,2}$} -- (4,5.5) node[right] {$Y_{a,2}$};
		\draw[thick](0,5) node[left] {$X_{a,3}$} -- (4,5) node[right] {$Y_{a,3}$};
		\draw[thick](0,4.5) node[left] {$X_{a,4}$} -- (4,4.5) node[right] {$Y_{a,4}$};
		\draw[thick](0,4) node[left] {$X_{a,5}$} -- (4,4) node[right] {$Y_{a,5}$};
		\draw[thick](0,3.5) node[left] {$X_{a,6}$} -- (4,3.5) node[right] {$Y_{a,6}$};
		\draw[thick](0,3) node[left] {$X_{a,7}$} -- (4,3) node[right] {$Y_{a,7}$};
		
		\draw (3.9,3) node[fill=white,draw,circle, inner sep=0pt]{$+$};
		\draw (3.9,3.5) node[fill=white,draw,circle, inner sep=0pt]{$+$};
		\draw (3.9,1) node[fill=white,draw,circle, inner sep=0pt]{$+$};
		\draw (3.9,0.5) node[fill=white,draw,circle, inner sep=0pt]{$+$};
		
%
\end{tikzpicture}
	\end{center}
	\caption{LDIC of Figure~\ref{fig:lindet_IC} Unwrapped into a Double Z-channel}
		\label{fig:unwrapped}
\end{figure}
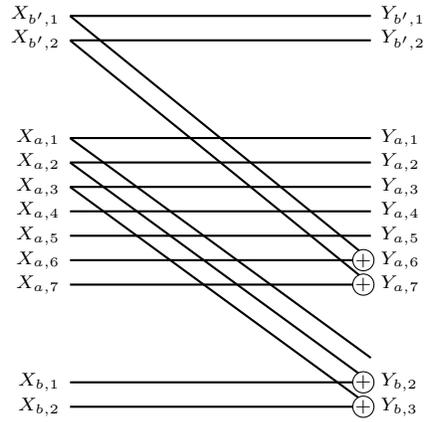

However, instead of considering a single Z-channel, we go one step further by ``unwrapping'' the two-user local view interference channel into a series of Z-channels, so as to simultaneously focus on effects of both outgoing and incoming interference on achievable user policy responses (Figure~\ref{fig:unwrapped}). 

In order to use the unwrapped IC as an outer bound, we impose constraints on the inputs of users in the unwrapped channel. In the case shown in Figure~\ref{fig:unwrapped}, if we assume full views of state, the constraint that the exact message and channel inputs of $b$ and $b^\prime$ are the same is enough to specify a double Z-channel whose capacity region characterizes (and thus bounds) the original IC. Notice that the restriction of $b$ and $b^\prime$ to the same messages and inputs results in $r_b=r_{b^\prime}$ as long as both receivers can decode. 

For full view, the resulting bound does not supply any intuition beyond that given by known genie-aided bounds. 
On the other hand, because a transmitter with a local view is uncertain of at least one link in the unwrapped channel, it must account for every possibility of unknown link states and resulting policy responses of the other transmitter.  Moreover, from the perspective of User~$a$, the input $\mathbf{X}_{b^\prime}(\widehat{G}_{b^\prime})$ most detrimental at Receiver~$a$ may result from one local view at Transmitter~$b$, $\widehat{G}_{b^\prime}$, while the input most sensitive to interference at Receiver~$b$ may result from a different local view, $\widehat{G}_{b}$. Therefore, the constraints we impose on inputs $\mathbf{X}_{b^\prime}(\widehat{G}_{b^\prime})$ and $\mathbf{X}_{b}(\widehat{G}_{b})$ are that the policy mapping view $\widehat{G}_b$ to scheme must be consistent.

We visualize design of local view policies by considering a series of virtual users arranged in a larger Z-channel with the following properties:
\begin{itemize}
 \item Each transmitter-receiver pair in an unwrapped channel is a \emph{virtual user} corresponding to a policy response of either User~$a$ or User~$b$ in the original local view IC.
 \item A virtual User~$a$ always interferes with a virtual User~$b$ (and vice-versa), or does not interfere at all (terminates the Z-channel). The Z-channel may be cyclic.
 \item Each virtual transmitter uses the policy governing channel inputs corresponding to its local view.
 \item Any link gain known to both transmitters must be consistent throughout the virtual channel.
\end{itemize}
By visualizing the design of local view policies in sets of expanded Z-channels, we can consider possibly many worst case virtual channel states, each of which corresponds to a sequence of dependencies between policy responses.

\subsubsection{Genie}

In the following, we show how a particular level-by-level application of the chain rule in LDIC analysis can be emulated for the GIC using a well designed genie. The genie we describe here occurs is implicit in analysis of the the LDIC, owing to the fact that the channel is noiseless, but construction of the GIC genie is motivated by intuitions drawn from how, in the LDIC, entropies of signals may be decomposed level-by-level through application of the chain rule and the entropy of each level may be examined conditioned on higher levels.

Consider an LDIC where $g_{aa}\geq g_{ba}>0$ (i.e., the direct link of $a$ has a higher gain than the impinging interference from $b$). In this case, the mutual information of Link~$a$ can be expressed as
\begin{align}
I(&\mathbf{X}_a^n;\mathbf{Y}_a^n)\nonumber\\
	={}& H(\mathbf{Y}_a^n) - H(\mathbf{Y}_a^n|\mathbf{X}_a^n) \label{eq:decom1}\\
	={}& H(Y_{a,1}^n,...,Y_{a,g_{aa}-g_{ba}}^n) + H(Y_{a,g_{aa}-g_{ba}+1}^n,...,Y_{a,g_{aa}}^N|Y_{a,1}^n,...,Y_{a,g_{aa}-g_{ba}}^n)- H(X_{b,1}^n,\ldots,X_{b,g_{ba}}^n) \label{eq:decom3}\\
	={}& H(X_{a,1}^n,...,X_{a,g_{aa}-g_{ba}}^n) + H(Y_{a,g_{aa}-g_{ba}+1}^n,...,Y_{a,g_{aa}}^n|X_{a,1}^n,...,X_{a,g_{aa}-g_{ba}}^n)- H(X_{b,1}^n,\ldots,X_{b,g_{ba}}^n). \label{eq:decom4}
\end{align}
Expression (\ref{eq:decom3}) results from an application of the chain rule, and (\ref{eq:decom4}) notes that the chain rule was applied at the boundary between interfered and un-interfered receive signal levels. If on the other hand $g_{ba} > g_{aa}$, we have
\begin{align}
I(&\mathbf{X}_a^n;\mathbf{Y}_a^n)
	={}& H(X_{b,1}^n,...,X_{b,g_{ba}-g_{aa}}^n) 
		+ H(Y_{a,g_{ba}-g_{aa}+1}^n,...,Y_{a,g_{ba}}^n|X_{b,1}^n,...,X_{b,g_{ba}-g_{aa}}^n)
		- H(X_{b,1}^n,\ldots,X_{b,g_{ba}}^n) \label{eq:decom5}.
\end{align}

Given the definitions
\begin{align}
 L_{a,i} \triangleq{}& H(X_{a,i}^n|X_{a,1}^n\ldots,X_{a,i-1}^n),\\
 L_{b,j} \triangleq{}& H(X_{b,j}^n|X_{b,1}^n\ldots,X_{b,j-1}^n),\\
 u_a^+ \triangleq{}& (g_{aa} - g_{ba})^+,\\
 u_a^- \triangleq{}& (g_{ba} - g_{aa})^+,
\end{align}
The relations (\ref{eq:decom4}) and (\ref{eq:decom5}) can be more generally stated in (\ref{eq:decom_bound_a}) as
\begin{align}
	I(\mathbf{X}_a^n;\mathbf{Y}_a^n)	
	\leq{}& \left(n\min\{g_{aa},g_{ba}\} - \sum_{k=1}^{g_{ba}}L_{b,k} \right)
		+ \left(\sum_{i=1}^{u_a^+}L_{a,i} + \sum_{j=1}^{u_a^-}L_{b,j}\right). \label{eq:decom_bound_a}
\end{align}

Similarly, for Link~$b$ if
\begin{align}
 u_b^+ \triangleq{}& (g_{bb} - g_{ab})^+,\\
 u_b^- \triangleq{}& (g_{ab} - g_{bb})^+,
\end{align}
then
\begin{align}
	I(\mathbf{X}_b^n;\mathbf{Y}_b^n)	
	\leq{}& \left(n\min\{g_{bb},g_{ab}\} - \sum_{k=1}^{g_{ab}}L_{a,k} \right)
		+ \left(\sum_{j=1}^{u_b^+}L_{b,j} + \sum_{i=1}^{u_b^-}L_{a,i}\right).  \label{eq:decom_bound_b}
\end{align}

The first two quantities in both decompositions (\ref{eq:decom_bound_a}) and (\ref{eq:decom_bound_b}) emphasize that if the strengths of incoming signals are not equal, the most significant bits of the combined received signal are easy to decode, and the contention occurs in those levels where the two signals overlap.

For the GIC, it is not apparent how to ``decode the most significant bits'' without restricting analysis to layered coding schemes. However, if the upper levels of the signal (those modeled as non-interfered bits in the LDIC) are ``easy to decode'', then there should be little benefit in supplying these levels of the signal separately to the receiver. This provides the intuition behind our genie, which provides a set of signals 
intended to emulate the layering of message content that is explicit in the LDIC model.

Our approach for constructing the genie is similar to \cite{RPV:09} in the sense that each signal is derived from a series of degraded signals, each representing a component of a received signal that may be received in a potential channel state. Furthermore, like \cite{RPV:09}, our genie signals are conditionally independent from the original received signal. 

Assume, for ease of explanation of the genie, that GIC gains result in integer LDIC gains without the use of the floor function:
\begin{align}
 g_{ij} ={}& \log\left(|h_{ij}|^2\right)^+,\ i,j \in \{a,b\}.
\end{align}
Let $Z_{a,\ell}^n \sim N(0,\mathbf{I})$ for $\ell\in\mathbb{N}$ be a series of independent (across $\ell$) length-$n$ i.i.d. zero-mean complex Gaussian random vectors. We define the maximum number of signals derived from Transmitter~$a$'s input as
\begin{align}
\ell_{a}^\star = \max\left\{g_{aa},g_{ab}\right\},
\end{align}
and the signal $U_{a,\ell_a^\star}^n$ be given as
\begin{align}
	U_{a,\ell_a^\star}^n = 
	\begin{cases}
		|h_{aa}| X_a^n + Z_{a,\ell_a^\star}^{n} & \text{ if }|h_{aa}| \geq |h_{ab}| \cr
		|h_{ab}| X_a^n + Z_{a,\ell_a^\star}^{n} & \text{ if }|h_{aa}| < |h_{ab}| \cr	 
	\end{cases}
\end{align}
From $U_{a,\ell_a^\star}^n$, we define our a collection of serially degraded signals $\{W_{a,\ell}\}_{\ell\in \mathbb{N}}$
\begin{align}
	U_{a,\ell_a^\star-1}^n = {}& U_{a,\ell_a^\star}^n + Z_{a,\ell_a^\star-1}^n,\\
	U_{a,\ell_a^\star-2}^n = {}& U_{a,\ell_a^\star-1}^n + \sqrt{2}Z_{a,\ell_a^\star-2}^n,\\
	\vdots {}&\nonumber\\
	U_{a,\ell}^n = {}& U_{a,\ell-1}^n + \sqrt{2^{\ell_a^\star - \ell - 1}}Z_{a,\ell}^n,\\
	\vdots {}&\nonumber\\
	U_{a,1}^n = {}& \frac{1}{\sqrt{2}}U_{a,2}^n + \sqrt{2^{\ell_a^\star - 2}}Z_{a,1}^n.
\end{align}
In each successive signal $ U_{a,\ell}^n$ the power in the total noise term doubles. Additionally, the following Markov relationship is formed
\begin{equation}
	U_{a,1}^n \text{ --- } U_{a,2}^n \text{ --- } \ldots \text{ --- } U_{a,\ell_a^\star}^n \text{ --- } X_a^n \text{ --- } Y_j^n,
\end{equation}
where $Y_j^n$ is either received signal. A similar collection of signals, $\{U_{b,\ell}\}_{\ell\in \mathbb{N}}$, is defined for Transmitter~$b$'s input as well. 

To these signals, we add a phase correction term of the form
\begin{align}
 W_{aa,\ell}^n \triangleq \Phi_{aa} U_{a,\ell}^n,\\
 W_{ab,\ell}^n \triangleq \Phi_{ab} U_{a,\ell}^n,\\
 W_{ba,\ell}^n \triangleq \Phi_{ba} U_{b,\ell}^n,\\
 W_{bb,\ell}^n \triangleq \Phi_{bb} U_{b,\ell}^n,
\end{align}
where $\Phi_{ij} = e^{j\angle h_{ij}}$ incorporates the appropriate phase into the genie signals.

At each receiver, the genie provides those signals that represent the ``easy to decode'' large scale variations. For instance, if $|h_{aa}| > |h_{ba}|$, then signals $\{W_{aa,\ell}\}_{\ell\in \{1,\ldots,g_{aa} - g_{ba}\}} $ are provided to Receiver~$a$. 

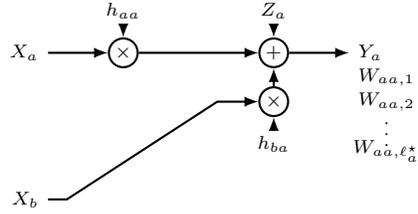
\begin{figure}[ht]
	\begin{center}
	\begin{tikzpicture}[font=\scriptsize,yscale=0.65]
		\node (times aa) at (1,3) [circle, draw, thick, inner sep=1pt] {$\times$};
		\node (times ba) at (3,2) [circle, draw, thick, inner sep=1pt] {$\times$};

		\node (plus a) at (3,3) [circle, draw, thick, inner sep=1pt] {$+$};
		
		\node (xa) at (0,3) [left]{$X_a$};
		\node (xb) at (0,0) [left]{$X_b$};
		
		\node (ya) at (4,3) [right]{$Y_a$};
		\node (va1) at (4.5,2.5) {$W_{aa,1}$};
		\node (va2) at (4.5,2) {$W_{aa,2}$};
		\node (vadots) at (4.5,1.5) {$\vdots$};
		\node (vadots) at (4.5,1) {$W_{aa,\ell_a^\star}$};
		
		\draw[thick,-latex] (xa) -- (times aa);
		\draw[thick,-latex] (times aa) -- (plus a);
		\draw[thick,-latex] (plus a) -- (ya);
		\draw[thick,-latex] (1,3.5) node[above] {$h_{aa}$} -- (times aa);
		\draw[thick,-latex] (3,1.5) node[below] {$h_{ba}$} -- (times ba);
		\draw[thick,-latex] (3,3.5) node[above] {$Z_a$} -- (plus a);
		
		
		\draw[thick,-latex] (xb) -- (0.25,0) -- (2.25,2) -- (times ba);

		\draw[thick,-latex] (times ba) -- (plus a);
	\end{tikzpicture}
	\end{center}
	\caption{Genie for Receiver~$a$ in the Gaussian Interference Channel}
	\label{fig:Genie}
\end{figure}

Assuming $|h_{aa}|>|h_{ba}|$, our newly defined genie allows us to arrive at
\begin{align}
 I(X_a^n;Y_a^n) \leq{}&  I(X_a^n;Y_a^n,W_{aa,1},W_{aa,2},\ldots,W_{aa,u_a^+})\\
	={}&  h(Y_a^n,W_{aa,1},W_{aa,2},\ldots,W_{aa,u_a^+}) 
		- h(Y_a^n,W_{aa,1},W_{aa,2},\ldots,W_{aa,u_a^+} | X_a^n)\\
	={}&  h(Y_a^n|W_{aa,u_a^+}) - h(Y_a^n|W_{aa,u_a^+},X_a^n) \\
		&  + \sum_{\ell=1}^{u_a^+} h(W_{aa,\ell}|W_{aa,1},\ldots,W_{aa,\ell-1}) - h(W_{aa,\ell}|W_{aa,1},\ldots,W_{aa,\ell-1},X_a^n)\\
	={}&  h(Y_a^n|W_{aa,u_a^+}) - h(Y_a^n|W_{aa,u_a^+},X_a^n) 
		+ \sum_{\ell=1}^{u_a^+} I(X_a^n;W_{aa,\ell}|W_{aa,1},\ldots,W_{aa,\ell-1}).\label{eq:Gauss_genie_dec_a}
\end{align}
As desired, the expression (\ref{eq:Gauss_genie_dec_a}) mimics (\ref{eq:decom_bound_a}) in its isolation of larger signal variations (more significant bits) from the variations which are contested by both direct and interference signals. Moreover, we notice that if $h_{ai}$ was the channel gain with larger magnitude, and using a substitution of statistically equivalent noise terms $Z^{n\prime}\sim N(0,\mathbf{I})$ and $Z^{n\prime\prime}\sim N(0,\mathbf{I})$
\begin{align}
 I(X_a^n&;W_{ai,\ell}|W_{ai,1},\ldots,W_{ai,\ell-1}) \\
	={}& I(X_a^n;W_{ai,\ell}|W_{ai,\ell-1})\\
	={}& h(W_{ai,\ell}|W_{ai,\ell-1}) - h(W_{ai,\ell}|X_a^n,W_{ai,\ell-1})\\
	={}& h( h_{ai} X_a^n + \sqrt{2^{\ell_a^\star - \ell}} Z^{n\prime} 
		| h_{ai} X_a^n + \sqrt{2^{\ell_a^\star - \ell}} Z^{n\prime} + \sqrt{2^{\ell_a^\star - \ell}} Z^{n\prime\prime})\nonumber\\
	&- h(h_{ai} X_a^n + \sqrt{2^{\ell_a^\star - \ell}} Z^{n\prime} 
		| X_a^n, h_{ai} X_i + \sqrt{2^{\ell_a^\star - \ell}} Z^{n\prime} + \sqrt{2^{\ell_a^\star - \ell}} Z^{n\prime\prime})\\
	={}& h(h_{ai} X_a^n + \sqrt{2^{\ell_a^\star - \ell}} Z^{n\prime} 
		| h_{ai} X_a^n + \sqrt{2^{\ell_a^\star - \ell}} Z^{n\prime} + \sqrt{2^{\ell_a^\star - \ell}} Z^{n\prime\prime})
	- h(\sqrt{2^{\ell_a^\star - \ell}} Z^{n\prime} 
		| \sqrt{2^{\ell_a^\star - \ell}} Z^{n\prime} + \sqrt{2^{\ell_a^\star - \ell}} Z^{n\prime\prime})\\
	={}& h(h_{ai} X_a^n + \sqrt{2^{\ell_a^\star - \ell}} Z^{n\prime} 
		| h_{ai} X_a^n + \sqrt{2^{\ell_a^\star - \ell}} Z^{n\prime} + \sqrt{2^{\ell_a^\star - \ell}} Z^{n\prime\prime})
	-\log\left(2\pi e \left[2^{\ell_a^\star - \ell-1}\right]\right)\\
	\leq{}& \log\left(1 + \frac{|h_{ai}|^2}
		{|h_{ai}|^2 + 2^{\ell_a^\star - \ell+1}}\right)\leq 1.
\end{align}
As in the LDIC, the payload of each genie signal level except for $\ell=1$\footnote{The topmost level of the signal $\ell=1$ is not subject to conditioning and thus is bounded by $\log(3)$.} is constrained to at most 1 bit:
\begin{align}
\Lambda_{a,\ell} \triangleq{}& I(X_a^n;W_{ai,\ell}|W_{ai,1},\ldots,W_{ai,\ell-1}) \leq 1.
\end{align}

Additionally, we note that
\begin{align}
 \sum_{\ell=k}^{K} \Lambda_{a,\ell} 
	={}& \sum_{\ell=k}^{K} I(X_a^n;W_{ai,\ell} | W_{ai,1},\ldots,W_{ai,\ell-1})\\
	={}& I(X_a^n;W_{ai,K} | W_{ai,1},\ldots,W_{ai,k}),
\end{align}
is the point-to-point rate of the $k^\text{th}$ through $K^\text{th}$ levels. If we consider the interference-noise term of (\ref{eq:Gauss_genie_dec_a}), we have
\begin{align}
 h(Y_a^n|W_{aa,u_a^+},X_a^n) ={}& h(h_{ba}X_b^n + Z_a^n)\\
 ={}& I(X_b^n;W_{ba,g_{ba}}) + h(Z_a^n)\\
 ={}& I(X_b^n;W_{ba,g_{ba}}) + n\log(2\pi e).
\end{align}
If only a portion of the interfering signal is considered (i.e., if $g_{ba} > g_{aa}$ and the genie supplies Receiver~$a$ with levels from Transmitter~$b$), we can also say
\begin{align}
 h(Y_a^n|W_{ba,u_a^-},X_a^n) ={}& h(h_{ba}X_b^n + Z_a^n|W_{ba,u_a^-})\\
 ={}& I(X_b^n;W_{ba,g_{ba}}|W_{ba,u_a^-}) + n\log(2\pi e).
\end{align}

Consequently, the genie-aided decomposition of mutual information at each receiver can be bounded by 
\begin{align}
 I(X_a^n;Y_a^n) \leq{}& \left( h(Y_a^n|W_{aa,u_a^+},W_{ba,u_a^-}) - n\log(2\pi e) 
		- \sum_{\ell=1}^{g_{ba}} \Lambda_{b,\ell} \right)
		+ \left(\sum_{\ell=1}^{u_a^+} \Lambda_{a,\ell}
		+ \sum_{\ell=1}^{u_a^-} \Lambda_{b,\ell}\right)
		\label{eq:Gauss_genie_dec_a_final},\\
 I(X_b^n;Y_b^n) \leq{}&  \left( h(Y_b^n|W_{bb,u_b^+},W_{ab,u_b^-}) - n\log(2\pi e) 
		- \sum_{\ell=1}^{g_{ab}} \Lambda_{a,\ell} \right)
		+ \left(\sum_{\ell=1}^{u_b^+} \Lambda_{b,\ell}
		+ \sum_{\ell=1}^{u_b^-} \Lambda_{a,\ell}\right)
		\label{eq:Gauss_genie_dec_b_final},
\end{align}
where $W_{ij,0}$ for $i,j\in\{a,b\}$ exist only as dummy (independent of the system or constant) signals.
The expressions (\ref{eq:Gauss_genie_dec_a_final}) and (\ref{eq:Gauss_genie_dec_b_final}) will later be used in Section~\ref{sec:gauss} to extend results for the LDIC to results for associated GICs. The similarities between the decoupling of signal levels in the LDIC and genie-aided decomposition for the GIC will permit similar analysis for both while clarifying how to account for the slack between the two models.

\pdfoutput=1
\section{Results for the Linear Deterministic Interference Channel}
\label{sec:lindet_results}

In this section we state results for the linear deterministic interference channel. The TDM-dominating capacity regions for each of the seven views shown in Figure~\ref{fig:VIEWS} falls in one of two categories. In the first category, we have Views~1 and 2 which enable opportunistic HK codes, thereby achieving rates dominating $\mathcal{R}_D^\mathsf{TDM}$. In the second category, containing Views~3--7, to achieve any point in the TDM-dominating capacity region, a TDM scheme is sufficient.

Before stating the results we draw the reader's attention to relationships between views shown in Figure~\ref{fig:VIEWS}. The chart from top to bottom depicts a partial ordering (considering the power set of the set of links and ordered by inclusion). Ordering relationships are depicted by directed edges (or paths along directed edges) with the view at the tail of the edge (or path) preferable to the view at the head of an edge: The transmitters lose knowledge of one link for each hop on a path.

Since a less preferable view has only a subset of the knowledge available to more preferable views, intuitively one might assume that, for a given channel state, the TDM-dominating capacity region of a particular local view IC may be bounded by the capacity region of any more preferable view. This is indeed the case, and simplifies the analysis of the many local views. Consequently, we need only analyze Views~1, 2, and 3 in full detail, and subsequently apply the results to Views~4--7.

Our capacity region characterizations are expressed as parameterizations based on potential policies, and highlight coupling of policy responses in different channel states. When appropriate, we also include a more concise set of inequalities stemming from the union over all such policies. Proofs are relegated to the appendices.

To further provide intuition as to why each view either enables or inhibits opportunities for advanced transmission schemes, for Views~1--3 we use the LDIC shown in Figure~\ref{fig:lindet_IC} and either define a policy that strictly dominates TDM, or demonstrate why strictly dominating TDM is impossible.

\subsection{Opportunity-Enabling Local Views}

\subsubsection{View 1}

The local view capacity region for View~1 may strictly dominate TDM, however the achievable region may be coupled across many states.

\begin{theorem}[View 1 LV-IC Capacity Region]
Let $\widehat{G}_a = (g_{aa},g_{ab},\varnothing,g_{bb})$ and $\widehat{G}_b = (g_{aa},\varnothing,g_{ba},g_{bb})$.
Additionally, WLOG let $g_{aa}\geq g_{bb}$, and define for a specific channel state $G = (g_{aa},g_{ab},g_{ba},g_{bb})$ the following value
\begin{align}
 \delta ={}& g_{aa} - g_{bb}.
\end{align}

The TDM-dominating capacity region for View~1 with channel state $G$ is the union of all regions indexed by $\tau_{b}(g_{aa},g_{bb}) \in[0,1]$, each region containing all tuples $(r_a(\widehat{G}_a),r_b(\widehat{G}_b))$ satisfying
\begin{align}
r_a(\widehat{G}_a) \leq r_a^c(\widehat{G}_a) + r_a^p(\widehat{G}_a),\\
r_b(\widehat{G}_b) \leq r_b^c(\widehat{G}_b) + r_b^p(\widehat{G}_b),
\end{align}
where
\begin{align}
 r_a^c(\widehat{G}_a)\geq{}&0\\
 r_a^p(\widehat{G}_a)\geq{}&0\\
 r_b^c(\widehat{G}_b)\geq{}&0\\
 r_b^p(\widehat{G}_b)\geq{}&0\\
 r_a\left(\widehat{G}_a\right)
	\leq{}& g_{aa} - g_{bb}\tau_b(g_{aa},g_{bb})\label{eq:view1_thm_1},\\
 r_b\left(\widehat{G}_b\right)
	\leq{}& g_{aa}\tau_b(g_{aa},g_{bb})\label{eq:view1_thm_2},\\
 {r}_a^c\left(\widehat{G}_a\right)
	\leq{}& \min_{\ell\geq0}[
		\max\{g_{bb}-\ell\delta,g_{ab}\} + \ell\delta\tau_b(g_{aa},g_{bb}) - g_{bb}\tau_b(g_{aa},g_{bb})
	]\label{eq:view1_thm_3},\\
 {r}_a^c\left(\widehat{G}_a\right) 
	\leq{}& g_{ab}\label{eq:view1_thm_4},\\
 {r}_b^c\left(\widehat{G}_b\right)
	\leq{}& g_{aa}\tau_b(g_{aa},g_{bb})\label{eq:view1_thm_5},\\
 {r}_b^c\left(\widehat{G}_b\right)
	\leq{}& \min_{\ell\geq0}[
		\max\{g_{ba}-\ell\delta,(g_{ba}-g_{aa})^+\} + \ell\delta\tau_b(g_{aa},g_{bb})
	]\label{eq:view1_thm_6},\\
 {r}_a^c\left(\widehat{G}_a\right) +r_b\left(\widehat{G}_b\right)
	\leq{}& \min_{\ell\geq0}[
		\max\{g_{bb}-\ell\delta,g_{ab}\} + \ell\delta\tau_b(g_{aa},g_{bb})
	]\label{eq:view1_thm_7},\\
 {r}_b^c\left(\widehat{G}_b\right) + r_a\left(\widehat{G}_a\right)
	\leq{}& \max\{g_{ba},g_{aa}\}\label{eq:view1_thm_8},\\
 {r}_b^c\left(\widehat{G}_b\right) + r_b\left(\widehat{G}_b\right)
	\leq{}& \min_{\ell\geq0}[
		\max\{g_{ba}-(\ell+1)\delta,(g_{ba}-g_{aa})^+\} + (g_{aa} + \ell\delta)\tau_b(g_{aa},g_{bb})
	]\label{eq:view1_thm_9},\\
 {r}_a^c\left(\widehat{G}_a\right) + r_b^p\left(\widehat{G}_b\right)
	\leq{}& \min_{\ell\geq0}[
		\max\{g_{ab},g_{bb}-g_{ba}-\ell\delta\} + \ell\delta\tau_b(g_{aa},g_{bb})
	]\label{eq:view1_thm_10},\\
 {r}_b^c\left(\widehat{G}_b\right) + r_a^p\left(\widehat{G}_a\right)
	\leq{}& \min_{\ell\geq0}[
		\max\{g_{ba}-\ell\delta,(g_{aa}-g_{ab})^+,g_{ba}-g_{ab},g_{ba}-g_{aa}\} + \ell\delta\tau_b(g_{aa},g_{bb})
	]\label{eq:view1_thm_11},\\
 r_a^p\left(\widehat{G}_a\right)
	\leq{}& \min_{\ell\geq0}[
		\max\{(g_{aa} - g_{ab})^+,g_{bb}-\ell\delta\} - g_{bb}\tau_b(g_{aa},g_{bb}) + \ell\delta\tau_b(g_{aa},g_{bb})
	]\label{eq:view1_thm_12},\\
 r_a^p\left(\widehat{G}_a\right)
	\leq{}& (g_{aa} - g_{ab})^+\label{eq:view1_thm_13},\\
 r_b^p\left(\widehat{G}_b\right)
	\leq{}& \min_{\ell\geq0}[
		(g_{bb} - g_{ba} - \ell\delta)^+ +\ell\delta\tau_b(g_{aa},g_{bb})
	].\label{eq:view1_thm_14}
\end{align}
\label{thm:view1}
\end{theorem}

Though the parameterized characterization of the region is somewhat unwieldy, each expression in (\ref{eq:view1_thm_1})--(\ref{eq:view1_thm_14}) results from a particular class of virtual Z-channels. Moreover, minimization over $\ell_i$ in an expression actually describes at most two ``worst-cases''. Which of the two cases is truly worst depends on the value of $\tau_b(g_{aa}, g_{bb})$ and the channel state $G$.

Though it is possible to state the capacity region in a non-parametric form (through Fourier-Motzkin elimination applied categorically for each of many different regimes), such a presentation is extremely unwieldy does little to illuminate the form of the capacity region. 
%
We find it more illustrative to provide an example of a policy strictly dominating TDM in the example channel (Figure~\ref{fig:v1_result_ex}). Transmitter~$b$ transmits at full rate ($r_b(\widehat{G}_b) = g_{bb} = 2$) while Transmitter~$a$ uses a HK code where the common message has rate $r_{a,c} = g_{ab} - g_{bb} = 1$. \emph{All three interfering signal levels} are used in a codebook drawn from a random distribution, which can be interpreted as the most significant bit carrying a one-bit message, and the next two signal levels providing parity.
A private message is encoded over all $g_{aa}-g_{ab}=4$ private signal levels at rate $r_{a,p} = 2$. Regardless of the value of $g_{ba}$ (which is unknown to Transmitter~$a$) the component rates at Receiver~$a$ satisfy (\ref{eq:rates_lindet_HK_1})--(\ref{eq:rates_lindet_HK_8}), so the rate point $(r_{a,c}+r_{a,p},r_{b,c}) = (3,2)$ is achievable. We see that 
\begin{align*}
 \frac{r_{a}(\widehat{G}_a)}{g_{aa}} +\frac{r_{b}(\widehat{G}_b)}{g_{bb}} ={}& \frac{3}{7} + \frac{2}{2}
	= \frac{10}{7} > 1.
\end{align*}

\subsubsection{View 2}

For View~2, each transmitter is aware of which of its signal levels may be causing interference, and which may be interfered with. Thus, each transmitter may opportunistically align bits to appropriate signal levels. Note that in the parametrized characterization, the constraint on rate for each transmitter is independent of the unknown link gain. This is because the worst case(s) --- which results in TDM even in the full view scenario --- has already been considered, and additional bits are gained through opportunism.
\begin{theorem}[View 2  LV-IC Capacity Region] 
If $\widehat{G}_a = (g_{aa},g_{ab},g_{ba},\varnothing)$ and $\widehat{G}_b = (\varnothing,g_{ab},g_{ba},g_{bb})$,
then the TDM-dominating capacity region is given by the union of all regions indexed by values $\tau_a$ and $\tau_b$, with $\tau_{a}(g_{ab},g_{ba}) + \tau_{b}(g_{ab},g_{ba}) = 1$, such that
\begin{align}
r_a(\widehat{G}_a) \leq{}& g_{aa}\label{eq:view2_parreg_1},\\
r_a(\widehat{G}_a) \leq{}& (g_{aa}-g_{ab})^+ +g_{ab}\tau_a(g_{ab},g_{ba})\label{eq:view2_parreg_2},\\
r_a(\widehat{G}_a) \leq{}& (g_{aa}-g_{ba})^+ +g_{ba}\tau_a(g_{ab},g_{ba})\label{eq:view2_parreg_3},\\
r_a(\widehat{G}_a) \leq{}& (g_{aa}-g_{ab}-g_{ba})^+ +(g_{ab}+ g_{ba})\tau_a(g_{ab},g_{ba})\label{eq:view2_parreg_4},\\
r_b(\widehat{G}_b) \leq{}& g_{bb},\\
r_b(\widehat{G}_b) \leq{}& (g_{bb}-g_{ba})^+ +g_{ba}\tau_b(g_{ab},g_{ba}),\\
r_b(\widehat{G}_b) \leq{}& (g_{bb}-g_{ab})^+ +g_{ab}\tau_b(g_{ab},g_{ba}),\\
r_b(\widehat{G}_b) \leq{}& (g_{bb}-g_{ab}-g_{ba})^+ +(g_{ab}+ g_{ba})\tau_b(g_{ab},g_{ba}).
\end{align}
\label{thm:view2}
\end{theorem}
In general, this region cannot be achieved with a simple orthogonalized scheme. The details of the scheme are more rigorously explained within the proof, but it is essentially a linear deterministic analogue of the approach used in \cite{ETW:08}. Knowledge of the outgoing link enables each transmitter to split its message into a public and private component, where the public message is coded using channel inputs that interfere with the other transmission. The private message is sent on the remaining inputs of the direct link, essentially ``hidden in the noise floor'' of the other receiver. Each receiver treats the desired public and private messages and the public message of the other user as a virtual MAC, and jointly decodes the components. 

The similarity to the result of \cite{ETW:08} also extends to the non-parametric characterization of the region.
\begin{corollary}
The View~2 capacity region consists of all non-negative rate points satisfying
\begin{align}
 r_a(\widehat{G}_a) \leq{}&g_{aa}\label{eq:reg_view2_1},\\
 r_b(\widehat{G}_b) \leq{}&g_{bb}\label{eq:reg_view2_2},\\
 r_a(\widehat{G}_a) + r_b(\widehat{G}_b) \leq{}& (g_{aa}-g_{ba})^+ + \max\{g_{bb},g_{ba}\}\label{eq:reg_view2_3},\\
 r_a(\widehat{G}_a) + r_b(\widehat{G}_b) \leq{}& (g_{bb}-g_{ab})^+ + \max\{g_{aa},g_{ab}\}\label{eq:reg_view2_4},\\
 r_a(\widehat{G}_a) + r_b(\widehat{G}_b) \leq{}& \max\{g_{ab},g_{aa}-g_{ba}\} + \max\{g_{ba},g_{bb}-g_{ab}\}\label{eq:reg_view2_5},\\
 \frac{g_{ab} + g_{ba}}{g_{ba}}r_a(\widehat{G}_a) + r_b(\widehat{G}_b) 
	\leq{}& \frac{g_{ab}}{g_{ba}}\max\{g_{aa},g_{ba}\} + (g_{aa}-g_{ba})^+ + \max\{g_{ba},g_{bb}-g_{ab}\}\label{eq:reg_view2_6},\\
 r_a(\widehat{G}_a) + \frac{g_{ab} + g_{ba}}{g_{ba}}r_b(\widehat{G}_b) 
	\leq{}& \frac{g_{ab}}{g_{ba}}\max\{g_{bb},g_{ba}\} + (g_{bb}-g_{ba})^+ + \max\{g_{ba},g_{aa}-g_{ab}\}\label{eq:reg_view2_7},\\
 \frac{g_{ab} + g_{ba}}{g_{ab}}r_a(\widehat{G}_a) + r_b(\widehat{G}_b)
	\leq{}& \frac{g_{ba}}{g_{ab}}\max\{g_{aa},g_{ab}\} + (g_{aa}-g_{ab})^+ + \max\{g_{ab},g_{bb}-g_{ba}\}\label{eq:reg_view2_8},\\
 r_a(\widehat{G}_a) + \frac{g_{ab} + g_{ba}}{g_{ab}}r_b(\widehat{G}_b) 
	\leq{}& \frac{g_{ba}}{g_{ab}}\max\{g_{bb},g_{ab}\} + (g_{bb}-g_{ab})^+ + \max\{g_{ab},g_{aa}-g_{ba}\}.\label{eq:reg_view2_9}
\end{align}
\end{corollary}
The region specified in (\ref{eq:reg_view2_1})--(\ref{eq:reg_view2_9}) results from Fourier-Motzkin elimination~\cite{NetwInfoTh:bk} of the time sharing parameters $\tau_a$ and $\tau_b$ and is not included for brevity. 

Notice that the Fourier-Motzkin eliminated region for View~2, (\ref{eq:reg_view2_1})--(\ref{eq:reg_view2_9}), is similar to the full view capacity region, (\ref{eq:reg_view0_1})--(\ref{eq:reg_view0_7}). Specifically, (\ref{eq:reg_view2_1})--(\ref{eq:reg_view2_5}) match (\ref{eq:reg_view0_1})--(\ref{eq:reg_view0_5}) exactly. Of the remaining inequalities, if $g_{ab} = g_{ba}$, then (\ref{eq:reg_view2_6}) and (\ref{eq:reg_view2_7}) are equivalent to (\ref{eq:reg_view2_8}) and (\ref{eq:reg_view2_9}) as well as the inequalities (\ref{eq:reg_view0_6}) and (\ref{eq:reg_view0_7}) of the full view case. These inequalities characterize the \emph{only loss from the full view capacity region} when transmitters are provided with View~2, and if either $g_{ab}=g_{ba}$ or bounds (\ref{eq:reg_view2_6})--(\ref{eq:reg_view2_9}) are dominated by the sum-rate bounds, then the View~2 region and the full view region coincide. However, this does not imply a lack of operational loss between a full view and View~2. The parametrized 
characterization of Theorem~\ref{thm:view2} highlights the lack of system flexibility needed to achieve points within this region. 

For an example of a policy outperforming TDM in the example channel from Figure~\ref{fig:lindet_IC}, allow Transmitter~$b$ to always transmit at full rate ($r_b(\widehat{G}_b) = g_{bb} = 2$). Transmitter~$a$ uses a HK code, however the common message is constrained to rate $r_{a,c} = 0$. 
The private message is encoded over all the top two of the non-interfering signal levels ($X_{a,4}$ and $X_{a,5}$) at rate $r_{a,p} = 2$. Receiver~$a$ treats the interference as noise and decodes the private message. Consequently, we have
\begin{align*}
 \frac{r_{a}(\widehat{G}_a)}{g_{aa}} +\frac{r_{b}(\widehat{G}_b)}{g_{bb}} ={}& \frac{2}{7} + \frac{2}{2}	= \frac{9}{7} > 1,
\end{align*}
as desired. The coding scheme described is depicted below in Figure~\ref{fig:v2_result_ex}.

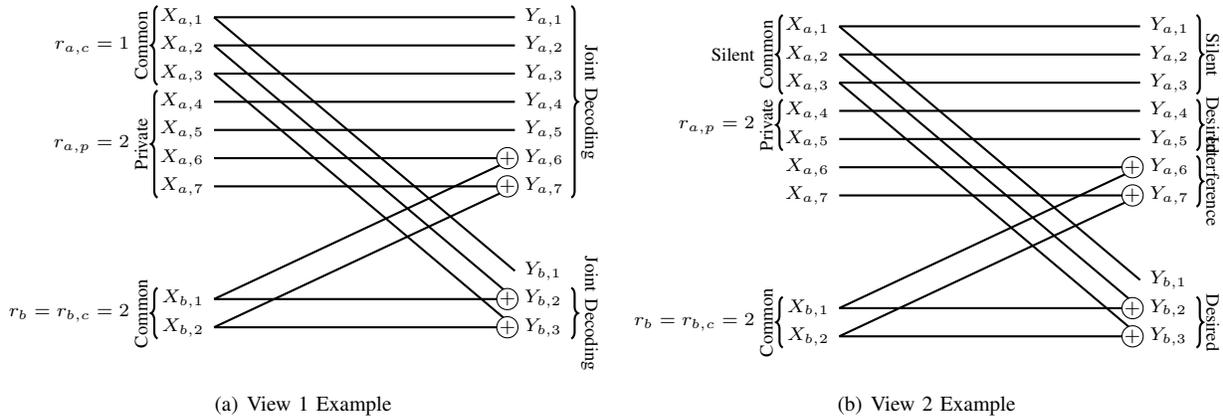
\begin{figure}[ht]
	\begin{center}
\subfigure[View 1 Example]{
\begin{tikzpicture}[font=\scriptsize,yscale=0.75]
\draw[white] (0,-0.25) -- (0,1);
		\draw[thick](0,6) -- (4,1.5);
		\draw[thick](0,5.5) -- (4,1);
		\draw[thick](0,5) -- (4,0.5);

		\draw[thick](0,1) -- (4,3.5);
		\draw[thick](0,0.5) -- (4,3);

		\draw[thick] (4,1.5) node[right] {$Y_{b,1}$};
		\draw[thick](0,1) node[left] {$X_{b,1}$} -- (4,1) node[right] {$Y_{b,2}$};
		\draw[thick](0,0.5) node[left] {$X_{b,2}$} -- (4,0.5) node[right] {$Y_{b,3}$};

		\draw[thick](0,6) node[left] {$X_{a,1}$} -- (4,6) node[right] {$Y_{a,1}$};
		\draw[thick](0,5.5) node[left] {$X_{a,2}$} -- (4,5.5) node[right] {$Y_{a,2}$};
		\draw[thick](0,5) node[left] {$X_{a,3}$} -- (4,5) node[right] {$Y_{a,3}$};
		\draw[thick](0,4.5) node[left] {$X_{a,4}$} -- (4,4.5) node[right] {$Y_{a,4}$};
		\draw[thick](0,4) node[left] {$X_{a,5}$} -- (4,4) node[right] {$Y_{a,5}$};
		\draw[thick](0,3.5) node[left] {$X_{a,6}$} -- (4,3.5) node[right] {$Y_{a,6}$};
		\draw[thick](0,3) node[left] {$X_{a,7}$} -- (4,3) node[right] {$Y_{a,7}$};
		
		\draw (3.9,3) node[fill=white,draw,circle, inner sep=0pt]{$+$};
		\draw (3.9,3.5) node[fill=white,draw,circle, inner sep=0pt]{$+$};
		\draw (3.9,1) node[fill=white,draw,circle, inner sep=0pt]{$+$};
		\draw (3.9,0.5) node[fill=white,draw,circle, inner sep=0pt]{$+$};
		
		\draw [thick,decorate,decoration={brace}] (-0.75,4.8) -- node[sloped,above]{Common} (-0.75,6.2);
		\draw [thick,decorate,decoration={brace}] (-0.75,2.8) -- node[sloped,above]{Private} (-0.75,4.7);

		\draw [thick,decorate,decoration={brace}] (-0.75,0.3) -- node[sloped,above]{Common} (-0.75,1.2);

		\draw (-1,0.75) node[left]{$r_b = r_{b,c} = 2$};
		\draw (-1,5.5) node[left]{$r_{a,c} = 1$};
		\draw (-1,3.75) node[left]{$r_{a,p} = 2$};
		
		\draw [thick,decorate,decoration={brace}] (4.75,6.2) -- node[sloped,above]{Joint Decoding} (4.75,2.8);
		\draw [thick,decorate,decoration={brace}] (4.75,1.2) -- node[sloped,above]{Joint Decoding} (4.75,0.3);
		
\end{tikzpicture}
\label{fig:v1_result_ex}
	}\subfigure[View 2 Example]{
\begin{tikzpicture}[font=\scriptsize,yscale=0.75]
\draw[white] (0,-0.25) -- (0,1);
		\draw[thick](0,6) -- (4,1.5);
		\draw[thick](0,5.5) -- (4,1);
		\draw[thick](0,5) -- (4,0.5);

		\draw[thick](0,1) -- (4,3.5);
		\draw[thick](0,0.5) -- (4,3);

		\draw[thick] (4,1.5) node[right] {$Y_{b,1}$};
		\draw[thick](0,1) node[left] {$X_{b,1}$} -- (4,1) node[right] {$Y_{b,2}$};
		\draw[thick](0,0.5) node[left] {$X_{b,2}$} -- (4,0.5) node[right] {$Y_{b,3}$};

		\draw[thick](0,6) node[left] {$X_{a,1}$} -- (4,6) node[right] {$Y_{a,1}$};
		\draw[thick](0,5.5) node[left] {$X_{a,2}$} -- (4,5.5) node[right] {$Y_{a,2}$};
		\draw[thick](0,5) node[left] {$X_{a,3}$} -- (4,5) node[right] {$Y_{a,3}$};
		\draw[thick](0,4.5) node[left] {$X_{a,4}$} -- (4,4.5) node[right] {$Y_{a,4}$};
		\draw[thick](0,4) node[left] {$X_{a,5}$} -- (4,4) node[right] {$Y_{a,5}$};
		\draw[thick](0,3.5) node[left] {$X_{a,6}$} -- (4,3.5) node[right] {$Y_{a,6}$};
		\draw[thick](0,3) node[left] {$X_{a,7}$} -- (4,3) node[right] {$Y_{a,7}$};
		
		\draw (3.9,3) node[fill=white,draw,circle, inner sep=0pt]{$+$};
		\draw (3.9,3.5) node[fill=white,draw,circle, inner sep=0pt]{$+$};
		\draw (3.9,1) node[fill=white,draw,circle, inner sep=0pt]{$+$};
		\draw (3.9,0.5) node[fill=white,draw,circle, inner sep=0pt]{$+$};
		
		\draw [thick,decorate,decoration={brace}] (-0.75,4.8) -- node[sloped,above]{Common} (-0.75,6.2);
		\draw [thick,decorate,decoration={brace}] (-0.75,3.8) -- node[sloped,above]{Private} (-0.75,4.7);

		\draw [thick,decorate,decoration={brace}] (-0.75,0.3) -- node[sloped,above]{Common} (-0.75,1.2);
		
		\draw (-1,0.75) node[left]{$r_b = r_{b,c} = 2$};
		\draw (-1,5.5) node[left]{Silent};
		\draw (-1,4.25) node[left]{$r_{a,p} = 2$};
		\draw [thick,decorate,decoration={brace}] (4.75,6.2) -- node[sloped,above]{Silent} (4.75,4.8);
		\draw [thick,decorate,decoration={brace}] (4.75,4.7) -- node[sloped,above]{Desired} (4.75,3.8);
		\draw [thick,decorate,decoration={brace}] (4.75,3.7) -- node[sloped,above]{Interference} (4.75,2.8);
		\draw [thick,decorate,decoration={brace}] (4.75,1.2) -- node[sloped,above]{Desired} (4.75,0.3);
		
\end{tikzpicture}
\label{fig:v2_result_ex}
	}
	\end{center}
	\caption{Policy dictated schemes responding to  \subref{fig:v1_result_ex} View~1, and \subref{fig:v2_result_ex} View~2. In each, the view provides enough information to achieve a rate point strictly dominating TDM.}
	\label{fig:beatTDM}
\end{figure}

\subsection{TDM-Optimal Local Views }

In each of the remaining 5 views considered, performance strictly dominating TDM is not possible. However, we separate our statements of the remaining results to emphasize the distinction between views that still facilitate some degree of transmission coordination. 

\subsubsection{Views 3 \& 5}
Although Views~3 and 5 do not permit use of strictly TDM-dominating policies, both may capitalize on common knowledge of $g_{aa}$ and $g_{bb}$ to adjust which point on the TDM region boundary is used.
\begin{theorem}[Views 3 \& 5  LV-IC  Capacity Region] 
If either $\widehat{G}_a = (g_{aa},\varnothing,g_{ba},g_{bb})$ and $\widehat{G}_b = (g_{aa},g_{ab},\varnothing,g_{bb})$ 
or $\widehat{G}_a = (g_{aa},\varnothing,\varnothing,g_{bb})$ and $\widehat{G}_b = (g_{aa},\varnothing,\varnothing,g_{bb})$,
then the TDM-dominating capacity region consists of all tuples $r_a(\widehat{G}_a),r_b(\widehat{G}_b))$ such that
\begin{align}
 r_a(\widehat{G}_a) \leq{}& g_{aa}\tau_{a}(g_{aa},g_{bb}),\\
 r_b(\widehat{G}_b) \leq{}& g_{bb}\tau_{b}(g_{aa},g_{bb}),
\end{align}
 with $\tau_{a}(g_{aa},g_{bb}) + \tau_{b}(g_{aa},g_{bb}) = 1$.
\label{thm:views35}
\end{theorem}

%
The TDM-dominating capacity region for View~3 may be the most negative finding of this work. Despite \emph{almost} complete knowledge of the network, transmitters are unable to strictly dominate TDM, which suggests that the costs associated with acquiring knowledge of the incoming interference and the other direct link were wasted.

To see why this is the case, we again refer to the channel in Figure~\ref{fig:lindet_IC}. First, we note that in order for a policy to strictly dominate TDM, there must exist some $\tau_a^\mathsf{min}$ and $\tau_b^\mathsf{min}$, where $\tau_a^\mathsf{min} + \tau_b^\mathsf{min}=1$, such that $r_a(\widehat{G}_a)\geq \tau_a^\mathsf{min}g_{aa}$ and $r_b(\widehat{G}_b)\geq \tau_b^\mathsf{min}g_{bb}$. 

Under View~3, each transmitter does not know its outgoing interference gain, but does know the direct gain of the other link.
Consider the POV of Transmitter~$a$ under the possibility $g_{ab}^\prime=1$ ($G^\prime = (g_{aa},g_{ab}^\prime=1,g_{ba},g_{bb})$). From (\ref{eq:decom_bound_b}), we have 
\begin{align}
n\tau_b^\mathsf{min}g_{bb} \leq{}& r_b(\widehat{G}_b^{(1)})\\
	\leq{}&I(\mathbf{X}_b^n;\mathbf{Y}_b^n)\\
	\leq{}& n \min\{g_{bb},g_{ab}\} - \sum_{k=1}^{g_{ab}}L_{a,k}(\widehat{G}_a) + \sum_{j=1}^{u_b^+}L_{b,j}(\widehat{G}_b^{(1)}) + \sum_{i=1}^{u_b^-}L_{a,i}(\widehat{G}_a)\\
	\leq{}& n \min\{2,1\} - \sum_{k=1}^{1}L_{a,k}(\widehat{G}_a) + \sum_{j=1}^{1}L_{b,1}(\widehat{G}_b^{(1)})\\
	\leq{}& n - L_{a,1}(\widehat{G}_a) + L_{b,1}(\widehat{G}_b^\prime).
\end{align}

Sweeping across a range of possible (from the POV of Transmitter~$a$) channels, the set $\{1,2,3,4,5,6,7\}$, we have
\begin{align}
2\tau_b^\mathsf{min} \leq{}& \frac{1}{n}\left( L_{b,1}\left(\widehat{G}_b^\prime\right) + n - L_{a,1}\left(\widehat{G}_a\right)\right) \label{eq:view3_ex_1},\\
2\tau_b^\mathsf{min} \leq{}& \frac{1}{n}\left(2 n - L_{a,1}\left(\widehat{G}_a\right) - L_{a,2}\left(\widehat{G}_a\right)\right) \label{eq:view3_ex_2},\\
2\tau_b^\mathsf{min} \leq{}& \frac{1}{n}\left(2 n - L_{a,2}\left(\widehat{G}_a\right) - L_{a,3}\left(\widehat{G}_a\right)\right) \label{eq:view3_ex_3},\\
2\tau_b^\mathsf{min} \leq{}& \frac{1}{n}\left(2 n - L_{a,3}\left(\widehat{G}_a\right) - L_{a,4}\left(\widehat{G}_a\right)\right) \label{eq:view3_ex_4},\\
2\tau_b^\mathsf{min} \leq{}& \frac{1}{n}\left(2 n - L_{a,4}\left(\widehat{G}_a\right) - L_{a,5}\left(\widehat{G}_a\right)\right) \label{eq:view3_ex_5},\\
2\tau_b^\mathsf{min} \leq{}& \frac{1}{n}\left(2 n - L_{a,5}\left(\widehat{G}_a\right) - L_{a,6}\left(\widehat{G}_a\right)\right) \label{eq:view3_ex_6},\\
2\tau_b^\mathsf{min} \leq{}& \frac{1}{n}\left(2 n - L_{a,6}\left(\widehat{G}_a\right) - L_{a,7}\left(\widehat{G}_a\right)\right). \label{eq:view3_ex_7}
\end{align}
The expressions (\ref{eq:view3_ex_1})--(\ref{eq:view3_ex_7}) already suggest a microcosm of an inability to strictly dominate TDM; every pair of signal levels consecutive in significance already simulate an orthogonalized scheme. By combining (\ref{eq:view3_ex_1}), (\ref{eq:view3_ex_3}), (\ref{eq:view3_ex_5}), and (\ref{eq:view3_ex_7}) gives us
\begin{equation}
\frac{1}{n}\sum_{i=1}^7 L_{a,i}\left(\widehat{G}_a\right) \leq \frac{1}{n}L_{b,1}\left(\widehat{G}_b^\prime\right) -\tau_b^\mathsf{min} + 7\tau_a^\mathsf{min},\label{eq:view3_ex_8}
\end{equation}
and combining (\ref{eq:view3_ex_2}), (\ref{eq:view3_ex_4}), and (\ref{eq:view3_ex_6}) yields
\begin{equation}
\frac{1}{n}\sum_{i=1}^6 L_{a,i}\left(\widehat{G}_a\right) \leq 6\tau_a^\mathsf{min}.\label{eq:view3_ex_9}
\end{equation}

If we recall that $n\tau_a^\mathsf{min}g_{aa}\leq nr_a(\widehat{G}_a) \leq \sum_i L_{a,i}$, (\ref{eq:view3_ex_9}) becomes
\begin{align}
 \frac{1}{n}L_{b,1}\left(\widehat{G}_b^\prime\right) \geq \tau_b^\mathsf{min}\label{eq:view3_ex_10}.
\end{align}
Since we have not used Transmitter~$a$'s knowledge of $g_{ba}$, (\ref{eq:view3_ex_8})--(\ref{eq:view3_ex_10}) also hold for any other values of $g_{ba}$ as well including $g_{ba}^{\prime\prime}=1$. Let $G^{\prime\prime} = (g_{aa},g_{ab}^\prime,g_{ba}^{\prime\prime},g_{bb}) = (7,1,1,2)$. Then in another channel state, to decode reliably at Receiver~$a$, 
\begin{align}
r_a(\widehat{G}_a^{\prime\prime})\leq{}& I(\mathbf{X}_a;\mathbf{Y}_a)\\ 
  \leq{}& \frac{1}{n}\sum_{i=1}^6 L_{a,i}\left(\widehat{G}_a^{\prime\prime}\right) + 1 - \frac{1}{n} L_{b,1}\left(\widehat{G}_b^\prime\right)\\
 \leq{}& 6\tau_a^\mathsf{min} + 1 - \frac{1}{n} L_{b,1}\left(\widehat{G}_b^\prime\right).
\end{align}
Combining the fact $r_a(\widehat{G}_a^{\prime\prime})\geq 7\tau_a^\mathsf{min}$ and (\ref{eq:view3_ex_10}) implies $\frac{1}{n} L_{b,1}\left(\widehat{G}_b^\prime\right) = \tau_b^\mathsf{min}$, which with (\ref{eq:view3_ex_8}) proves
\begin{equation}
 r_a(\widehat{G}_a) = 7\tau_a^\mathsf{min} = \tau_a^\mathsf{min}g_{aa}.
\end{equation}

This example demonstrates how Transmitter~$a$'s inability to effectively align its interference signal prevents performance strictly dominating TDM. A similar series of arguments confirms that Transmitter~$b$ also is limited to the TDM rate, however we omit this in lieu of the more general proof in the Appendix.

In View~5, transmitters have even less knowledge than in View~3, and thus is bounded by the same level of performance. However, it is interesting to note that the knowledge common to both transmitters is all the knowledge available to each transmitter. This not only allows them to synchronize their decision, but also suggests that View~5 models a centralized compound IC. It is therefore worth mentioning that the extension to multiple states discussed in~\cite{RPV:09} results in the same conclusion for the View~5 LDIC.

\subsubsection{Views 4, 6, \& 7} As the three local views with the least knowledge of network state and no links commonly known to both transmitters, the following result confirms the intuition that TDM is a good policy.
\begin{theorem}[Views 4, 6, \& 7  LV-IC Capacity Regions] 
If the views are any of the following three cases
\begin{itemize}
 \item $\widehat{G}_a = (g_{aa},\varnothing,g_{ba},\varnothing)\text{ and }\widehat{G}_b = (\varnothing,g_{ab},\varnothing,g_{bb})$
 \item $\widehat{G}_a = (g_{aa},g_{ab},\varnothing,\varnothing)\text{ and }\widehat{G}_b = (\varnothing,\varnothing,g_{ba},g_{bb})$
 \item $\widehat{G}_a = (g_{aa},\varnothing,\varnothing,\varnothing)\text{ and }\widehat{G}_b = (\varnothing,\varnothing,\varnothing,g_{bb}),$
\end{itemize}
then the TDM-dominating capacity region consists of all tuples $r_a(\widehat{G}_a),r_b(\widehat{G}_b))$ such that
\begin{align}
 r_a(\widehat{G}_a) \leq{}& g_{aa}\tau_{a},\\
 r_b(\widehat{G}_b) \leq{}& g_{bb}\tau_{b},
\end{align}
with $\tau_{a} + \tau_{b} = 1$.

%
\label{thm:views467}
\end{theorem}

As we described while referencing Figure~\ref{fig:VIEWS}, policies relying on Views~4--7 can perform no better than views containing more information. Therefore, it is not surprising that no policy can strictly dominate TDM in any of these views.

because this occurs for both transmitters every signal level of Transmitter~$a$ is possibly aligned with every signal level of Transmitter~$b$, and therefore for such cases orthogonalization is optimal. 

\pdfoutput=1
\section{Results for the Gaussian IC}
\label{sec:gauss}
In this section, we extend our results to the Gaussian interference channel and characterize local view capacity regions to within a bounded gap, and comment upon the local view GDoF of the GIC. 

\subsection{Main Result}

\begin{theorem}[Approximate Capacity Regions of Local View Gaussian ICs]
Given the relationship between LDIC and GIC gains presented in (\ref{eq:GtoLD1})--(\ref{eq:GtoLD4}),
and assuming WLOG $g_{aa}\geq g_{bb}$, for Local View~$k$, the per-user gap between the TDM-dominating GIC capacity region and the TDM-dominating capacity region of its LDIC analogue is less than $\Delta_k$ bits where $\Delta_k$ is given in Table~\ref{tab:gauss_result}.
\label{thm:gauss}
\begin{table}[ht]
\centering
\begin{tabular}{ | c | c | c | c | } 
	\hline
	\textbf{View ($k$)} & \textbf{View Diagram} & $\Delta_k$ & Common Knowledge\\
	\hline 
	\hline
	\centering 1 &
		\begin{tikzpicture}[scale=0.25]
		\draw[blue,thick,-latex,shorten >=3pt,shorten <=2pt](0.125,6) -- (4.875,6);
		\draw[blue,thick,-latex,shorten >=3pt,shorten <=2pt](0.125,4) -- (4.875,4);
		\draw[blue,thick,-latex,shorten >=3pt,shorten <=2pt](0.125,6) -- (4.875,4);
		\draw[white](0,3.96)--(0,6.5);
		\end{tikzpicture}
		& 
		$\begin{cases}
			\log(6) +4 & \text{ if } g_{aa} = g_{bb}\cr
			\log(9) + 2\max\left\{2\left\lceil\frac{g_{bb}}{g_{aa}-g_{bb}}\right\rceil +1 , \left\lceil\frac{g_{ba}}{g_{aa}-g_{bb}}\right\rceil + \left\lceil\frac{(g_{bb}-g_{ba})^+}{g_{aa}-g_{bb}}\right\rceil \right\} + 4
			& \text{ else } \cr
		\end{cases}$ 
		& $g_{aa}$, $g_{bb}$
	\\
	\hline
	\centering 2 &
		\begin{tikzpicture}[scale=0.25]
		\draw[blue,thick,-latex,shorten >=3pt,shorten <=2pt](0.125,6) -- (4.875,6);
		\draw[blue,thick,-latex,shorten >=3pt,shorten <=2pt](0.125,4) -- (4.875,6);
		\draw[blue,thick,-latex,shorten >=3pt,shorten <=2pt](0.125,6) -- (4.875,4);
		\draw[white](0,3.96)--(0,6.5);
		\end{tikzpicture}
		& 
		$2\log(6) + \log(3) + 4$ 
		& $g_{ab}$, $g_{ba}$
	\\
	\hline
	\centering 3 &
		\begin{tikzpicture}[scale=0.25]
		\draw[blue,thick,-latex,shorten >=3pt,shorten <=2pt](0.125,6) -- (4.875,6);
		\draw[blue,thick,-latex,shorten >=3pt,shorten <=2pt](0.125,4) -- (4.875,6);
		\draw[blue,thick,-latex,shorten >=3pt,shorten <=2pt](0.125,4) -- (4.875,4);
		\draw[white](0,3.96)--(0,6.5);
		\end{tikzpicture}
		& 
		$\left(\frac{\lcm(g_{aa},g_{bb})}{g_{aa}} + \frac{\lcm(g_{aa},g_{bb})}{g_{bb}} - 1\right)\log(6)$ 
		& $g_{aa}$, $g_{bb}$
	\\
	\hline
	\centering 4 &
		\begin{tikzpicture}[scale=0.25]
		\draw[blue,thick,-latex,shorten >=3pt,shorten <=2pt](0.125,6) -- (4.875,6);
		\draw[blue,thick,-latex,shorten >=3pt,shorten <=2pt](0.125,6) -- (4.875,4);
		\draw[white](0,3.96)--(0,6.5);
		\end{tikzpicture}
		& 
		$\log(6)$ 
		& $\emptyset$
	\\
	\hline
	\centering 5 &
		\begin{tikzpicture}[scale=0.25]
		\draw[blue,thick,-latex,shorten >=3pt,shorten <=2pt](0.125,6) -- (4.875,6);
		\draw[blue,thick,-latex,shorten >=3pt,shorten <=2pt](0.125,4) -- (4.875,4);
		\draw[white](0,3.96)--(0,6.5);
		\end{tikzpicture}
		& 
		$\left(\frac{\lcm(g_{aa},g_{bb})}{g_{aa}} + \frac{\lcm(g_{aa},g_{bb})}{g_{bb}} - 1\right)\log(6)$ 
		& $g_{aa}$, $g_{bb}$
	\\
	\hline
	\centering 6 &
		\begin{tikzpicture}[scale=0.25]
		\draw[blue,thick,-latex,shorten >=3pt,shorten <=2pt](0.125,6) -- (4.875,6);
		\draw[blue,thick,-latex,shorten >=3pt,shorten <=2pt](0.125,4) -- (4.875,6);
		\draw[white](0,3.96)--(0,6.5);
		\end{tikzpicture}
		& 
		$\log(6)$ 
		& $\emptyset$
	\\
	\hline
	\centering 7 &
		\begin{tikzpicture}[scale=0.25]
		\draw[blue,thick,-latex,shorten >=3pt,shorten <=2pt](0.125,6) -- (4.875,6);
		\draw[white](0,3.96)--(0,6.5);
		\end{tikzpicture}
		&
		$\log(6)$ 
		& $\emptyset$
	\\
	\hline
	\end{tabular} 
	\caption{Per-user gap between GIC and LDIC TDM-dominating capacity regions.}
	\label{tab:gauss_result}
\end{table}
\end{theorem}

Note that for Views~1, 3, and 5 the gap is dependent on channel state and not universal in sense of prior work on approximate capacity. This stems from the interdependence between possible states resulting from nodes having local views and an outer bound based on Z-channel expansion. However, the gap is dependent on only relative channel gain, and under an appropriate scaling of link gains discussed in Section~\ref{sec:GDoF} the gap remains bounded by the same constant as SNR goes to infinity.

Detailed proofs for each view are located in Appendix~\ref{append:gauss}. In the remainder of this section we comment on the sources of gaps between models and intuitions that can be drawn from our result. Prior to explaining calculation of gaps between the regions given in Theorems~\ref{thm:view1}--\ref{thm:views467} and their GIC counterparts, we first recall from Section~\ref{sec:inner} that each scheme prescribed for every local view LDIC (either TDM or a simple HK based code) approximates an analogous achievable scheme for the GIC. At the same time, the rates achieved using each scheme dictated by LDIC policies may be approximately achieved using the analogous GIC scheme. Therefore, we do not explicitly define the policy for each local view GIC, but instead rely on results of the LDIC to characterize the capacity approximation gaps.

\subsection{Approximate Capacity}
\label{sec:approx}
To account for the gaps shown in Table~\ref{tab:gauss_result}, we use the results of the LDIC as an intermediate step in determining the gap from GIC achievable policy to GIC outer bound. Note that if TDM is the prescribed policy for the LDIC case (Views~3--7), the GIC TDM fully contains the analogous LDIC TDM region, and thus the rate prescribed by the LDIC policy is achievable. If the prescribed policy is based on a HK scheme (Views~1 and 2), we apply Lemma~\ref{lem:codegap} to bound the gap between rates achievable by analogous schemes in GIC and LDIC models to 4 bits per user. 

Because the LDIC characterization is tight, we proceed to bound the gap between LDIC and GIC outer bounds. We illustrate the impact of each of the two main features of the linear deterministic approximation of the Gaussian channel (quantization of a complex gain $h$ into an integer value $g$ and representation of a superposition of signals with modulo addition) by comparing the level-by-level decompositions of mutual information bounds for the LDIC and GIC.

Consider the first term in both (\ref{eq:decom_bound_a}) and (\ref{eq:Gauss_genie_dec_a_final}). In (\ref{eq:decom_bound_a}) the maximum entropy of the signal levels affected by both the desired and interfering signals is no higher than if maximizing the entropy of just one of the two signals. On the other hand, the analogous term in the Gaussian IC is not so easily bounded. If $g_{aa} > g_{ba} > 0$ we have
\begin{align}
 h\Big(Y_a^n & \Big|W_{aa,u_a^+},W_{ba,u_a^-} \Big) - n\log(2\pi e) \nonumber\\
	={}& h\left( Y_a^n \middle| W_{aa,u_a^+} \right)- n\log(2\pi e) \\
	={}& h\left( h_{aa}X_a^n + h_{ba}X_b^n + Z_a^n \middle| \max\left\{|h_{aa}|,|h_{ab}|\right\}\Phi_{aa} X_a^n + \sqrt{2^{\max(g_{aa},g_{ab})- (g_{aa} - g_{ba})}}Z^{n\prime}\right) - n\log(2\pi e) \\
%
	\stackrel{(a)}{\leq}{}& h\left( \left(|h_{aa}| - \sqrt{2^{g_{aa}}}\frac{\max\{|h_{aa}|,|h_{ab}|\}}{\sqrt{2^{\max\{g_{aa},g_{ab}\}}}}\right)\Phi_{aa}X_a^n + h_{ba}X_b^n + Z_a^n - \sqrt{2^{g_{ba}}}Z^{n\prime} \right)
	 - n\log(2\pi e) \\
	\stackrel{(b)}{\leq}{}& n\log\left(2\pi e\left[\left||h_{aa}| - \sqrt{2^{g_{aa}}}\frac{\max\{|h_{aa}|,|h_{ab}|\}}{\sqrt{2^{\max\{g_{aa},g_{ab}\}}}}\right|^2+ |h_{ba}|^2 + 1 + 2^{g_{ba}}\right]\right)- n\log(2\pi e) \\
	\stackrel{(c)}{\leq}{}& n\log\left(2\pi e\left[2+ |h_{ba}|^2 + 1 + 2^{g_{ba}}\right]\right)- n\log(2\pi e) \\
	\leq{}& n\log\left(2\pi e\left[2^{g_{ba}+1} + 3 + 2^{g_{ba}}\right]\right)- n\log(2\pi e) \\
	\leq{}& n\log\left(3(2^{g_{ba}}) + 3\right),
\end{align}
where in (a) we subtract the genie signal scaled by $\frac{\sqrt{2^{g_{aa}}}}{\sqrt{2^{\max\{g_{aa},g_{ab}\}}}}$, in (b) we invoke a maximum entropy argument, and in (c) we bound the variance of the uncanceled part of $X_a^n$. If $g_{aa} < g_{ba}$ then
\begin{align}
 h\Big(Y_a^n&\Big|W_{aa,u_a^+},W_{ba,u_a^-}\Big) - n\log(2\pi e) \nonumber\\
	={}& h\left(Y_a^n\middle|W_{ba,u_a^-}\right)- n\log(2\pi e) \\
	={}& h\left(h_{aa}X_a^n + h_{ba}X_b^n + Z_a^n \middle| \max\left\{|h_{bb}|,|h_{ba}|\right\}\Phi_{ba} X_b^n + \sqrt{2^{\max\{g_{bb},g_{ba}\}- (g_{ba} - g_{aa})}}Z^{n\prime}\right) 
		- n\log(2\pi e) \\
%
	\leq{}& h\left( h_{aa}X_a^n + \left(|h_{ba}| - \sqrt{2^{g_{ba}}}\frac{\max\{|h_{ba}|,|h_{bb}|\}}{\sqrt{2^{\max\{g_{ba},g_{bb}\}}}}\right)\Phi_{ba}X_b^n + Z_a^n - \sqrt{2^{g_{aa}}}Z^{n\prime} \right)
	 - n\log(2\pi e) \\
	\leq{}& n\log\left(2\pi e\left[|h_{aa}|^2 + \left||h_{ba}| - \sqrt{2^{g_{ba}}}\frac{\max\{|h_{ba}|,|h_{bb}|\}}{\sqrt{2^{\max\{g_{ba},g_{bb}\}}}}\right|^2 + 1 + 2^{g_{aa}}\right]\right)- n\log(2\pi e) \\
	\leq{}& n\log\left(2\pi e\left[|h_{aa}|^2 + 2 + 1 + 2^{g_{aa}}\right]\right)- n\log(2\pi e) \\
	\leq{}& n\log\left(2\pi e\left[2^{g_{aa}+1} + 3 + 2^{g_{aa}}\right]\right)- n\log(2\pi e) \\
	\leq{}& n\log\left(3(2^{g_{aa}}) + 3\right).
\end{align} 
Finally if $g_{aa} = g_{ba}$ then
\begin{align}
 h(Y_a^n|W_{aa,u_a^+},W_{ba,u_a^-}) - n\log(2\pi e) 
	={}& h(Y_a^n)- n\log(2\pi e) \\
	={}& h\left(h_{aa}X_a^n + h_{ba}X_b^n + Z_a^n\right) - n\log(2\pi e) \\
	\leq{}& n\log\left(|h_{aa}|^2 + |h_{ba}|^2 + 1\right)\\
	\leq{}& n\log\left(4(2^{g_{aa}}) + 1\right).
\end{align} 
We can upper bound the gap between this term and its LDIC counterpart over all channels:
\begin{align}
 h(Y_a^n & |W_{aa,u_a^+},W_{ba,u_a^-}) - n\log(2\pi e) - n\min\left\{g_{aa},g_{ba}\right\} \nonumber\\
	\leq{}& n\max\left\{\log\left(3\left(2^{\min\{g_{aa},g_{ba}\}}\right)+3\right) ,\log\left(4\left(2^{\min(g_{aa},g_{ba})}\right)+1\right)\right\} - n\min\left\{g_{aa},g_{ba}\right\}\\
	\leq{}& n\log\left(6\right),\label{eq:per_vir_user_gap}
\end{align} 
which implies that in each interference scenario considered, there may be up to $\log(6)$ bits per channel use that is not utilized by rates prescribed by the LDIC region.

This extra headroom is partially the result of the power gain (a multiple-access channel type of gain) that occurs when adding signals in the Gaussian model. Additionally, the quantized channel magnitudes in the linear deterministic model also incur a reduction in represented signal strength of both desired and interference signal components.

In the context of local view capacity analysis, this gap between the (tight) linear deterministic bound and our Gaussian outer bound exists for each interference scenario considered. Therefore, the larger the number of channel states that jointly constrain the rate of a transmitter (i.e., the number of virtual Z-channel links considered in establishing an outer bound) the larger the gap between LDIC and GIC capacity region boundaries. This is reflected in Views~4, 6 and 7, where due to extremely limited knowledge, only a single channel state can be established as a ``worst case'', and our gap is relatively small.

Our bound is admittedly not tight, and for certain cases (Views~2 and 5) application of existing techniques (\cite{ETW:08} and \cite{RPV:09} respectively) may result in smaller gaps. However, our analysis better parallels intuition imparted by the linear deterministic model applied in the local view setting.

\subsection{Generalized Degrees of Freedom}
\label{sec:GDoF}

Our capacity approximation gaps for GIC depend both on local view as well as channel state and thus are not universal. However, the notion of \emph{generalized degrees of freedom} (GDoF) characterizes the high SNR behavior of each view-channel scenario, thereby skirting the non-universality of gaps and presenting a clearer comparison between local view Gaussian interference channels at high SNR.

Let the parameter $\alpha = (\alpha_1,\alpha_2,\alpha_3)$ be defined as
\begin{align}
 \alpha_1 \triangleq {}& \frac{\log\left(|h_{bb}|^2\right)}{\log\left(|h_{aa}|^2\right)},\\
 \alpha_2 \triangleq {}& \frac{\log\left(|h_{ba}|^2\right)}{\log\left(|h_{aa}|^2\right)},\\
 \alpha_3 \triangleq {}& \frac{\log\left(|h_{ab}|^2\right)}{\log\left(|h_{aa}|^2\right)}.
\end{align}
and $\mathcal{C}(h_{aa},h_{ab},h_{ba},h_{bb})$ be the capacity region of the complex Gaussian IC given by gains $H = (h_{aa},h_{ab},h_{ba},h_{bb})$. The GDoF region was defined in~\cite{ETW:08} as
\begin{equation}
 \mathcal{D}(\alpha) = \lim_{\substack{|h|\rightarrow\infty \\ \alpha\text{ fixed}\\ \angle h_{ij}\text{ fixed }\forall i,j}\\}\left\{\left(\frac{r_a}{\log\left(|h_{aa}|^2\right)},\frac{r_b}{\log\left(|h_{bb}|^2\right)}:(r_a,r_b)\in \mathcal{C}\left(h_{aa},h_{ab},h_{ba},h_{bb}\right) \right)\right\}
\end{equation}
Although the exact impact of phase on the capacity region is unknown, because the one bit characterization of \cite{ETW:08} provided a phase-independent approximation of capacity the high-SNR characterization of GDoF can be tight without considering phase.

In order to make comments with regard to the GDoF of local view ICs, we note the following property regarding integer multiples channel gains of LDICs. The claim is due to the linearity of the model and can be verified by examination of expressions defining each region:
\begin{property}[Integer Multiples of Channel States]
If all the channel gains of a channel state $G^\prime$ can be expressed as an integer multiple of another state, $G^\prime = cG$ where $c$ is a positive integer, then the capacity region of the $G^\prime$ is the integer multiple of the capacity region of $G$. 
\begin{equation}
\mathcal{C}(G^\prime) = c\mathcal{C}(G). 
\end{equation}
\label{prop:lin}
\end{property}

With respect to the Gaussian channel, this property coupled with the gap analysis results allows us to comment on the generalized degrees of freedom (GDoF) region for each view:
\begin{corollary}[Local View GDoF Regions]
	Let $\alpha_1$, $\alpha_2$, and $\alpha_3$, be positive rational values. The GDoF of View~$k$ is given by
	\begin{equation}
	 \mathcal{D}_k(\alpha) = \left\{\left(\frac{r_a}{g_{aa}}, \frac{r_b}{g_{bb}}\right) : (r_a,r_b)\in \mathcal{C}_{D,k}(G)\right\},
	\end{equation}
	where $G$ is such that $\frac{g_{bb}}{g_{aa}} = \alpha_1$, $\frac{g_{ba}}{g_{aa}} = \alpha_2$, $\frac{g_{ab}}{g_{aa}} = \alpha_3$.
\end{corollary}
While this does not define the local view GDoF for all values of $\alpha$, because the rationals are dense in the reals, the local view GDoF can be found for an arbitrarily precise approximation of the parameter $\alpha$.
\pdfoutput=1
\section{Summary}
\label{sec:summary}

We studied the impact of incomplete, mismatched views of channel state on the capacity region of the two-user interference channel. We proposed a new formalization of capacity regions for local view ICs, and presented exact characterizations of the capacity region of the linear deterministic IC. One of our major conclusions shows the critical importance of each transmitter knowing its outgoing interference link in order to use advanced coding schemes to achieve performance strictly dominating TDM, providing critical insight into the engineering of wireless networks. Finally, we extended our work to the two-user Gaussian IC, where we found bounds on the gap between the Gaussian IC capacity region and its linear deterministic analogue, and used these results to comment on the generalized degrees of freedom region for the Gaussian IC with local views.

\appendix
\pdfoutput=1
\subsection{Proof for LV-MAC}
\label{append:LV-MAC}
\begin{IEEEproof}
%
From the inequalities defining the boundary of the MAC capacity region, we have
\begin{equation}
 \sum_{k=1}^K r_{k}\left(\widehat{G}_k\middle)\right|_{g_k=d} \leq d,
\end{equation}
where $h$ is any potential channel gain. Applying the minimum performance criterion yields
\begin{equation}
 1\leq \sum_{k=1}^K \frac{r_{k}\left(\widehat{G}_k\middle)\right|_{g_k=d}}{d} \leq 1,
\end{equation}
or
\begin{equation}
 \sum_{k=1}^K \frac{r_{k}\left(\widehat{G}_k\middle)\right|_{g_k=d}}{d} = 1.\label{eq:MACproof_1}
\end{equation}

Select $K$ non-negative integer values, $d_1,\ldots,d_K$, and notice
\begin{align}
	\sum_{k=1}^K \frac{r_{k}\left(\widehat{G}_k\middle)\right|_{g_k=d_k}}{d_k}
	={}&\sum_{k=1}^K \frac{r_{k}\left(\widehat{G}_k\middle)\right|_{g_k=d_k}}{d_k} 
		+ K - \sum_{\ell=1}^K\left( \sum_{k=1}^K \frac{r_{k}\left(\widehat{G}_k\middle)\right|_{g_k=d_\ell}}{d_\ell} \right)\\
	={}&K - \sum_{\ell=1}^{K-1}\left(
		\left(\sum_{k=1}^{\ell-1} \frac{r_{k}\left(\widehat{G}_k\middle)\right|_{g_k=d_\ell}}{d_\ell}\right) 
		+ \frac{r_{\ell}\left(\widehat{G}_\ell\middle)\right|_{g_\ell=d_K}}{d_K}
	+\left(\sum_{k=\ell+1}^K \frac{r_{k}\left(\widehat{G}_k\middle)\right|_{g_k=d_\ell}}{d_\ell}\right)\right).\label{eq:MACproof_2}
\end{align}
Applying the minimum performance criterion to the left hand side, as well as to each of the terms indexed by $\ell$ on the right hand side in (\ref{eq:MACproof_2}), we have
\begin{equation}
 1\leq \sum_{k=1}^K \frac{r_{k}\left(\widehat{G}_k\middle)\right|_{g_k=d_k}}{d_{k}} \leq K-(K-1) = 1.
\end{equation}
Since this holds for any non-negative values of $d_1,\ldots,d_K$ the theorem holds.
\end{IEEEproof}

\subsection{Proof for View 1}
\label{append:view1}
\begin{IEEEproof}
First, we clarify that the capacity-achieving policy is specifically catered to the channel state considered, and may require the rate point achieved in other channel states to be on the TDM boundary. 


\subsubsection*{Outer Bound}
For any policy satisfying the minimum performance criteria, by definition there must exist non-negative $\tau_a$ and $\tau_b$ such that for all $G$,
\begin{equation}
 \tau_{a}(g_{aa},g_{bb}) + \tau_{b}(g_{aa},g_{bb}) = 1,
\end{equation}
and
\begin{align}
 r_a(\widehat{G}_a)\geq g_{aa}\tau_{a}(g_{aa},g_{bb})\label{eq:view1_mpc_lb_a},\\
 r_b(\widehat{G}_b)\geq g_{bb}\tau_{b}(g_{aa},g_{bb}),\label{eq:view1_mpc_lb_b}
\end{align}
where system-wide parameters are allowed to depend on the common knowledge ($g_{aa}$ and $g_{bb}$).

Two virtual single Z-channels immediately result in bounds (\ref{eq:view1_thm_1}) and (\ref{eq:view1_thm_3}). At Receiver~$a$, if $g_{ba}=g_{bb}$ we apply (\ref{eq:decom_bound_a}) and find
\begin{align}
 nr_{a}\left(\widehat{G}_a\right) 
	\leq{}& \left( ng_{ba} - \sum_{j=1}^{g_{ba}}L_{b,j}\left(\widehat{G}_b\right)
		+ \sum_{i=1}^{g_{aa}-g_{ba}}L_{a,i}\left(\widehat{G}_a\right)\middle)\right|_{g_{ba}=g_{bb}}\\
	\leq{}& n\left( g_{bb} - r_b\left(\widehat{G}_b\middle)\right|_{g_{ba}=g_{bb}}\right) 
		+ \sum_{i=1}^{g_{aa}-g_{bb}}L_{a,i}\left(\widehat{G}_a\right)\\
	\leq{}& n\left[g_{aa} - \tau_b(g_{aa},g_{bb})g_{bb}\right].\label{eq:view1_proof_ob1}
\end{align}
Similarly, at Receiver~$b$ if $g_{ab}=g_{aa}$ we apply (\ref{eq:decom_bound_b})
\begin{align}
 nr_{b}\left(\widehat{G}_b\right) 
 	\leq{}& \left( ng_{bb} - \sum_{j=1}^{g_{ab}}L_{a,j}\left(\widehat{G}_a\right)
		+ \sum_{i=1}^{g_{ab}-g_{bb}}L_{a,i}\left(\widehat{G}_a\right)\middle)\right|_{g_{ab}=g_{aa}}\\
	\leq{}& n\left( g_{bb} - r_a\left(\widehat{G}_a\middle)\right|_{g_{ab}=g_{aa}}\right) 
		+ \sum_{i=1}^{g_{aa}-g_{bb}}L_{a,i}\left(\widehat{G}_a\right)\\
	\leq{}& n\left(g_{aa}-r_a\left(\widehat{G}_a\middle)\right|_{g_{ab}=g_{aa}}\right)\\
	\leq{}& n\left(g_{aa}-g_{aa}\tau_a(g_{aa},g_{bb})\right)\\
	\leq{}& n g_{aa}\tau_b(g_{aa},g_{bb}).\label{eq:view1_proof_ob2}
\end{align}

To arrive at the other bounds, we note in (\ref{eq:decom_bound_a}) that regardless of the incoming interference gain, $g_{ba}$, the following statement is necessary for achievability:
\begin{equation}
 nr_{a}(\widehat{G}_a) \leq \sum_{i=1}^{g_{ab}}L_{a,i}\left(\widehat{G}_a\right) + n\max(g_{aa}-g_{ab},g_{ba}) 
 - \sum_{j=1}^{g_{ba}}L_{b,j}\left(\widehat{G}_b\right).\label{eq:view1proof_1}
\end{equation}

In this expression, we draw a distinction between the entropy of the interference component and the entropy of the non-interfering component of the signal. To clarify analysis, we define the average entropies of the interference components of each transmitter's input as
\begin{align}
\overline{r}_{a}^{c}\left(\widehat{G}_a\right) 
	\triangleq{}& \frac{1}{n}\sum_{i=1}^{g_{ab}}L_{a,i}\left(\widehat{G}_a\right)\label{eq:view1proof_comp_a},\\
\overline{r}_{b}^{c}\left(\widehat{G}_b\right)
	\triangleq{}& \frac{1}{n}\sum_{j=1}^{g_{ba}}L_{b,j}\left(\widehat{G}_b\right).\label{eq:view1proof_comp_b}
\end{align}

Each transmitter's interference component and the non-interference component separately by constructing Z-channels both in the forward (adding virtual users that receive interference) and backward (adding virtual users that may induce interference) directions, as described in Section~\ref{sec:virtualZ}. 
Examples of the virtual Z-channels and their relation to specific bounds are shown in Figure~\ref{fig:view1_virtualZ}. 

First, consider Transmitter~$a$'s interference component. If the Z-channel terminates at the next signal, the following are two necessary conditions to guarantee achievability of rate $r_b(\widehat{G}_b^\prime)$.
\begin{align}
\overline{r}_a^c\left(\widehat{G}_a\right)
	\leq{}& \max\{g_{bb},g_{ab}\} - r_b\left(\widehat{G}_b^\prime\right)\label{eq:view1proof_2},\\
\overline{r}_a^c\left(\widehat{G}_a\right)
	\leq{}& \frac{1}{n} \sum_{j=1}^{g_{ba}^\prime}L_{b,j}\left(\widehat{G}_b^\prime\right) 
		+ \max\{g_{bb}-g_{ba}^\prime,g_{ab}\} - r_b\left(\widehat{G}_b^\prime\right)\\
	={}& \overline{r}_b^c\left(\widehat{G}_b^\prime\right) 
		+ \max\{g_{bb}-g_{ba}^\prime,g_{ab}\} - r_b\left(\widehat{G}_b^\prime\right).\label{eq:view1proof_3}
\end{align}
Similarly,
\begin{align}
\overline{r}_b^c\left(\widehat{G}_b\right) 
	\leq{}& \max\{g_{aa},g_{ba}\} - r_a\left(\widehat{G}_a^\prime\right)\label{eq:view1proof_4},\\
\overline{r}_b^c\left(\widehat{G}_b\right) 
	\leq{}& \overline{r}_a^c\left(\widehat{G}_a^\prime\right) 
		+ \max\{g_{aa}-g_{ab}^\prime,g_{ba}\} - r_a\left(\widehat{G}_a^\prime\right).\label{eq:view1proof_5}
\end{align}
We may expand the virtual Z-channel in the forward direction by substitution of (\ref{eq:view1proof_4}) or (\ref{eq:view1proof_5}) into (\ref{eq:view1proof_3}), or (\ref{eq:view1proof_2}) or (\ref{eq:view1proof_3}) into (\ref{eq:view1proof_5}). This expansion may be repeated indefinitely resulting in a virtual Z-channel that incorporates the possibility of many different channel states $G^{(1)}, G^{(2)}, \ldots$ and local views $\widehat{G}_a^{(1)},\widehat{G}_a^{(2)},\ldots$ and $\widehat{G}_b^{(1)},\widehat{G}_b^{(2)},\ldots$ seen by the respective virtual users. Expansion of (\ref{eq:view1proof_3}) in the forward direction, and indexing with $\theta$ each of $\Theta$ considered channel states, yields the following four families of bounds where $M\in\mathbb{Z}^+$:
\begin{align}
 \overline{r}_a^c\left(\widehat{G}_a\right) 
	\leq{}& \max\{g_{bb}-g_{ba}^{(1)},g_{ab}\} - r_b\left(\widehat{G}_b^{(1)}\right)\nonumber\\
		{}& + \sum_{\theta=1}^{\Theta-2}\left[ \max\{g_{aa}-g_{ab}^{(\theta)},g_{ba}^{(\theta)}\} -  \overline{r}_a\left(\widehat{G}_a^{(\theta)}\right)
		+ \max\{g_{bb}-g_{ba}^{(\theta+1)},g_{ab}^{(\theta)}\} - r_b\left(\widehat{G}_b^{(\theta+1)}\right)\right]\nonumber\\
		{}& + \max\{g_{aa}-g_{ab}^{(\Theta-1)},g_{ba}^{(\Theta-1)}\} -  \overline{r}_a\left(\widehat{G}_a^{(\Theta-1)}\right)
		+ \max\{g_{bb},g_{ab}^{(\Theta-1)}\} - r_b\left(\widehat{G}_b^{(\Theta)}\right)
		\label{eq:view1_exp_1},\\
 \overline{r}_a^c\left(\widehat{G}_a\right) 
	\leq{}& \max\{g_{bb}-g_{ba}^{(1)},g_{ab}\} - r_b\left(\widehat{G}_b^{(1)}\right)\nonumber\\
		{}& + \sum_{\theta=1}^{\Theta-2}\left[ \max\{g_{aa}-g_{ab}^{(\theta)},g_{ba}^{(\theta)}\} -  \overline{r}_a\left(\widehat{G}_a^{(\theta)}\right)
		+ \max\{g_{bb}-g_{ba}^{(\theta+1)},g_{ab}^{(\theta)}\} - r_b\left(\widehat{G}_b^{(\theta+1)}\right)\right]\nonumber\\
		{}& + \max\{g_{aa}-g_{ab}^{(M-1)},g_{ba}^{(\Theta-1)}\} -  \overline{r}_a\left(\widehat{G}_a^{(\Theta-1)}\right)
		+ \max\{g_{bb}-g_{ba}^{(\Theta)},g_{ab}^{(\Theta-1)}\} - r_b\left(\widehat{G}_b^{(\Theta)}\right) + \overline{r}_b^c\left(\widehat{G}_b^{(\Theta)}\right)\label{eq:view1_exp_2},\\
 \overline{r}_a^c\left(\widehat{G}_a\right) 
	\leq{}& \max\{g_{bb}-g_{ba}^{(1)},g_{ab}\} - r_b\left(\widehat{G}_b^{(1)}\right)\nonumber\\
		{}& + \sum_{\theta=1}^{\Theta-2}\left[ \max\{g_{aa}-g_{ab}^{(\theta)},g_{ba}^{(\theta)}\} -  \overline{r}_a\left(\widehat{G}_a^{(\theta)}\right)
		+ \max\{g_{bb}-g_{ba}^{(\theta+1)},g_{ab}^{(\theta)}\} - r_b\left(\widehat{G}_b^{(\theta+1)}\right)\right]\nonumber\\
		{}& + \max\{g_{aa},g_{ba}^{(\Theta)}\} -  \overline{r}_a\left(\widehat{G}_a^{(\Theta)}\right)
		\label{eq:view1_exp_3},\\
 \overline{r}_a^c\left(\widehat{G}_a\right) 
	\leq{}& \max\{g_{bb}-g_{ba}^{(1)},g_{ab}\} - r_b\left(\widehat{G}_b^{(1)}\right)\nonumber\\
		{}& + \sum_{\theta=1}^{\Theta-2}\left[ \max\{g_{aa}-g_{ab}^{(\theta)},g_{ba}^{(\theta)}\} -  \overline{r}_a\left(\widehat{G}_a^{(\theta)}\right)
		+ \max\{g_{bb}-g_{ba}^{(\theta+1)},g_{ab}^{(\theta)}\} - r_b\left(\widehat{G}_b^{(\theta+1)}\right)\right]\nonumber\\
		{}& + \max\{g_{aa}-g_{ab}^{(\Theta)},g_{ba}^{(\Theta)}\} -  \overline{r}_a\left(\widehat{G}_a^{(\Theta)}\right) + \overline{r}_a^c\left(\widehat{G}_a^{(\Theta)}\right),
		\label{eq:view1_exp_4}
\end{align}
which hold for any $\Theta\in\mathbb{Z}^+$, and arbitrary values of $G^{(\theta)}$. 
We tighten the bounds, by applying (\ref{eq:view1_mpc_lb_a}) and (\ref{eq:view1_mpc_lb_b}) and considering the values of $\Theta$ and $G^{(\theta)}$ that minimize the right hand sides of (\ref{eq:view1_exp_1})--(\ref{eq:view1_exp_4}). 

Applying (\ref{eq:view1_mpc_lb_a}) and (\ref{eq:view1_mpc_lb_b}), expression (\ref{eq:view1_exp_1}) becomes
\begin{align}
 \overline{r}_a^c\left(\widehat{G}_a\right) 
	\leq{}& \max\{g_{bb}-g_{ba}^{(1)},g_{ab}\} - g_{bb}\tau_b(g_{aa},g_{bb})\nonumber\\
		{}& + \sum_{\theta=1}^{\Theta-2}[ \max\{g_{aa}-g_{ab}^{(\theta)},g_{ba}^{(\theta)}\} 
		+ \max\{g_{bb}-g_{ba}^{(\theta+1)},g_{ab}^{(\theta)}\} - (g_{bb} + \delta\tau_a(g_{aa},g_{bb}))]\nonumber\\
		{}& + \max\{g_{aa}-g_{ab}^{(\Theta-1)},g_{ba}^{(\Theta-1)}\} - g_{aa}\tau_a(g_{aa},g_{bb})
		+ \max\{g_{bb},g_{ab}^{(\Theta-1)}\} - g_{bb}\tau_b(g_{aa},g_{bb})\\
	= {}& \max\{g_{bb}-g_{ba}^{(1)},g_{ab}\} - (\Theta-1)(g_{bb} + \delta\tau_a(g_{aa},g_{bb})) 
		- g_{bb}\tau_b(g_{aa},g_{bb})\nonumber\\
		{}& + \sum_{\theta=1}^{\Theta-2}[ \max\{g_{aa}-g_{ab}^{(\theta)},g_{ba}^{(\theta)}\} + \max\{g_{bb}-g_{ba}^{(\theta+1)},g_{ab}^{(\theta)}\}]\nonumber\\
		{}& + \max\{g_{aa}-g_{ab}^{(\Theta-1)},g_{ba}^{(\Theta-1)}\} + \max\{g_{bb},g_{ab}^{(\Theta-1)}\}.\label{eq:view1_proof_1}
\end{align}
In order to minimize this expression for a given $\Theta$, we assign the values of $g_{ab}^{(\theta)}$ and $g_{ab}^{(\theta)}$ as
\begin{align}
 g_{ab}^{(\Theta-1)} ={}& g_{bb},\\
 g_{ba}^{(\theta)} ={}& g_{aa}-g_{ab}^{(\theta)},\\
 g_{ab}^{(\theta-1)} ={}& \left(g_{bb}=g_{ba}^{(\theta)}\right)^+.
\end{align}
Substituting $\ell = \Theta-1$ and noting $g_{ab}\geq0$,we arrive at the first bound on interference component entropy
\begin{align}
 \overline{r}_a^c\left(\widehat{G}_a\right) 
	\leq{}& 
	\max\{g_{bb}-\ell\delta,g_{ab}\} + \ell\delta\tau_b(g_{aa},g_{bb}) - g_{bb}\tau_b(g_{aa},g_{bb}).\label{eq:view1_proof_ob3}
\end{align}

We can also consider (\ref{eq:view1_exp_1}) and (\ref{eq:view1_exp_3}) where the terminal link represents the response to the optimized channel state (i.e., $g_{ba}^{(\Theta)}=g_{ba}$), and derive from (\ref{eq:view1_exp_1}) the bound
\begin{align}
 \overline{r}_a^c\left(\widehat{G}_a\right) +r_b\left(\widehat{G}_b\right)
	\leq{}& \max\{g_{bb}-\ell\delta,g_{ab}\} + \ell\delta\tau_b(g_{aa},g_{bb}).\label{eq:view1_proof_ob4}
\end{align}

For (\ref{eq:view1_exp_2}), selecting $g_{ba}^{(\Theta)} = g_{ba}$ yields
\begin{align}
 \overline{r}_a^c\left(\widehat{G}_a\right) + r_b\left(\widehat{G}_b\right) - \overline{r}_b^c\left(\widehat{G}_b\right)
	\leq{}& \max\left\{g_{bb}-g_{ba}^{(1)},g_{ab}\right\} 
		- (\Theta-1)\left(g_{bb} + \delta\tau_a(g_{aa},g_{bb})\right)\nonumber\\
		{}& + \sum_{\theta=1}^{\Theta-1}\left[ \max\left\{g_{aa}-g_{ab}^{(\theta)},g_{ba}^{(\theta)}\right\} + \max\left\{g_{bb}-g_{ba}^{(\theta+1)},g_{ab}^{(\theta)}\right\}\right]\nonumber\\
		{}& + \max\left\{g_{aa}-g_{ab}^{(\Theta-1)},g_{ba}^{(\Theta-1)}\right\} + \max\left\{g_{bb}-g_{ba},g_{ab}^{(\Theta-1)}\right\}.
\end{align}
Selecting possible interference gains in the manner,
\begin{align}
 g_{ab}^{(\Theta-1)} ={}& (g_{bb}-g_{ba})^+,\\
 g_{ba}^{(\theta)} ={}& g_{aa}-g_{ab}^{(\theta)},\\
 g_{ab}^{(\theta-1)} ={}& (g_{bb}-g_{ba}^{(\theta)})^+,
\end{align}
results in the bound (given $\ell\geq0$)
\begin{align}
 \overline{r}_a^c\left(\widehat{G}_a\right) + r_b\left(\widehat{G}_b\right) - \overline{r}_b^c\left(\widehat{G}_b\right)
	\leq{}& \max\{g_{bb}-g_{ba}-\ell\delta,g_{ab}\} + \ell\delta\tau_b(g_{aa},g_{bb}).\label{eq:view1_proof_ob5}
\end{align}

Similar analysis and choices for free parameters from the expressions (\ref{eq:view1_exp_3}) and (\ref{eq:view1_exp_4}) yield the following bounds (given $\ell\geq0$):
\begin{align}
 \overline{r}_a^c\left(\widehat{G}_a\right) 
	\leq{}& g_{ab} + \ell\delta\tau_b(g_{aa},g_{bb})\label{eq:view1_proof_ob6},\\
 \overline{r}_a^c\left(\widehat{G}_a\right) + r_a\left(\widehat{G}_a\right)
	\leq{}& g_{ab} + (\ell+1)\delta\tau_b(g_{aa},g_{bb}) + g_{aa}\tau_a(g_{aa},g_{bb})\label{eq:view1_proof_ob7},\\
 r_a\left(\widehat{G}_a\right)
	\leq{}& \max\{g_{aa},g_{ab}\} + \ell\delta\tau_b(g_{aa},g_{bb}) - g_{bb}\tau_b(g_{aa},g_{bb}).\label{eq:view1_proof_ob8}
\end{align}
Analogously, from expansion from the interference component of $b$, we have (given $\ell\geq0$)
\begin{align}
 \overline{r}_b^c\left(\widehat{G}_b\right) 
	\leq{}& \max\{g_{ba},g_{aa}\} - g_{aa}\tau_a(g_{aa},g_{bb}) + \ell\delta\tau_b(g_{aa},g_{bb})\label{eq:view1_proof_ob9},\\
 \overline{r}_b^c\left(\widehat{G}_b\right) + r_a\left(\widehat{G}_a\right)
	\leq{}& \max\{g_{ba},g_{aa}\} + \ell\delta\tau_b(g_{aa},g_{bb})\label{eq:view1_proof_ob10},\\
 \overline{r}_b^c\left(\widehat{G}_a\right) + r_a\left(\widehat{G}_b\right) - \overline{r}_a^c\left(\widehat{G}_b\right)
	\leq{}& \max\{g_{ba}-\ell\delta,(g_{aa}-g_{ab})^+,(g_{ba}-g_{ab})^+,(g_{ba}-g_{aa})^+\} + \ell\delta\tau_b(g_{aa},g_{bb})\label{eq:view1_proof_ob11},\\
 \overline{r}_b^c\left(\widehat{G}_b\right) 
	\leq{}& \max\{g_{ba}-\ell\delta,(g_{ba}-g_{aa})^+\} + \ell\delta\tau_b(g_{aa},g_{bb})\label{eq:view1_proof_ob12},\\
 \overline{r}_b^c\left(\widehat{G}_a\right) + r_b\left(\widehat{G}_b\right)
	\leq{}& \max\{g_{ba}-(\ell+1)\delta,(g_{ba}-g_{aa})^+\} + (g_{bb} + (\ell+1)\delta)\tau_b(g_{aa},g_{bb})\label{eq:view1_proof_ob13},\\
 r_b\left(\widehat{G}_b\right)
	\leq{}& \max\{g_{ba},g_{aa}\} - g_{aa}\tau_a(g_{aa},g_{bb}) + \ell\delta\tau_b(g_{aa},g_{bb}).\label{eq:view1_proof_ob14}
\end{align}

Although these bounds were derived by expanding a Z-channel forward from each transmitter's interference component, expressions (\ref{eq:view1_proof_ob5}) and (\ref{eq:view1_proof_ob11}) also account for the message component not contained in the interference signal: $r_b(\widehat{G}_b) - \overline{r}_b^c(\widehat{G}_b)$ and $r_a(\widehat{G}_a) - \overline{r}_a^c(\widehat{G}_a)$ respectively. 

We now establish an additional pair of bounds for this component --- what is representative of the `private' message component in HK coding ---
derived from extension of a Z-channel in the reverse direction
(Figure~\ref{fig:view1_virtualZ-rev1}). To terminate each chain, we assume the interference gain of the final link is equal to the direct link gain. Consequently, we have (given $\ell\geq0$)
\begin{align}
 r_a\left(\widehat{G}_a\right) - \overline{r}_a^c\left(\widehat{G}_a\right) 
	\leq{}& \max\{(g_{aa} - g_{ab})^+,g_{bb}-\ell\delta\} - g_{bb}\tau_b(g_{aa},g_{bb}) + \ell\delta\tau_b(g_{aa},g_{bb})\label{eq:view1_proof_ob15},\\
 r_a\left(\widehat{G}_a\right) - \overline{r}_a^c\left(\widehat{G}_a\right) 
	\leq{}& (g_{aa} - g_{ab})^+ + \ell\delta\tau_b(g_{aa},g_{bb})\label{eq:view1_proof_ob16},\\
 r_b\left(\widehat{G}_b\right) - \overline{r}_b^c\left(\widehat{G}_b\right) 
	\leq{}& (g_{aa} + \ell\delta)\tau_b(g_{aa},g_{bb})\label{eq:view1_proof_ob17},\\
 r_b\left(\widehat{G}_b\right) - \overline{r}_b^c\left(\widehat{G}_b\right) 
	\leq{}& (g_{bb} - g_{ba} - \ell\delta)^+ +\ell\delta\tau_b(g_{aa},g_{bb}).\label{eq:view1_proof_ob18}
\end{align}

%
%

We remove redundant bounds from the signal component bounds derived thus far. For instance, (\ref{eq:view1_proof_ob8}), (\ref{eq:view1_proof_ob14}), and (\ref{eq:view1_proof_ob17}) are undeniably looser bounds than (\ref{eq:view1_proof_ob1}), (\ref{eq:view1_proof_ob2}), and (\ref{eq:view1_proof_ob18}) respectively. Additionally, (\ref{eq:view1_proof_ob7}) is the sum of (\ref{eq:view1_proof_ob1}) and (\ref{eq:view1_proof_ob3}). In addition to redundancies, the inequality (\ref{eq:view1_proof_ob9}) can be tightened by observing its relationship to (\ref{eq:view1_proof_ob2}): as a bound on a ``public'' component of the signal, the entropy bounded in (\ref{eq:view1_proof_ob9}) must be less than the entropies of the full transmitted signals. 

In summary, we have the following set of bounds
\begin{align}
 r_a\left(\widehat{G}_a\right)
	\leq{}& g_{aa} - g_{bb}\tau_b(g_{aa},g_{bb})\label{eq:view1_proof_comp1},\\
 r_b\left(\widehat{G}_b\right)
	\leq{}& g_{aa}\tau_b(g_{aa},g_{bb})\label{eq:view1_proof_comp2},\\
 \overline{r}_a^c\left(\widehat{G}_a\right)
	\leq{}& \min_{\ell\geq0}[
		\max\{g_{bb}-\ell\delta,g_{ab}\} + \ell\delta\tau_b(g_{aa},g_{bb}) - g_{bb}\tau_b(g_{aa},g_{bb})
	]\label{eq:view1_proof_comp3},\\
 \overline{r}_a^c\left(\widehat{G}_a\right) 
	\leq{}& g_{ab}\label{eq:view1_proof_comp4},\\
 \overline{r}_b^c\left(\widehat{G}_b\right)
	\leq{}& g_{aa}\tau_b(g_{aa},g_{bb})\label{eq:view1_proof_comp5},\\
 \overline{r}_b^c\left(\widehat{G}_b\right)
	\leq{}& \min_{\ell\geq0}[
		\max\{g_{ba}-\ell\delta,(g_{ba}-g_{aa})^+\} + \ell\delta\tau_b(g_{aa},g_{bb})
	]\label{eq:view1_proof_comp6},\\
 \overline{r}_a^c\left(\widehat{G}_a\right) +r_b\left(\widehat{G}_b\right)
	\leq{}& \min_{\ell\geq0}[
		\max\{g_{bb}-\ell\delta,g_{ab}\} + \ell\delta\tau_b(g_{aa},g_{bb})
	]\label{eq:view1_proof_comp7},\\
 \overline{r}_b^c\left(\widehat{G}_b\right) + r_a\left(\widehat{G}_a\right)
	\leq{}& \max\{g_{ba},g_{aa}\}\label{eq:view1_proof_comp8},\\
 \overline{r}_b^c\left(\widehat{G}_b\right) + r_b\left(\widehat{G}_b\right)
	\leq{}& \min_{\ell\geq0}[
		\max\{g_{ba}-(\ell+1)\delta,(g_{ba}-g_{aa})^+\} + (g_{aa} + \ell\delta)\tau_b(g_{aa},g_{bb})
	]\label{eq:view1_proof_comp9},\\
 \overline{r}_a^c\left(\widehat{G}_a\right) + r_b\left(\widehat{G}_b\right) - \overline{r}_b^c\left(\widehat{G}_b\right)
	\leq{}& \min_{\ell\geq0}[
		\max\{g_{ab},g_{bb}-g_{ba}-\ell\delta\} + \ell\delta\tau_b(g_{aa},g_{bb})
	]\label{eq:view1_proof_comp10},\\
 \overline{r}_b^c\left(\widehat{G}_b\right) + r_a\left(\widehat{G}_a\right) - \overline{r}_a^c\left(\widehat{G}_a\right)
	\leq{}& \min_{\ell\geq0}[
		\max\{g_{ba}-\ell\delta,(g_{aa}-g_{ab})^+,g_{ba}-g_{ab},g_{ba}-g_{aa}\} + \ell\delta\tau_b(g_{aa},g_{bb})
	]\label{eq:view1_proof_comp11},\\
 r_a\left(\widehat{G}_a\right) - \overline{r}_a^c\left(\widehat{G}_a\right)
	\leq{}& \min_{\ell\geq0}[
		\max\{(g_{aa} - g_{ab})^+,g_{bb}-\ell\delta\} - g_{bb}\tau_b(g_{aa},g_{bb}) + \ell\delta\tau_b(g_{aa},g_{bb})
	]\label{eq:view1_proof_comp12},\\
 r_a\left(\widehat{G}_a\right) - \overline{r}_a^c\left(\widehat{G}_a\right)
	\leq{}& (g_{aa} - g_{ab})^+\label{eq:view1_proof_comp13},\\
 r_b\left(\widehat{G}_b\right) - \overline{r}_b^c\left(\widehat{G}_b\right) 
	\leq{}& \min_{\ell\geq0}[
		(g_{bb} - g_{ba} - \ell\delta)^+ +\ell\delta\tau_b(g_{aa},g_{bb})
	].\label{eq:view1_proof_comp14}
\end{align}

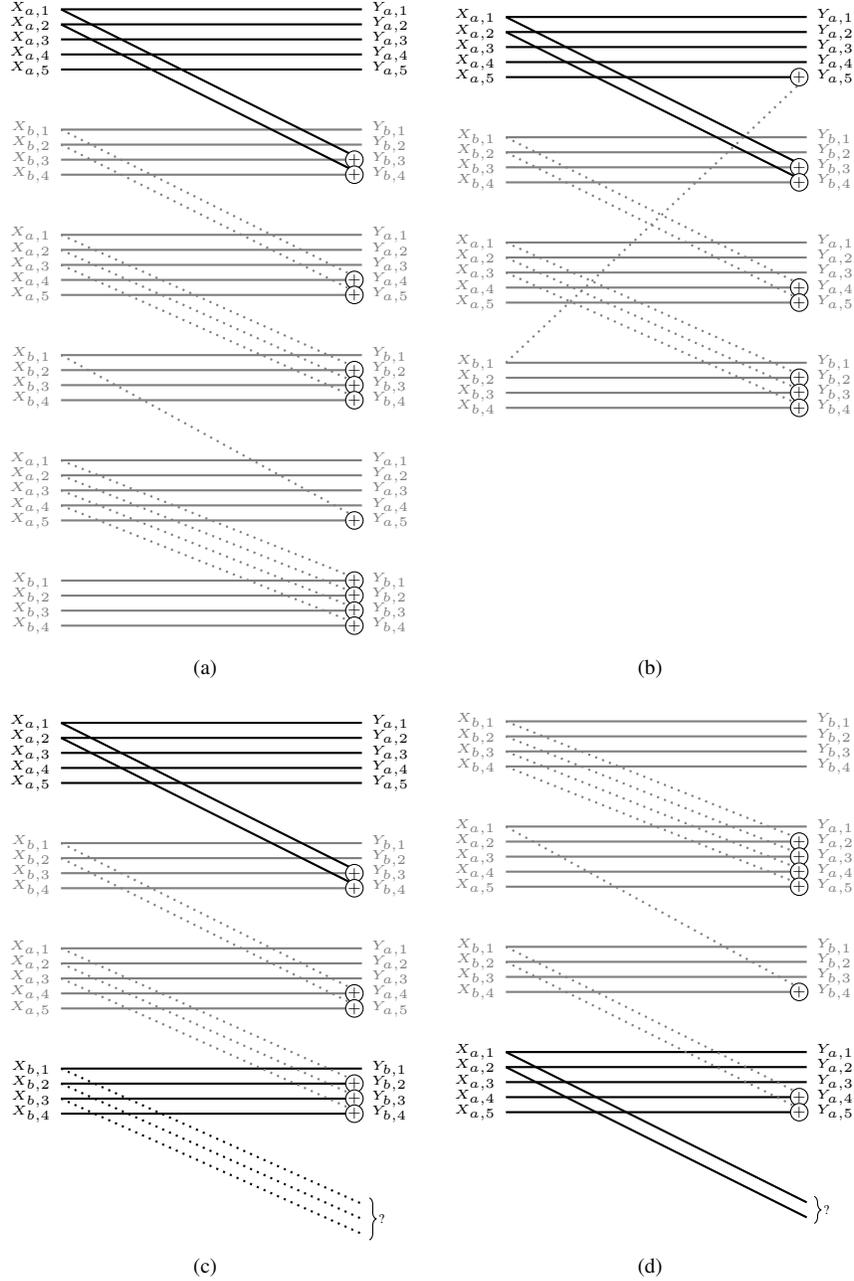
\begin{figure}[ht]
	\begin{center}
\subfigure[]{
	\begin{tikzpicture}[font=\tiny,yscale=0.4]
\draw[white] (0,-0.25) -- (0,1);
		\draw[thick](0,6) node[left] {$X_{a,1}$} -- (4,6) node[right] {$Y_{a,1}$};
		\draw[thick](0,5.5) node[left] {$X_{a,2}$} -- (4,5.5) node[right] {$Y_{a,2}$};
		\draw[thick](0,5) node[left] {$X_{a,3}$} -- (4,5) node[right] {$Y_{a,3}$};
		\draw[thick](0,4.5) node[left] {$X_{a,4}$} -- (4,4.5) node[right] {$Y_{a,4}$};
		\draw[thick](0,4) node[left] {$X_{a,5}$} -- (4,4) node[right] {$Y_{a,5}$};

		\draw[thick,black!50](0,2) node[left] {$X_{b,1}$} -- (4,2) node[right] {$Y_{b,1}$};
		\draw[thick,black!50](0,1.5) node[left] {$X_{b,2}$} -- (4,1.5) node[right] {$Y_{b,2}$};
		\draw[thick,black!50](0,1) node[left] {$X_{b,3}$} -- (4,1) node[right] {$Y_{b,3}$};
		\draw[thick,black!50](0,0.5) node[left] {$X_{b,4}$} -- (4,0.5) node[right] {$Y_{b,4}$};

		\draw[thick,black!50](0,-1.5) node[left] {$X_{a,1}$} -- (4,-1.5) node[right] {$Y_{a,1}$};
		\draw[thick,black!50](0,-2) node[left] {$X_{a,2}$} -- (4,-2) node[right] {$Y_{a,2}$};
		\draw[thick,black!50](0,-2.5) node[left] {$X_{a,3}$} -- (4,-2.5) node[right] {$Y_{a,3}$};
		\draw[thick,black!50](0,-3) node[left] {$X_{a,4}$} -- (4,-3) node[right] {$Y_{a,4}$};
		\draw[thick,black!50](0,-3.5) node[left] {$X_{a,5}$} -- (4,-3.5) node[right] {$Y_{a,5}$};

		\draw[thick,black!50](0,-5.5) node[left] {$X_{b,1}$} -- (4,-5.5) node[right] {$Y_{b,1}$};
		\draw[thick,black!50](0,-6) node[left] {$X_{b,2}$} -- (4,-6) node[right] {$Y_{b,2}$};
		\draw[thick,black!50](0,-6.5) node[left] {$X_{b,3}$} -- (4,-6.5) node[right] {$Y_{b,3}$};
		\draw[thick,black!50](0,-7) node[left] {$X_{b,4}$} -- (4,-7) node[right] {$Y_{b,4}$};

		\draw[thick,black!50](0,-9) node[left] {$X_{a,1}$} -- (4,-9) node[right] {$Y_{a,1}$};
		\draw[thick,black!50](0,-9.5) node[left] {$X_{a,2}$} -- (4,-9.5) node[right] {$Y_{a,2}$};
		\draw[thick,black!50](0,-10) node[left] {$X_{a,3}$} -- (4,-10) node[right] {$Y_{a,3}$};
		\draw[thick,black!50](0,-10.5) node[left] {$X_{a,4}$} -- (4,-10.5) node[right] {$Y_{a,4}$};
		\draw[thick,black!50](0,-11) node[left] {$X_{a,5}$} -- (4,-11) node[right] {$Y_{a,5}$};

		\draw[thick,black!50](0,-13) node[left] {$X_{b,1}$} -- (4,-13) node[right] {$Y_{b,1}$};
		\draw[thick,black!50](0,-13.5) node[left] {$X_{b,2}$} -- (4,-13.5) node[right] {$Y_{b,2}$};
		\draw[thick,black!50](0,-14) node[left] {$X_{b,3}$} -- (4,-14) node[right] {$Y_{b,3}$};
		\draw[thick,black!50](0,-14.5) node[left] {$X_{b,4}$} -- (4,-14.5) node[right] {$Y_{b,4}$};

		\draw[thick](0,6) -- (4,1);
		\draw[thick](0,5.5) -- (4,0.5);
		\draw (3.9,1) node[fill=white,draw,circle, inner sep=0pt]{$+$};
		\draw (3.9,0.5) node[fill=white,draw,circle, inner sep=0pt]{$+$};

		\draw[thick,dotted,black!50](0,2) -- (4,-3);
		\draw[thick,dotted,black!50](0,1.5) -- (4,-3.5);
		\draw (3.9,-3) node[fill=white,draw,circle, inner sep=0pt]{$+$};
		\draw (3.9,-3.5) node[fill=white,draw,circle, inner sep=0pt]{$+$};

		\draw[thick,dotted,black!50](0,-1.5) -- (4,-6);
		\draw[thick,dotted,black!50](0,-2) -- (4,-6.5);
		\draw[thick,dotted,black!50](0,-2.5) -- (4,-7);
		\draw (3.9,-6) node[fill=white,draw,circle, inner sep=0pt]{$+$};
		\draw (3.9,-6.5) node[fill=white,draw,circle, inner sep=0pt]{$+$};
		\draw (3.9,-7) node[fill=white,draw,circle, inner sep=0pt]{$+$};

		\draw[thick,dotted,black!50](0,-5.5) -- (4,-11);
		\draw (3.9,-11) node[fill=white,draw,circle, inner sep=0pt]{$+$};

		\draw[thick,dotted,black!50](0,-9) -- (4,-13);
		\draw[thick,dotted,black!50](0,-9.5) -- (4,-13.5);
		\draw[thick,dotted,black!50](0,-10) -- (4,-14);
		\draw[thick,dotted,black!50](0,-10.5) -- (4,-14.5);
		\draw (3.9,-13) node[fill=white,draw,circle, inner sep=0pt]{$+$};
		\draw (3.9,-13.5) node[fill=white,draw,circle, inner sep=0pt]{$+$};
		\draw (3.9,-14) node[fill=white,draw,circle, inner sep=0pt]{$+$};
		\draw (3.9,-14.5) node[fill=white,draw,circle, inner sep=0pt]{$+$};
\end{tikzpicture}
	\label{fig:view1_virtualZ-for1}
	}
\subfigure[]{
	\begin{tikzpicture}[font=\tiny,yscale=0.4]
\draw[white] (0,-0.25) -- (0,1);
		\draw[thick](0,6) node[left] {$X_{a,1}$} -- (4,6) node[right] {$Y_{a,1}$};
		\draw[thick](0,5.5) node[left] {$X_{a,2}$} -- (4,5.5) node[right] {$Y_{a,2}$};
		\draw[thick](0,5) node[left] {$X_{a,3}$} -- (4,5) node[right] {$Y_{a,3}$};
		\draw[thick](0,4.5) node[left] {$X_{a,4}$} -- (4,4.5) node[right] {$Y_{a,4}$};
		\draw[thick](0,4) node[left] {$X_{a,5}$} -- (4,4) node[right] {$Y_{a,5}$};

		\draw[thick,black!50](0,2) node[left] {$X_{b,1}$} -- (4,2) node[right] {$Y_{b,1}$};
		\draw[thick,black!50](0,1.5) node[left] {$X_{b,2}$} -- (4,1.5) node[right] {$Y_{b,2}$};
		\draw[thick,black!50](0,1) node[left] {$X_{b,3}$} -- (4,1) node[right] {$Y_{b,3}$};
		\draw[thick,black!50](0,0.5) node[left] {$X_{b,4}$} -- (4,0.5) node[right] {$Y_{b,4}$};

		\draw[thick,black!50](0,-1.5) node[left] {$X_{a,1}$} -- (4,-1.5) node[right] {$Y_{a,1}$};
		\draw[thick,black!50](0,-2) node[left] {$X_{a,2}$} -- (4,-2) node[right] {$Y_{a,2}$};
		\draw[thick,black!50](0,-2.5) node[left] {$X_{a,3}$} -- (4,-2.5) node[right] {$Y_{a,3}$};
		\draw[thick,black!50](0,-3) node[left] {$X_{a,4}$} -- (4,-3) node[right] {$Y_{a,4}$};
		\draw[thick,black!50](0,-3.5) node[left] {$X_{a,5}$} -- (4,-3.5) node[right] {$Y_{a,5}$};

		\draw[thick,black!50](0,-5.5) node[left] {$X_{b,1}$} -- (4,-5.5) node[right] {$Y_{b,1}$};
		\draw[thick,black!50](0,-6) node[left] {$X_{b,2}$} -- (4,-6) node[right] {$Y_{b,2}$};
		\draw[thick,black!50](0,-6.5) node[left] {$X_{b,3}$} -- (4,-6.5) node[right] {$Y_{b,3}$};
		\draw[thick,black!50](0,-7) node[left] {$X_{b,4}$} -- (4,-7) node[right] {$Y_{b,4}$};

		\draw[thick](0,6) -- (4,1);
		\draw[thick](0,5.5) -- (4,0.5);
		\draw (3.9,1) node[fill=white,draw,circle, inner sep=0pt]{$+$};
		\draw (3.9,0.5) node[fill=white,draw,circle, inner sep=0pt]{$+$};

		\draw[thick,dotted,black!50](0,2) -- (4,-3);
		\draw[thick,dotted,black!50](0,1.5) -- (4,-3.5);
		\draw (3.9,-3) node[fill=white,draw,circle, inner sep=0pt]{$+$};
		\draw (3.9,-3.5) node[fill=white,draw,circle, inner sep=0pt]{$+$};

		\draw[thick,dotted,black!50](0,-1.5) -- (4,-6);
		\draw[thick,dotted,black!50](0,-2) -- (4,-6.5);
		\draw[thick,dotted,black!50](0,-2.5) -- (4,-7);
		\draw (3.9,-6) node[fill=white,draw,circle, inner sep=0pt]{$+$};
		\draw (3.9,-6.5) node[fill=white,draw,circle, inner sep=0pt]{$+$};
		\draw (3.9,-7) node[fill=white,draw,circle, inner sep=0pt]{$+$};

		\draw[thick,dotted,black!50](0,-5.5) -- (4,4);
		\draw (3.9,4) node[fill=white,draw,circle, inner sep=0pt]{$+$};

		\draw (3.9,-14.5) node[white,draw,circle, inner sep=0pt]{$+$};

\end{tikzpicture}
	\label{fig:view1_virtualZ-for2}
	}\\
\subfigure[]{
	\begin{tikzpicture}[font=\tiny,yscale=0.4]
\draw[white] (0,-0.25) -- (0,1);
		\draw[thick](0,6) node[left] {$X_{a,1}$} -- (4,6) node[right] {$Y_{a,1}$};
		\draw[thick](0,5.5) node[left] {$X_{a,2}$} -- (4,5.5) node[right] {$Y_{a,2}$};
		\draw[thick](0,5) node[left] {$X_{a,3}$} -- (4,5) node[right] {$Y_{a,3}$};
		\draw[thick](0,4.5) node[left] {$X_{a,4}$} -- (4,4.5) node[right] {$Y_{a,4}$};
		\draw[thick](0,4) node[left] {$X_{a,5}$} -- (4,4) node[right] {$Y_{a,5}$};

		\draw[thick,black!50](0,2) node[left] {$X_{b,1}$} -- (4,2) node[right] {$Y_{b,1}$};
		\draw[thick,black!50](0,1.5) node[left] {$X_{b,2}$} -- (4,1.5) node[right] {$Y_{b,2}$};
		\draw[thick,black!50](0,1) node[left] {$X_{b,3}$} -- (4,1) node[right] {$Y_{b,3}$};
		\draw[thick,black!50](0,0.5) node[left] {$X_{b,4}$} -- (4,0.5) node[right] {$Y_{b,4}$};

		\draw[thick,black!50](0,-1.5) node[left] {$X_{a,1}$} -- (4,-1.5) node[right] {$Y_{a,1}$};
		\draw[thick,black!50](0,-2) node[left] {$X_{a,2}$} -- (4,-2) node[right] {$Y_{a,2}$};
		\draw[thick,black!50](0,-2.5) node[left] {$X_{a,3}$} -- (4,-2.5) node[right] {$Y_{a,3}$};
		\draw[thick,black!50](0,-3) node[left] {$X_{a,4}$} -- (4,-3) node[right] {$Y_{a,4}$};
		\draw[thick,black!50](0,-3.5) node[left] {$X_{a,5}$} -- (4,-3.5) node[right] {$Y_{a,5}$};

		\draw[thick](0,-5.5) node[left] {$X_{b,1}$} -- (4,-5.5) node[right] {$Y_{b,1}$};
		\draw[thick](0,-6) node[left] {$X_{b,2}$} -- (4,-6) node[right] {$Y_{b,2}$};
		\draw[thick](0,-6.5) node[left] {$X_{b,3}$} -- (4,-6.5) node[right] {$Y_{b,3}$};
		\draw[thick](0,-7) node[left] {$X_{b,4}$} -- (4,-7) node[right] {$Y_{b,4}$};

		\draw[thick](0,6) -- (4,1);
		\draw[thick](0,5.5) -- (4,0.5);
		\draw (3.9,1) node[fill=white,draw,circle, inner sep=0pt]{$+$};
		\draw (3.9,0.5) node[fill=white,draw,circle, inner sep=0pt]{$+$};

		\draw[thick,dotted,black!50](0,2) -- (4,-3);
		\draw[thick,dotted,black!50](0,1.5) -- (4,-3.5);
		\draw (3.9,-3) node[fill=white,draw,circle, inner sep=0pt]{$+$};
		\draw (3.9,-3.5) node[fill=white,draw,circle, inner sep=0pt]{$+$};

		\draw[thick,dotted,black!50](0,-1.5) -- (4,-6);
		\draw[thick,dotted,black!50](0,-2) -- (4,-6.5);
		\draw[thick,dotted,black!50](0,-2.5) -- (4,-7);
		\draw (3.9,-6) node[fill=white,draw,circle, inner sep=0pt]{$+$};
		\draw (3.9,-6.5) node[fill=white,draw,circle, inner sep=0pt]{$+$};
		\draw (3.9,-7) node[fill=white,draw,circle, inner sep=0pt]{$+$};

		\draw[thick,dotted](0,-5.5) -- (4,-10);
		\draw[thick,dotted](0,-6) -- (4,-10.5);
		\draw[thick,dotted](0,-6.5) -- (4,-11);
		\draw[decorate,decoration=brace] (4.1,-9.8)--node[right]{?}(4.1,-11.2);
\end{tikzpicture}
	\label{fig:view1_virtualZ-for3}
	}
	\subfigure[]{
	\begin{tikzpicture}[font=\tiny,yscale=0.4]
\draw[white] (0,-0.25) -- (0,1);
		\draw[thick](0,6) node[left] {$X_{a,1}$} -- (4,6) node[right] {$Y_{a,1}$};
		\draw[thick](0,5.5) node[left] {$X_{a,2}$} -- (4,5.5) node[right] {$Y_{a,2}$};
		\draw[thick](0,5) node[left] {$X_{a,3}$} -- (4,5) node[right] {$Y_{a,3}$};
		\draw[thick](0,4.5) node[left] {$X_{a,4}$} -- (4,4.5) node[right] {$Y_{a,4}$};
		\draw[thick](0,4) node[left] {$X_{a,5}$} -- (4,4) node[right] {$Y_{a,5}$};

		\draw[thick,black!50](0,9.5) node[left] {$X_{b,1}$} -- (4,9.5) node[right] {$Y_{b,1}$};
		\draw[thick,black!50](0,9) node[left] {$X_{b,2}$} -- (4,9) node[right] {$Y_{b,2}$};
		\draw[thick,black!50](0,8.5) node[left] {$X_{b,3}$} -- (4,8.5) node[right] {$Y_{b,3}$};
		\draw[thick,black!50](0,8) node[left] {$X_{b,4}$} -- (4,8) node[right] {$Y_{b,4}$};

		\draw[thick,black!50](0,13.5) node[left] {$X_{a,1}$} -- (4,13.5) node[right] {$Y_{a,1}$};
		\draw[thick,black!50](0,13) node[left] {$X_{a,2}$} -- (4,13) node[right] {$Y_{a,2}$};
		\draw[thick,black!50](0,12.5) node[left] {$X_{a,3}$} -- (4,12.5) node[right] {$Y_{a,3}$};
		\draw[thick,black!50](0,12) node[left] {$X_{a,4}$} -- (4,12) node[right] {$Y_{a,4}$};
		\draw[thick,black!50](0,11.5) node[left] {$X_{a,5}$} -- (4,11.5) node[right] {$Y_{a,5}$};

		\draw[thick,black!50](0,17) node[left] {$X_{b,1}$} -- (4,17) node[right] {$Y_{b,1}$};
		\draw[thick,black!50](0,16.5) node[left] {$X_{b,2}$} -- (4,16.5) node[right] {$Y_{b,2}$};
		\draw[thick,black!50](0,16) node[left] {$X_{b,3}$} -- (4,16) node[right] {$Y_{b,3}$};
		\draw[thick,black!50](0,15.5) node[left] {$X_{b,4}$} -- (4,15.5) node[right] {$Y_{b,4}$};

		\draw[thick](0,6) -- (4,1);
		\draw[thick](0,5.5) -- (4,0.5);
		\draw[decorate,decoration=brace] (4.1,1.2)--node[right]{?}(4.1,0.3);

		\draw[thick,dotted,black!50](0,9.5) -- (4,4.5);
		\draw[thick,dotted,black!50](0,9) -- (4,4);
		\draw (3.9,4.5) node[fill=white,draw,circle, inner sep=0pt]{$+$};
		\draw (3.9,4) node[fill=white,draw,circle, inner sep=0pt]{$+$};

		\draw[thick,dotted,black!50](0,13.5) -- (4,8);
		\draw (3.9,8) node[fill=white,draw,circle, inner sep=0pt]{$+$};

		\draw[thick,dotted,black!50](0,17) -- (4,13);
		\draw[thick,dotted,black!50](0,16.5) -- (4,12.5);
		\draw[thick,dotted,black!50](0,16) -- (4,12);
		\draw[thick,dotted,black!50](0,15.5) -- (4,11.5);
		\draw (3.9,13) node[fill=white,draw,circle, inner sep=0pt]{$+$};
		\draw (3.9,12.5) node[fill=white,draw,circle, inner sep=0pt]{$+$};
		\draw (3.9,12) node[fill=white,draw,circle, inner sep=0pt]{$+$};
		\draw (3.9,11.5) node[fill=white,draw,circle, inner sep=0pt]{$+$};

\end{tikzpicture}
	\label{fig:view1_virtualZ-rev1}
	}
\end{center}
	\caption{Some virtual Z channels considered in deriving outer bound for View~1. Optimized channel state is $g_{aa}=5$, $g_{ab}=2$, $g_{ba}=3$, $g_{bb}=4$, and the expansion is from Transmitter~$a$'s POV. Black lines depict optimized channel states, solid lines reflect `known' link gains. 
\subref{fig:view1_virtualZ-for1} Virtual Z-channel forward expansion to arrive at bound of type (\ref{eq:view1_proof_ob3}),
\subref{fig:view1_virtualZ-for2} Virtual Z-channel forward expansion to arrive at bound of type (\ref{eq:view1_proof_ob8}),
\subref{fig:view1_virtualZ-for3} Virtual Z-channel forward expansion to arrive at bound of type (\ref{eq:view1_proof_ob4}) or (\ref{eq:view1_proof_ob5}),
\subref{fig:view1_virtualZ-rev1} Virtual Z-channel backward expansion to arrive at bound of type (\ref{eq:view1_proof_ob15}).
	}
	\label{fig:view1_virtualZ}
\end{figure}

\subsubsection*{Achievable Scheme}
To complete the proof we have three tasks:
\begin{enumerate}
 \item Define a policy that specifies the transmission scheme not only for the channel state at hand, but for all states with the same direct link gains
 \item Show that any rate point on the outer bound can be achieved by such a scheme
 \item Confirm that the rates prescribed for other channel states are achievable and satisfy the minimum specified rate.
\end{enumerate}

As in View~2 and \cite{ETW:08}, the achievable scheme relies on each message being divided into common and a private components. For each channel state, the common component of Sender~$a$'s codebook is generated by randomly selecting $n r_a^c$ codewords from the set of all $n\times \max\{g_{ab},g_{aa}\}$ binary matrices. The private message codebook is generated by randomly selecting $n r_a^p$ codewords from the set of $n\times(g_{aa}-g_{ab})^+$ matrices. On outgoing links that interfere with Link~$b$, the $g_{ab}$ most significant levels of the common message are sent. If $g_{aa}-g_{ab}>0$, then the modulo addition of the private message and lower levels of the common message is transmitted. Similarly for Sender~$b$, $r_b^c$ and $r_b^p$ govern the number of codewords (randomly drawn) in the common and private codebooks of Link~$b$.


The size of component codebooks varies for different channel states, and is a function of each sender's local view. For the channel state being considered, 
the number of codewords in each component codebook are chosen 
such that ${r}_a^c(\widehat{G}_a)$, ${r}_a^p(\widehat{G}_a)$, $r_a(\widehat{G}_a)$, ${r}_b^c(\widehat{G}_b)$,${r}_b^p(\widehat{G}_b)$, and $r_b(\widehat{G}_b)$ obey (\ref{eq:view1_thm_1})--(\ref{eq:view1_thm_14}).

We assume joint decoding of all received components --- each receiver perceives a virtual three-user MAC --- which implies that the proposed policy is achievable if nonnegative $r_a^c$,$r_a^p$,$r_b^c$,$r_b^p$ satisfy (\ref{eq:rates_lindet_HK_1})--(\ref{eq:rates_lindet_HK_x}).
Noting that the restrictions imposed in (\ref{eq:view1_thm_1})--(\ref{eq:view1_thm_14}) are actually stricter than (\ref{eq:rates_lindet_HK_1})--(\ref{eq:rates_lindet_HK_x}), the policy proposed thus far is achievable, 
thus completing Step~2.

For the responses to other channel states with local views $\widehat{G}_a^\prime\neq\widehat{G}_a$ and $\widehat{G}_b^\prime\neq\widehat{G}_b$, the public and private codebook sizes must also conform to a similar set of bounds such that the remain consistent with the responses of the considered channel state. Applying a similar virtual Z-channel expansion to arbitrary channel states, and assuming
%
\begin{align}
 r_a(\widehat{G}_a^\prime) = r_a^c(\widehat{G}_a^\prime)+r_a^p(\widehat{G}_a^\prime) \geq g_{aa}\tau_a(g_{aa},g_{bb}),\\
 r_b(\widehat{G}_b^\prime) = r_b^c(\widehat{G}_b^\prime)+r_b^p(\widehat{G}_b^\prime) \geq g_{bb}\tau_b(g_{aa},g_{bb}),
\end{align}
we find for local views $\widehat{G}_a^{\prime\prime}\neq\widehat{G}_a$ and $\widehat{G}_b^{\prime\prime}\neq\widehat{G}_b$ and $\ell \geq 0$
\begin{align}
r_a^c(\widehat{G}_a^{\prime}) \leq{}& g_{ab}^{\prime}\label{eq:view1_ach_beg},\\
r_a^c(\widehat{G}_a^{\prime}) \leq{}& \max\{g_{ab}^{\prime},g_{bb} - g_{ba}-\ell\delta\} 
	+ \ell\delta\tau_b(g_{aa},g_{bb})  
	- r_b^p(\widehat{G}_b),\\
r_a^c(\widehat{G}_a^{\prime}) \leq{}& \max\{g_{ab}^{\prime},g_{bb} + g_{ab} -g_{aa} - \ell\delta\} 
	+ \ell\delta\tau_b(g_{aa},g_{bb})  
	+(g_{aa}-g_{ab})^+ - g_{bb}\tau_b(g_{aa},g_{bb}) - r_a^p(\widehat{G}_a),\\
r_a^c(\widehat{G}_a^{\prime}) \leq{}& \max\{g_{ab}^{\prime},g_{bb}-\ell\delta\} 
	+ \ell\delta\tau_b(g_{aa},g_{bb})  
	- r_b(\widehat{G}_b),\\
r_a^c(\widehat{G}_a^{\prime}) \leq{}& \max\{g_{ab}^{\prime},g_{bb} - \ell\delta\} 
	+ \ell\delta\tau_b(g_{aa},g_{bb})  
	+ g_{aa} - g_{bb}\tau_b(g_{aa},g_{bb}) - r_a(\widehat{G}_a),\\
r_a^p(\widehat{G}_a^{\prime}) \leq{}& \left(g_{aa} - g_{ab}^{\prime}\right)^+,\\
r_a^p(\widehat{G}_a^{\prime}) \leq{}& \max\{g_{ba}-\ell\delta,g_{aa} - g_{ab}^{\prime}\} 
	+ \ell\delta\tau_b(g_{aa},g_{bb})  
	- r_b^c(\widehat{G}_b),\\
r_a^p(\widehat{G}_a^{\prime}) \leq{}& (g_{aa} - g_{ab}^{\prime})^+ - g_{bb}\tau_b(g_{aa},g_{bb})  
	+ \max\{g_{ab},g_{bb}-(g_{aa}-g_{ab}^{\prime}\} - \ell\delta) + \ell\delta\tau_b(g_{aa},g_{bb})
	- r_a^c(\widehat{G}_a),\\
r_a^p(\widehat{G}_a^{\prime}) + r_a^c(\widehat{G}_a^{\prime}) \leq{}& g_{aa},\\
r_a^p(\widehat{G}_a^{\prime}) + r_a^c(\widehat{G}_a^{\prime})\leq{}& 
	\max\{g_{ba}-\ell\delta,g_{aa}\} + \ell\delta\tau_b(g_{aa},g_{bb})  
	- r_b^c(\widehat{G}_b),\\
r_a^p(\widehat{G}_a^{\prime}) + r_a^c(\widehat{G}_a^{\prime})\leq{}& c
	g_{ab} + g_{aa} - r_a^c(\widehat{G}_a) - g_{bb}\tau_b(g_{aa},g_{bb}),\\
r_b^c(\widehat{G}_a^{\prime}) \leq{}& g_{ba}^{\prime},\\
r_b^c(\widehat{G}_a^{\prime}) \leq{}& 
	\max\{g_{ba}^{\prime}-\ell\delta,g_{aa} - g_{ab}\} + \ell\delta\tau_b(g_{aa},g_{bb})  
	- r_a^p(\widehat{G}_a),\\
r_b^c(\widehat{G}_a^{\prime}) \leq{}& 
	\max\{g_{ba}^{\prime}-\ell\delta,g_{aa} - (g_{bb} - g_{ba})\} + \ell\delta\tau_b(g_{aa},g_{bb})  
	- g_{aa}\tau_a(g_{aa},g_{bb}) + (g_{bb} - g_{ba})^+ - r_b^p(\widehat{G}_b),\\
r_b^c(\widehat{G}_a^{\prime}) \leq{}& 
	\max\{g_{ba}^{\prime}-\ell\delta,g_{aa}\} + \ell\delta\tau_b(g_{aa},g_{bb})  
	- r_a(\widehat{G}_a),\\
r_b^c(\widehat{G}_a^{\prime}) \leq{}& 
	\max\{g_{ba}^{\prime}-\ell\delta,\delta\} + \ell\delta\tau_b(g_{aa},g_{bb})  
	- g_{aa}\tau_a(g_{aa},g_{bb}) + g_{bb} - r_b(\widehat{G}_b),\\
r_b^p(\widehat{G}_a^{\prime}) \leq{}& \left(g_{bb}-g_{ba}^{\prime}\right)^+,\\
r_b^p(\widehat{G}_b^{\prime}) \leq{}& 
\max\{g_{ab},g_{bb} - g_{ba}^{\prime}-\ell\delta\} + \ell\delta\tau_b(g_{aa},g_{bb})  
	- r_a^c(\widehat{G}_a),\\
r_b^p(\widehat{G}_b^{\prime}) \leq{}& \max\{g_{aa} - g_{ba},g_{bb} - g_{ba}^{\prime}\} 
	- g_{aa}\tau_a(g_{aa},g_{bb})  
	+ g_{ba} - r_b^c(\widehat{G}_b),\\
r_b^p(\widehat{G}_b^{\prime}) + r_b^c(\widehat{G}_b^{\prime})\leq{}& 
	g_{bb},\\
r_b^p(\widehat{G}_b^{\prime}) + r_b^c(\widehat{G}_b^{\prime})\leq{}& 
	\max\{g_{ab},g_{bb}-\ell\delta\} + \ell\delta\tau_b(g_{aa},g_{bb})  
	- r_a^c(\widehat{G}_a),\\
r_b^p(\widehat{G}_b^{\prime}) + r_b^c(\widehat{G}_b^{\prime})\leq{}& 
	\max\{g_{ba} + g_{bb} - \ell\delta,g_{aa}\} + \ell\delta\tau_b(g_{aa},g_{bb}) - r_b^c(\widehat{G}_b) - g_{aa}\tau_a(g_{aa},g_{bb}).\label{eq:view1_ach_end}
\end{align}
Using these expressions along with (\ref{eq:view1_thm_1})--(\ref{eq:view1_thm_14}), we define the rates (and by proxy size) of codebooks in each policy response. Moreover, substitution of (\ref{eq:view1_thm_1})--(\ref{eq:view1_thm_14}) into (\ref{eq:view1_ach_beg})--(\ref{eq:view1_ach_end}) we see that the rate satisfies the minimum performance criterion as desired.
\end{IEEEproof}

\subsection{Proof for View 2}
\subsubsection*{Outer Bound}

We first consider two limiting cases. If $g_{aa}=g_{bb}=g_{ab}$, by considering (\ref{eq:reg_view0_4}), we have
\begin{align}
r_a\left(\widehat{G}_a\middle)\right|_{g_{aa}=g_{ab}} + r_b\left(\widehat{G}_b\middle)\right|_{g_{bb}=g_{ab}} \leq{}& g_{ab}\\
	={}&g_{ab},
\end{align}
where the equality is enforced in order to satisfy the minimum performance criterion. Similarly, if $g_{aa}=g_{bb}=g_{ba}$, we have
\begin{align}
r_a\left(\widehat{G}_a\middle)\right|_{g_{aa}=g_{ba}} + r_b\left(\widehat{G}_b\middle)\right|_{g_{bb}=g_{ba}} ={}&g_{ba}.
\end{align}

These two cases can be restated as 
\begin{align}
 \frac{r_a\left(\widehat{G}_a\middle)\right|_{g_{aa}=g_{ab}}}{g_{ab}} + \frac{r_b\left(\widehat{G}_b\middle)\right|_{g_{bb}=g_{ab}}}{g_{ab}} ={}&1,\\
 \frac{r_a\left(\widehat{G}_a\middle)\right|_{g_{aa}=g_{ba}}}{g_{ba}} + \frac{r_b\left(\widehat{G}_b\middle)\right|_{g_{bb}=g_{ba}}}{g_{ba}} ={}&1.
\end{align}
and when summed we have
\begin{align}
  \frac{r_a\left(\widehat{G}_a\middle)\right|_{g_{aa}=g_{ab}}}{g_{ab}} + \frac{r_b\left(\widehat{G}_b\middle)\right|_{g_{bb}=g_{ba}}}{g_{ba}} ={}& 2 - \frac{r_a\left(\widehat{G}_a\middle)\right|_{g_{aa}=g_{ba}}}{g_{ba}} - \frac{r_b\left(\widehat{G}_b\middle)\right|_{g_{bb}=g_{ab}}}{g_{ab}}
  \leq{} 1,\label{eq:view2proof_1}
\end{align}
where (\ref{eq:view2proof_1}) is due to the minimum performance criterion. By applying the same constraint on the other side, we have 
\begin{equation}
 1\leq \frac{r_a\left(\widehat{G}_a\middle)\right|_{g_{aa}=g_{ab}}}{g_{ab}} + \frac{r_b\left(\widehat{G}_b\middle)\right|_{g_{bb}=g_{ba}}}{g_{ba}} \leq{} 1,\label{eq:view2proof_1b}
\end{equation}
which implies that the two cases discussed are not only both constrained to a region where TDM is sufficient, but the operating points must be consistent, i.e., 
\begin{align}
 \frac{r_a\left(\widehat{G}_a\middle)\right|_{g_{aa}=g_{ab}}}{g_{ab}} = \frac{r_a\left(\widehat{G}_a\middle)\right|_{g_{aa}=g_{ba}}}{g_{ba}} ={}& \tau_a\left(g_{ab},g_{ba}\right),\\
 \frac{r_b\left(\widehat{G}_b\middle)\right|_{g_{bb}=g_{ab}}}{g_{ab}} = \frac{r_b\left(\widehat{G}_b\middle)\right|_{g_{bb}=g_{ba}}}{g_{ba}} ={}& \tau_b\left(g_{ab},g_{ba}\right),\\
 \tau_a\left(g_{ab},g_{ba}\right) + \tau_b\left(g_{ab},g_{ba}\right) ={}&1.
\end{align}

For other cases of direct link gain, we assume the viewpoint of Transmitter~$a$ in considering its policy options. As we have shown, there must exist $\tau_a(g_{ab},g_{ba})$ and $\tau_a(g_{ab},g_{ba})$ summing to one, such that 
\begin{align}
 r_a\left(\widehat{G}_a\right)\geq g_{aa}\tau_{a}\left(g_{ab},g_{ba}\right),\\
 r_b\left(\widehat{G}_b\right)\geq g_{bb}\tau_{b}\left(g_{ab},g_{ba}\right).
\end{align}

\begin{figure}[ht]
 	\begin{center}
\begin{tikzpicture}[font=\scriptsize,scale=0.75]
		\draw[thick](0,6) -- (4,1.5);
		\draw[thick](0,5.5) -- (4,1);
		\draw[thick](0,5) -- (4,0.5);

		\draw[thick](0,8.5) -- (4,3.5);
		\draw[thick](0,8) -- (4,3);

		\draw[thick,dotted](0,8.5) node[left] {$X_{b^\prime,1}$} -- (4,8.5) node[right] {$Y_{b^\prime,1}$};
		\draw[thick,dotted](0,8) node[left] {$X_{b^\prime,2}$} -- (4,8) node[right] {$Y_{b^\prime,2}$};

		\draw[thick,dotted](0,1.5) node[left] {$X_{b,1}$} -- (4,1.5) node[right] {$Y_{b,1}$};
		\draw[thick,dotted](0,1) node[left] {$X_{b,2}$} -- (4,1) node[right] {$Y_{b,2}$};
		\draw[thick,dotted](0,0.5) node[left] {$X_{b,3}$} -- (4,0.5) node[right] {$Y_{b,3}$};

		\draw[thick](0,6) node[left] {$X_{a,1}$} -- (4,6) node[right] {$Y_{a,1}$};
		\draw[thick](0,5.5) node[left] {$X_{a,2}$} -- (4,5.5) node[right] {$Y_{a,2}$};
		\draw[thick](0,5) node[left] {$X_{a,3}$} -- (4,5) node[right] {$Y_{a,3}$};
		\draw[thick](0,4.5) node[left] {$X_{a,4}$} -- (4,4.5) node[right] {$Y_{a,4}$};
		\draw[thick](0,4) node[left] {$X_{a,5}$} -- (4,4) node[right] {$Y_{a,5}$};
		\draw[thick](0,3.5) node[left] {$X_{a,6}$} -- (4,3.5) node[right] {$Y_{a,6}$};
		\draw[thick](0,3) node[left] {$X_{a,7}$} -- (4,3) node[right] {$Y_{a,7}$};
		
		\draw (3.9,3) node[fill=white,draw,circle, inner sep=0pt]{$+$};
		\draw (3.9,3.5) node[fill=white,draw,circle, inner sep=0pt]{$+$};
		\draw (3.9,1.5) node[fill=white,draw,circle, inner sep=0pt]{$+$};
		\draw (3.9,1) node[fill=white,draw,circle, inner sep=0pt]{$+$};
		\draw (3.9,0.5) node[fill=white,draw,circle, inner sep=0pt]{$+$};
\end{tikzpicture}
	\end{center}
	\caption{Virtual double Z-channel for Transmitter~$a$'s in View~2. Dotted segments represent unknown link gains.}
		\label{fig:view2_virtualZ}
\end{figure}
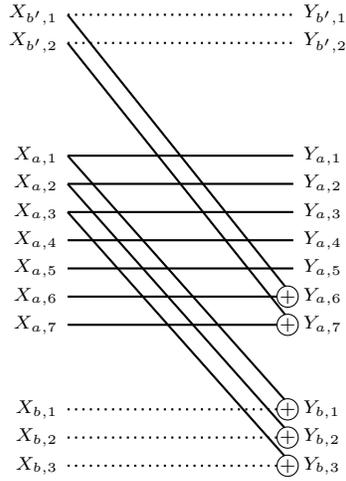

When $g_{aa}\notin\{g_{ab},g_{ba}\}$, there still exists a possibility of the other direct link being fully interfering/interfered, $g_{bb}\in\{g_{ab},g_{ba}\}$. Therefore, regardless of the known direct link, the channel input and decoding process must both accommodate the constraints imposed by these two limiting possibilities, resulting in the virtual Z-channel shown in Figure~\ref{fig:view2_virtualZ} for Transmitter~$a$'s view. 

This provides us with three constraints related to the limiting cases, and one which is essentially the point-to-point capacity:
\begin{align}
  r_a\left(\widehat{G}_a\right)	\leq{}& \frac{1}{n} \sum_{i=1}^{g_{aa}}L_{a,i}\left(\widehat{G}_a\right) \\
	\leq{}& g_{aa}.
\end{align}

Of the remaining three bounds, we begin with the first with one resulting from an adaptation of (\ref{eq:decom_bound_b}) and isolating the lower of the two Z-channels depicted in the three-user Z-channel, 
\begin{align}
   \sum_{i=1}^{g_{ab}}L_{a,i}\left(\widehat{G}_a\right) \leq{}& \sum_{j=1}^{g_{bb}-g_{ab}}L_{b,j}\left(\widehat{G}_b\right)  + ng_{ab} - nr_b(\widehat{G}_b)\\
		\leq{}& ng_{ab} - nr_b\left(\widehat{G}_b\middle)\right|_{g_{bb} = g_{ab}}\\
		\leq{}& ng_{ab}\tau_a\left(g_{ab},g_{ba}\right),
\end{align}
which we apply in
\begin{align}
  r_a\left(\widehat{G}_a\right)	\leq{}& \frac{1}{n} \sum_{i=1}^{g_{aa}}L_{a,i}\left(\widehat{G}_a\right) \\
	={}& \frac{1}{n} \left(\sum_{i=1}^{g_{ab}}L_{a,i}\left(\widehat{G}_a\right) + 
		\sum_{i=g_{ab}+1}^{g_{aa}}L_{a,i}\left(\widehat{G}_a\right)\right) \\
	\leq{}& g_{ab}\tau_a\left(g_{ab},g_{ba}\right) + \left(g_{aa} - g_{ab}\right)^+.
\end{align}

For the third bound on $r_a(\widehat{G}_a)$, we recall (\ref{eq:pol_ach_constr_a}) and isolate only the upper of the two Z-channels:
\begin{align}
  r_a\left(\widehat{G}_a\right)	\leq{}& \frac{1}{n} \left( 
	\sum_{i=1}^{g_{aa}-g_{ba}}L_{a,i}\left(\widehat{G}_a\right) 
	- \sum_{j=1}^{g_{ba}}L_{b,j}\left(\widehat{G}_b\right) \right) + g_{ba}\\
	\leq{}& \frac{1}{n} \left( \sum_{i=1}^{g_{aa}-g_{ba}}l_{a,i}\left(\widehat{G}_a\right) +ng_{ba} - \sum_{j=1}^{g_{ba}}L_{b,j}\left(\widehat{G}_b\middle)\right|_{g_{bb}=g_{ba}} \right) \\
	\leq{}& \frac{1}{n} \left( \sum_{i=1}^{g_{aa}-g_{ba}}L_{a,i}\left(\widehat{G}_a\right) +ng_{ba} - nr_b\left(\widehat{G}_b\middle)\right|_{g_{bb}=g_{ba}} \right) \\
	={}& \frac{1}{n} \left( \sum_{i=1}^{g_{aa}-g_{ba}}L_{a,i} +ng_{ba} - ng_{ba}\tau_b\left(g_{ab},g_{ba}\right) \right) \label{eq:view2proof_2}\\
  \leq{}& \left(g_{aa}-g_{ba}\right)^+ +g_{ba}\tau_a\left(g_{ab},g_{ba}\right).
\end{align}

For the fourth bound we also apply (\ref{eq:pol_ach_constr_a}) but deviate at (\ref{eq:view2proof_2}):
\begin{align}
  r_a\left(\widehat{G}_a\right)	\leq{}& \frac{1}{n} \left( \sum_{i=1}^{g_{aa}-g_{ba}}L_{a,i} +ng_{ba} - ng_{ba}\tau_b\left(g_{ab},g_{ba}\right) \right)\\
  \leq{}& \frac{1}{n}\sum_{i=1}^{g_{aa}-g_{ba}}L_{a,i}\left(\widehat{G}_a\right) +g_{ba}\tau_a\left(g_{ab},g_{ba}\right)\\
  \leq{}& \frac{1}{n}\sum_{i=1}^{g_{ab}}L_{a,i} +\frac{1}{n}\sum_{i=g_{ab}+1}^{g_{aa}-g_{ba}}L_{a,i} +g_{ba}\tau_a(g_{ab},g_{ba})\\
  \leq{}& g_{ab}\tau_a\left(g_{ab},g_{ba}\right) +\left(g_{aa}-g_{ab}-g_{ba}\right)^+ +g_{ba}\tau_a\left(g_{ab},g_{ba}\right).
\end{align}

In summary, we have for User~$a$:
\begin{align}
r_a\left(\widehat{G}_a\right) \leq{}& g_{aa}\label{eq:view2proof_3},\\
r_a\left(\widehat{G}_a\right) \leq{}& \left(g_{aa}-g_{ab}\right)^+ +g_{ab}\tau_a\left(g_{ab},g_{ba}\right),\\
r_a\left(\widehat{G}_a\right) \leq{}& \left(g_{aa}-g_{ba}\right)^+ +g_{ba}\tau_a\left(g_{ab},g_{ba}\right),\\
r_a\left(\widehat{G}_a\right) \leq{}& \left(g_{aa}-g_{ab}-g_{ba}\right)^+ +\left(g_{ab}+ g_{ba}\right)\tau_a\left(g_{ab},g_{ba}\right).\label{eq:view2proof_4}
\end{align}

The analogous bounds on the rate chosen by Sender~$b$ follow the same process.

\subsubsection*{Achievable Scheme}

The scheme used in this scenario is the linear deterministic model version of the simple Han-Kobayashi scheme proposed in \cite{ETW:08}, and described in Section~\ref{sec:HK}.

We generate public and private codebooks using random codes. For the public message of Sender~$a$, let the $\nu_a = \min\{g_{aa},g_{ab}\}$. We choose $n\nu_ar_a^c(\widehat{G}_a)$ codewords randomly from the set of all $n \times \nu_a$ binary vectors using a uniform distribution over the set. If $g_{ab}<g_{aa}$ we also choose $n\nu_ar_a^p(\widehat{G}_a)$ codewords randomly from the set of all $n \times (g_{aa}-g_{ab})$  binary vectors again using a uniform distribution. At Sender~$b$ we do the same for $n\nu_b = \min\{g_{bb},g_{ba}\}$, $r_b^c(\widehat{G}_b)$ and $r_b^c(\widehat{G}_b)$. 

The set of decodable rates $r_a^c(\widehat{G}_a)$, $r_a^p(\widehat{G}_a)$, $r_b^c(\widehat{G}_b)$, and $r_b^c(\widehat{G}_b)$ at Receiver~$a$ is given by (\ref{eq:rates_lindet_HK_1})--(\ref{eq:rates_lindet_HK_x}).
Since it is necessary for Sender~$a$ to know the rate of Sender~$b$'s public message in order to determine limits on its own public and private rates, we impose the constraints
\begin{align}
 r_a^c\left(\widehat{G}_a\right) \leq{}& g_{ab}\tau_a\left(g_{ab},g_{ba}\right),\\
 r_b^c\left(\widehat{G}_b\right) \leq{}& g_{ba}\tau_b\left(g_{ab},g_{ba}\right),
\end{align}
chosen based on our understanding of the two limiting cases in the outer bound. Furthermore, we note that in order to satisfy the minimum performance criterion
\begin{align}
 r_a^c\left(\widehat{G}_a\right)|_{g_{aa}=g_{ab}} = r_a\left(\widehat{G}_a\right)|_{g_{aa}=g_{ab}} ={}& g_{ab}\tau_a\left(g_{ab},g_{ba}\right),\\
 r_b^c\left(\widehat{G}_b\right)|_{g_{bb}=g_{ba}} = r_b\left(\widehat{G}_a\right)|_{g_{bb}=g_{ba}} ={}& g_{ba}\tau_b\left(g_{ab},g_{ba}\right).
\end{align}

The resulting region of rates achievable for the public and private messages of $a$ are
\begin{align}
 r_a^c\left(\widehat{G}_a\right) \leq{}&\min\left\{g_{aa},g_{ab}\right\},\\
 r_a^p\left(\widehat{G}_a\right) \leq{}&\left(g_{aa}-g_{ab}\right)^+,\\
 r_a^c\left(\widehat{G}_a\right) + r_a^p\left(\widehat{G}_a\right) \leq{}&g_{aa},\\
 r_a^c\left(\widehat{G}_a\right)  \leq{}& \max\left\{g_{aa},g_{ba}\right\} - r_b^c\left(\widehat{G}_b\right)\nonumber\\
	\leq{}& \max\left\{g_{aa},g_{ba}\right\} - g_{ba}\tau_b\left(g_{ab},g_{ba}\right)\\
	={}& \left(g_{aa} - g_{ba}\right)^+ + g_{ba}\tau_a\left(g_{ab},g_{ba}\right),\\
 r_a^p\left(\widehat{G}_a\right) \leq{}& \max\left\{g_{aa}-g_{ab},g_{ba}\right\}- r_b^c\left(\widehat{G}_b\right) \\ 
	\leq{}& \max\left\{g_{aa}-g_{ab},g_{ba}\right\}- g_{ba}\tau_b\left(g_{ab},g_{ba}\right)\\ 
	={}& \left(g_{aa}-g_{ab}-g_{ba}\right)^+ + g_{ba}\tau_a\left(g_{ab},g_{ba}\right),\\ 
 r_a^c\left(\widehat{G}_a\right) + r_a^p\left(\widehat{G}_a\right) \leq{}& \max\left\{g_{aa},g_{ba}\right\} - r_b^c\left(\widehat{G}_b\right)\\
	\leq{}& \max\left\{g_{aa},g_{ba}\right\} - g_{ba}\tau_b\left(g_{ab},g_{ba}\right)\\
	={}& \left(g_{aa} - g_{ba}\right)^+ + g_{ba}\tau_a\left(g_{ab},g_{ba}\right).
\end{align}
Which when simplified under the assumption $r_a(\widehat{G}_a) = r_a^c(\widehat{G}_a)+r_a^p(\widehat{G}_a)$, corresponds with the outer bounds of (\ref{eq:view2proof_3})--(\ref{eq:view2proof_4}). Similar analysis of Sender~$b$'s scheme yields the analogous result.

\subsection{Proof for Views 3 \& 5}
\label{append:views35}
\begin{IEEEproof}
As in View~1, by definition of the problem and noting the knowledge common to both transmitters, the following must hold for some $\tau_{a}(g_{aa},g_{bb})$, $\tau_{b}(g_{aa},g_{bb})$ summing to one.
\begin{align}
 r_a(\widehat{G}_a)\geq g_{aa}\tau_{a}(g_{aa},g_{bb})\label{eq:view3_assum_a},\\
 r_b(\widehat{G}_b)\geq g_{bb}\tau_{b}(g_{aa},g_{bb}).\label{eq:view3_assum_b}
\end{align}

Our proof relies upon consideration of virtual Z-channels that provide structure to the uncertainty of each transmitter.

Let us first consider the POV of Transmitter~$a$. Recalling the assumption that $g_{aa}\geq g_{bb}$, we consider all possible weak interference gain values for Transmitter~$a$'s out-going interference: $g_{ab} \in \{0,1,\ldots, g_{aa}\}$. At Receiver~$b$, the achievability of desired rates $r_b(\widehat{G}_b)$ is dependent on the following conditions
\begin{align}
r_b\left(\widehat{G}_b\middle)\right|_{g_{ab}=0}
	\leq{}& \frac{1}{n} \sum_{j=1}^{g_{bb}}L_{b,j}\left(\widehat{G}_b\middle)\right|_{g_{ab}=0}\label{eq:view3_proof1},\\
r_b\left(\widehat{G}_b\middle)\right|_{g_{ab}=1}
	\leq{}& \frac{1}{n} \left( \sum_{j=1}^{g_{bb}-1}L_{b,j}\left(\widehat{G}_b\middle)\right|_{g_{ba}=1} + n - L_{a,1}\left(\widehat{G}_a\right)\right)\label{eq:view3_proof2},\\
	\vdots{}&\nonumber\\
r_b\left(\widehat{G}_b\middle)\right|_{g_{ab}=g_{aa}-1}
	\leq{}& \frac{1}{n} \left( ng_{bb} - \sum_{i=g_{aa}-g_{bb}}^{g_{aa}-1}L_{a,i}\left(\widehat{G}_a\right) \right)\label{eq:view3_proof4},\\
r_b\left(\widehat{G}_b\middle)\right|_{g_{ab}=g_{aa}}
	\leq{}& \frac{1}{n} \left( ng_{bb} - \sum_{i=g_{aa}-g_{bb}+1}^{g_{aa}-1}L_{a,i}\left(\widehat{G}_a\right) \right).\label{eq:view3_proof5}
\end{align}
Combining (\ref{eq:view3_proof1})--(\ref{eq:view3_proof5}) with expression (\ref{eq:view3_assum_b}) implies more generally
\begin{align}
\sum_{i=\kappa+1}^{\kappa+g_{bb}}L_{a,i}\left(\widehat{G}_a\right) 
	\leq{}& n\left(g_{bb} -  g_{bb}\tau_b(g_{aa},g_{bb})\right)\\
	={}& n\left(g_{bb}\tau_a(g_{aa},g_{bb})\right),\label{eq:view3_proof6}
\end{align}
for $\kappa \in \{0,\ldots,g_{aa}-g_{bb}\}$; i.e. any $g_{bb}$ successive signal levels of Transmitter~$a$'s input are constrained to a TDM-like rate. Notice that if $g_{aa}$ is a multiple of $g_{bb}$, we can select disjoint sets of successive signal level that span Transmitter~$a$'s input, thus completing the proof.

On the other hand, if $g_{aa}$ is not evenly divisible by $g_{bb}$, we construct a virtual Z-channel with the actual Transmitter~$a$ as the initial interferer (the top link in Figure~\ref{fig:view3_virtualZb}). Notice that in doing so, we neglect the incoming interference link gain $g_{ab}$, however this can be rationalized as a genie providing the interference signal to Receiver~$a$. Moreover, we will demonstrate that it is in fact the objective of not inhibiting the transmission of the other link that provides the active constraint on Transmitter~$a$'s input.

Let $\theta_0 = g_{aa}\mod g_{bb}$, and notice that $\theta_0 < g_{bb} \leq g_{aa}$ --- in Figure~\ref{fig:view3_virtualZb}, $\theta_0=1$.
By properly selecting inequalities of the form (\ref{eq:view3_proof6}) and also including the bound from (\ref{eq:view3_proof1})--(\ref{eq:view3_proof5}) for $r_b(\widehat{G}_b)|_{g_{ab}=\theta_0}$, we have
\begin{align}
 nr_a(\widehat{G}_a) \leq{}& \sum_{i=1}^{g_{aa}}L_{a,i}\left(\widehat{G}_a\right)\\
	\leq{}& n\left(g_{aa}-\theta_0\right)\tau_a\left(g_{aa},g_{bb}\right) + 
		\sum_{i=1}^{\theta_0}L_{a,i}\left(\widehat{G}_{a}\right)\label{eq:view3_proof7}\\
	\leq{}& n\left(g_{aa}-\theta_0\right)\tau_a\left(g_{aa},g_{bb}\right) + 
		\sum_{j=1}^{g_{bb}-\theta_0}L_{b,j}\left(\widehat{G}_b\middle)\right|_{g_{ab}=\theta_0} 
		+ n\theta_0 - ng_{bb}\tau_b\left(g_{aa},g_{bb}\right).\label{eq:view3_proof8}
\end{align}
In our virtual Z-channel, we now have a virtual $b$ link where $g_{ba}=\theta_0$. We must now consider the constraints on the un-interfered signal levels of the virtual link, $\sum_{j=1}^{g_{bb}-\theta_0}L_{b,j}\left(\widehat{G}_b\middle)\right|_{g_{ab}=\theta_0}$.
We bound the summation over $j$ using a bound adapted from (\ref{eq:decom_bound_a}) coupled again with bounds of the form (\ref{eq:view3_proof6}) and arrive at
\begin{align}
\sum_{j=1}^{g_{bb}-\theta_0}L_{b,j}\left(\widehat{G}_b\middle)\right|_{g_{ab}=\theta_0}
	\leq{}& \sum_{i=1}^{g_{aa}-(g_{bb}-\theta_0)}L_{a,i}\left(\widehat{G}_a\middle)\right|_{g_{ba}=g_{bb}-\theta_0} 
		+ n\left(g_{bb}-\theta_0\right) - nr_a\left(\widehat{G}_a\middle)\right|_{g_{ba}=g_{bb}-\theta_0}\\
	\leq{}& \sum_{i=1}^{g_{aa}-(g_{bb}-\theta_0)}L_{a,i}\left(\widehat{G}_a\middle)\right|_{g_{ba}=g_{bb}-\theta_0} 
		+ n\left(g_{bb}-\theta_0\right) - ng_{aa}\tau_a\left(g_{aa},g_{bb}\right)\\
	\leq{}& \sum_{i=1}^{\theta_1}L_{a,i}\left(\widehat{G}_a\middle)\right|_{g_{ba}=g_{bb}-\theta_0} 
		+ n\left(g_{aa}-\left(g_{bb}-\theta_0\right)-\theta_1\right)\tau_a\left(g_{aa},g_{bb}\right)
		+ n\left(g_{bb}-\theta_0\right) - ng_{aa}\tau_a\left(g_{aa},g_{bb}\right)\\
	={}& \sum_{i=1}^{\theta_1}L_{a,i}\left(\widehat{G}_a\middle)\right|_{g_{ba}=g_{bb}-\theta_0} 
		- n\theta_1\tau_a\left(g_{aa},g_{bb}\right) + n\left(g_{bb}-\theta_0\right)\tau_b\left(g_{aa},g_{bb}\right)
\end{align}
where
\begin{align}
 \theta_1 ={}& \left(g_{aa}-\left(g_{bb}-\theta_0\right)\right)\mod g_{bb}\\
	={}&  \left(g_{aa}+\theta_0\right)\mod g_{bb}.
\end{align}
If $\theta_1 = 0$, then we arrive a scenario like that in Figure~\ref{fig:view3_virtualZb}, where the second Link~$a$ (first virtual Link~$a$) considered has a number of non-interfered signal levels that is evenly divisible by $g_{bb}$. If this is not the case, we may continue the growth of the virtual Z-channel and arrive at a bound on the remaining levels of the virtual Link~$a$:
\begin{align}
 \sum_{i=1}^{\theta_1}L_{a,i}\left(\widehat{G}_a\middle)\right|_{g_{ba}=g_{bb}-\theta_0} 
	\leq{}& \sum_{j=1}^{g_{bb}-\theta_1}L_{b,j}\left(\widehat{G}_b\middle)\right|_{g_{ab}=\theta_1} 
		+ n\theta_1 - ng_{bb}\tau_b\left(g_{aa},g_{bb}\right)\\
	\leq{}& \sum_{j=1}^{g_{bb}-\theta_1}L_{b,j}\left(\widehat{G}_b\middle)\right|_{g_{ab}=\theta_1}- n\left(g_{bb}-\theta_1\right)\tau_b\left(g_{aa},g_{bb}\right) 
		+ n\theta_1\tau_a\left(g_{aa},g_{bb}\right),
\end{align}
and
\begin{equation}
 \sum_{j=1}^{g_{bb}-\theta_2}L_{b,j}\left(\widehat{G}_b\middle)\right|_{g_{ab}=\theta_1} 
	\leq \sum_{i=1}^{\theta_2}L_{a,i}\left(\widehat{G}_a\middle)\right|_{g_{ba}=g_{bb}-\theta_1} 
		- n\theta_2\tau_a\left(g_{aa},g_{bb}\right) + n\left(g_{bb}-\theta_2\right)\tau_b\left(g_{aa},g_{bb}\right) 
\end{equation}
where
\begin{equation}
 \theta_2 = \left(g_{aa}+\theta_1\right)\mod g_{bb}.
\end{equation}
Though this process may seem cyclic, we note that 
\begin{equation}
 \theta_\ell = \left(g_{aa}+\ell\theta_0\right)\mod g_{bb},
\end{equation}
and that there exists some value for $\ell$ such that $\theta_\ell = 0$. When this is the case
\begin{align}
 \sum_{j=1}^{g_{bb}-\theta_{\ell-1}}L_{b,j}\left(\widehat{G}_b\middle)\right|_{g_{ab}=\theta_{\ell-1}} 
	\leq{}& n\theta_{\ell}\tau_a\left(g_{aa},g_{bb}\right) - n\theta_{\ell}\tau_a\left(g_{aa},g_{bb}\right) + n\left(g_{bb}-\theta_\ell\right)\tau_b\left(g_{aa},g_{bb}\right) \\
	={}& n\left(g_{bb}-\theta_{\ell-1}\right)\tau_b\left(g_{aa},g_{bb}\right),
\end{align}
and
\begin{align}
 \sum_{i=1}^{\theta_1}L_{a,i}\left(\widehat{G}_a\middle)\right|_{g_{ba}=g_{bb}-\theta_0} 
	\leq{}& n\theta_{\ell-1}\tau_a\left(g_{aa},g_{bb}\right).
\end{align}
Carrying this process down to $\ell=0$ and substituting into (\ref{eq:view3_proof8}) yields
\begin{equation}
 nr_a\left(\widehat{G}_a\right) \leq Ng_{aa}\tau_a\left(g_{aa},g_{bb}\right)
\end{equation}
as desired.

To show that the input at Transmitter~$b$ is also constrained to its TDM allotted rate requires a single extension to the previously constructed Z-channel. If we now assume that the top link is a $b$-link and consider the possibility of $g_{ba}=g_{bb}$, then achievability of the TDM-like rate is reliant on
\begin{align}
nr_b\left(\widehat{G}_b\right) 
	\leq{}& \sum_{j=1}^{g_{bb}}L_{b,j}\left(\widehat{G}_b\right)\\
	\leq{}& \sum_{i=1}^{g_{aa}-g_{bb}} L_{a,i}\left(\widehat{G}_a\middle)\right|_{g_{ba}=g_{bb}} 
		+ ng_{bb} - nr_a\left(\widehat{G}_a\middle)\right|_{g_{ba}=g_{bb}}\\
	\leq{}& n\left[ \left(g_{aa}-g_{bb}\right)\tau_a\left(g_{aa},g_{bb}\right) + g_{bb} 
		- r_a\left(\widehat{G}_a\middle)\right|_{g_{ba}=g_{bb}}\right]\\
	\leq{}& n\left[\left(g_{aa}-g_{bb}\right)\tau_a\left(g_{aa},g_{bb}\right) + g_{bb} 
		- g_{aa}\tau_a\left(g_{aa},g_{bb}\right)\right]\\
	={}& n\left[g_{bb} 
		- g_{bb}\tau_a\left(g_{aa},g_{bb}\right)\right]\\
	={}& ng_{bb}\tau_b\left(g_{aa},g_{bb}\right).
\end{align}
as desired.

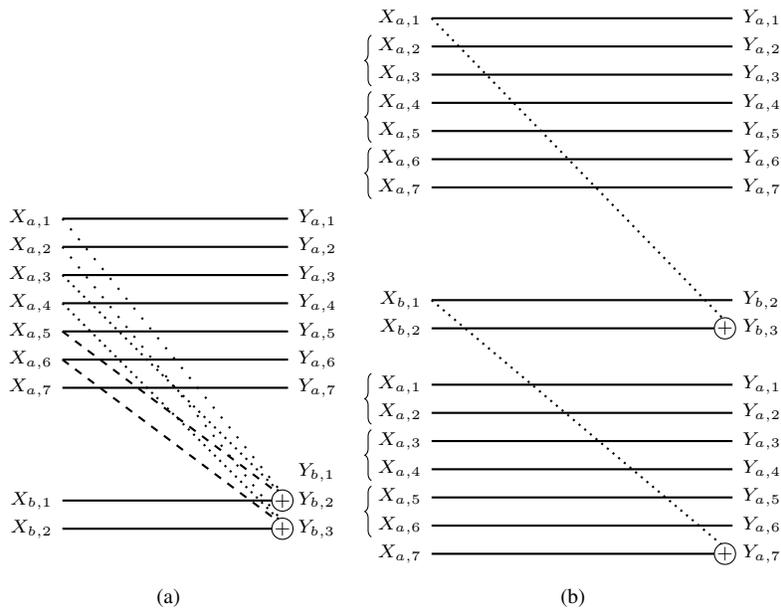
\begin{figure}[ht]
	\begin{center}
\subfigure[]{
	\begin{tikzpicture}[font=\scriptsize,scale=0.75]
\draw[white] (0,-0.25) -- (0,1);

		\draw[thick,loosely dotted](0,6) -- (4,1);
		\draw[thick,loosely dotted](0,5.5) -- (4,0.5);

		\draw[thick,dotted](0,5) -- (4,1);
		\draw[thick,dotted](0,4.5) -- (4,0.5);

		\draw[thick,dashed](0,4) -- (4,1);
		\draw[thick,dashed](0,3.5) -- (4,0.5);

		\draw[thick] (4,1.5) node[right] {$Y_{b,1}$};
		\draw[thick](0,1) node[left] {$X_{b,1}$} -- (4,1) node[right] {$Y_{b,2}$};
		\draw[thick](0,0.5) node[left] {$X_{b,2}$} -- (4,0.5) node[right] {$Y_{b,3}$};

		\draw[thick](0,6) node[left] {$X_{a,1}$} -- (4,6) node[right] {$Y_{a,1}$};
		\draw[thick](0,5.5) node[left] {$X_{a,2}$} -- (4,5.5) node[right] {$Y_{a,2}$};
		\draw[thick](0,5) node[left] {$X_{a,3}$} -- (4,5) node[right] {$Y_{a,3}$};
		\draw[thick](0,4.5) node[left] {$X_{a,4}$} -- (4,4.5) node[right] {$Y_{a,4}$};
		\draw[thick](0,4) node[left] {$X_{a,5}$} -- (4,4) node[right] {$Y_{a,5}$};
		\draw[thick](0,3.5) node[left] {$X_{a,6}$} -- (4,3.5) node[right] {$Y_{a,6}$};
		\draw[thick](0,3) node[left] {$X_{a,7}$} -- (4,3) node[right] {$Y_{a,7}$};
		
		\draw (3.9,1) node[fill=white,draw,circle, inner sep=0pt]{$+$};
		\draw (3.9,0.5) node[fill=white,draw,circle, inner sep=0pt]{$+$};
\end{tikzpicture}
	\label{fig:view3_virtualZa}
	}\subfigure[]{
	\begin{tikzpicture}[font=\scriptsize,yscale=0.75]
\draw[white] (0,-0.25) -- (0,1);
		\draw[thick,dotted](0,6) -- (4,0.5);
		\draw[thick,dotted](0,1) -- (4,-3.5);

		\draw[thick](0,1) node[left] {$X_{b,1}$} -- (4,1) node[right] {$Y_{b,2}$};
		\draw[thick](0,0.5) node[left] {$X_{b,2}$} -- (4,0.5) node[right] {$Y_{b,3}$};

		\draw[thick](0,6) node[left] {$X_{a,1}$} -- (4,6) node[right] {$Y_{a,1}$};
		\draw[thick](0,5.5) node[left] {$X_{a,2}$} -- (4,5.5) node[right] {$Y_{a,2}$};
		\draw[thick](0,5) node[left] {$X_{a,3}$} -- (4,5) node[right] {$Y_{a,3}$};
		\draw[thick](0,4.5) node[left] {$X_{a,4}$} -- (4,4.5) node[right] {$Y_{a,4}$};
		\draw[thick](0,4) node[left] {$X_{a,5}$} -- (4,4) node[right] {$Y_{a,5}$};
		\draw[thick](0,3.5) node[left] {$X_{a,6}$} -- (4,3.5) node[right] {$Y_{a,6}$};
		\draw[thick](0,3) node[left] {$X_{a,7}$} -- (4,3) node[right] {$Y_{a,7}$};

		\draw[decorate,decoration=brace] (-0.8,4.8)--(-0.8,5.7);
		\draw[decorate,decoration=brace] (-0.8,3.8)--(-0.8,4.7);
		\draw[decorate,decoration=brace] (-0.8,2.8)--(-0.8,3.7);
		
		\draw[thick](0,-0.5) node[left] {$X_{a,1}$} -- (4,-0.5) node[right] {$Y_{a,1}$};
		\draw[thick](0,-1) node[left] {$X_{a,2}$} -- (4,-1) node[right] {$Y_{a,2}$};
		\draw[thick](0,-1.5) node[left] {$X_{a,3}$} -- (4,-1.5) node[right] {$Y_{a,3}$};
		\draw[thick](0,-2) node[left] {$X_{a,4}$} -- (4,-2) node[right] {$Y_{a,4}$};
		\draw[thick](0,-2.5) node[left] {$X_{a,5}$} -- (4,-2.5) node[right] {$Y_{a,5}$};
		\draw[thick](0,-3) node[left] {$X_{a,6}$} -- (4,-3) node[right] {$Y_{a,6}$};
		\draw[thick](0,-3.5) node[left] {$X_{a,7}$} -- (4,-3.5) node[right] {$Y_{a,7}$};

		\draw[decorate,decoration=brace] (-0.8,-1.2)--(-0.8,-0.3);
		\draw[decorate,decoration=brace] (-0.8,-2.2)--(-0.8,-1.3);
		\draw[decorate,decoration=brace] (-0.8,-3.2)--(-0.8,-2.3);

		\draw (3.9,-3.5) node[fill=white,draw,circle, inner sep=0pt]{$+$};
		\draw (3.9,0.5) node[fill=white,draw,circle, inner sep=0pt]{$+$};
\end{tikzpicture}
	\label{fig:view3_virtualZb}
	}
	\end{center}
	\caption{Virtual Z channels used to derive outer bound for View~3: \subref{fig:view3_virtualZa} Some of the two-user virtual Z channels used to derive (\ref{eq:view3_proof1})--(\ref{eq:view3_proof5}), and \subref{fig:view3_virtualZb} larger virtual Z to bound relatively prime direct link channels. Bracketed inputs are bounded by constraints from (\ref{eq:view3_proof1})--(\ref{eq:view3_proof5}) and other inputs are constrained by potential interference interactions.}
	\label{fig:view3_virtualZ}
\end{figure}

The statement for View~5 is a direct result of the result from View~3. In designing a policy let it be assumed that a genie will provide Transmitter~$a$ with knowledge of $g_{ba}$, and Transmitter~$b$ with knowledge of $g_{ab}$. The resulting genie-aided view is exactly the same as View~1, and thus the result that the capacity region is confined to that of TDM also holds.

\end{IEEEproof}

\subsection{Proofs for Views 4, 6, \& 7}
\label{append:views467}
\begin{IEEEproof}

We prove the results for Views 4, 6, \& 7 by applying the results and intuitions gained from View~2. Whereas View~2 had two bottleneck cases --- namely the channel state where the unknown direct link was equal to either interference gain --- to prove the statement for Views 4, 6, \& 7, we only require one worst case potential channel state for each: the case where the unknown links form a fully contested Z-channel. In the case of View~4 at Sender~$a$, we apply the possibility that 
\begin{align}
 g_{ab} = g_{bb} = g_{aa},
\end{align}
which requires
\begin{align}
 r_a\left(\widehat{G}_a\right) + r_b\left(\widehat{G}_b\middle)\right|_{ g_{ab} = g_{bb} = g_{aa}} \leq{}& g_{aa}.
\end{align}
Notice that in order to satisfy the minimum performance criterion, this inequality must be an equality which also implies 
\begin{align}
 \frac{r_a\left(\widehat{G}_a\right)}{g_{aa}} \leq{}& 1- \frac{r_b(\widehat{G}_b)|_{ g_{ab} = g_{bb} = g_{aa}}}{g_{aa}}.
\end{align}
We define 
\begin{align}
 \tau_a^{g_{aa}} \triangleq{}& \frac{r_a\left(\widehat{G}_a\right)}{g_{aa}},\\
 \tau_b^{g_{aa}} \triangleq{}& \frac{r_b\left(\widehat{G}_b^\prime\right)|_{ g_{ab}^\prime = g_{bb}^\prime = g_{aa}}}{g_{aa}},
\end{align}
where $\tau_a^{g_{aa}} + \tau_b^{g_{aa}} =1$, and now consider the response of Sender~$b$ to its view of the channel. By considering an analogous Z-channel (where the direct link is fully interfered), we have
\begin{align}
 \tau_a^{g_{bb}} \triangleq{}& \frac{r_a\left(\widehat{G}_a^\prime\right)|_{ g_{ba}^\prime = g_{aa}^\prime = g_{bb}}}{g_{bb}},\\
 \tau_b^{g_{bb}} \triangleq{}& \frac{r_b\left(\widehat{G}_b\right)}{g_{bb}},
\end{align}
where $\tau_a^{g_{bb}} + \tau_b^{g_{bb}} =1$. The final step to completing the proof is to note that the minimum performance criterion
\begin{equation}
 \tau_a^{g_{aa}} + \tau_b^{g_{bb}} \geq 1,
\end{equation}
for all $G$, requires that the inequality be an equality. Therefore the View~4 region is exactly that of TDM.

To demonstrate the theorem for Views~6 and 7, we need only consider the proper worst case Z-channels and apply the same logic. 
\end{IEEEproof}

\subsection{Gap between LDIC and GIC Capacity Regions}
\label{append:gauss}
 
We use heavily the result (\ref{eq:per_vir_user_gap}) from Section~\ref{sec:approx}.
Additionally, we make the following observation
\begin{align}
 \sum_{\ell_1}^{\ell_2}\Lambda_a\left(\widehat{G}_a\right) - (\ell_2 - \ell_1 )^+ \leq{}& \log(3).
\end{align}  

\subsubsection{View 1}

If $g_{aa} = g_{bb}$, then we refer the reader to the gap analysis for View~4, for which the worst case channel state assumes $g_{bb} = g_{aa}$. Otherwise, bounding the gap $\Delta_1$ proceeds as follows. 

By Lemma~\ref{lem:codegap}, a Gaussian policy based on HK schemes that uses rates prescribed by the associated LDIC policy less 4 bits per user is achievable.

In order to bound the gap between the LDIC and GIC outer bounds, we manipulate expression (\ref{eq:Gauss_genie_dec_b_final}) while noting (\ref{eq:per_vir_user_gap}) and arrive at a bound on the interference component of Transmitter~$a$'s signal. 
\begin{align}
  \sum_{\ell=1}^{g_{ab}} \Lambda_{a,\ell} \left(\widehat{H}_a\right)
  \leq{}& \left( h(Y_b^n|W_{bb,u_b^+},W_{ab,u_b^-}) - n\log(2\pi e) 
		- I(X_b^n;Y_b^n) \right)
		+ \left(\sum_{\ell=1}^{u_b^+} \Lambda_{b,\ell}\left(\widehat{H}_b^\prime\right)
		+ \sum_{\ell=1}^{u_b^-} \Lambda_{a,\ell}\right)\\
  \leq{}&  n\min\left\{g_{bb},g_{ab}\right\} + n\log(6)
		- nr_b\left(\widehat{H}_b^\prime\right) 
		+ \sum_{\ell=1}^{u_b^+} \Lambda_{b,\ell}\left(\widehat{H}_b^\prime\right)
		+ \sum_{\ell=1}^{u_b^-} \Lambda_{a,\ell}\left(\widehat{H}_a\right).\label{eq:v1_g_int_a}
\end{align}
Similarly, a constraint on the amount of interference in Transmitter $b$'s signal is given by
\begin{align}
  \sum_{\ell=1}^{g_{ba}} \Lambda_{b,\ell} \left(\widehat{H}_b\right)
  \leq{}&  n\min\left\{g_{aa},g_{ba}\right\} + n\log(6)
		- nr_a\left(\widehat{H}_a^\prime\right) 
		+ \sum_{\ell=1}^{u_a^+} \Lambda_{a,\ell}\left(\widehat{H}_a^\prime\right)
		+ \sum_{\ell=1}^{u_a^-} \Lambda_{b,\ell}\left(\widehat{H}_b\right).\label{eq:v1_g_int_b}
\end{align}
In Appendix~\ref{append:view1}, expressions analogous to (\ref{eq:v1_g_int_a}) and (\ref{eq:v1_g_int_a}) --- namely (\ref{eq:view1proof_3}) and (\ref{eq:view1proof_5}) --- were used to add virtual users representing different policy responses to a virtual Z-channel in constructing an outer bound for View~1 of the LDIC. As alluded to in Section~\ref{sec:gauss}, each additional interference event considered (each additional virtual user added to the virtual Z-channel) increases the gap between the LDIC and GIC outer bounds by $\log(6)$. 

Adding additional virtual users accounts for any remaining sum of un-interfered signal levels (e.g., $\sum_{\ell=1}^{u_a^+} \Lambda_{a,\ell}\left(\widehat{H}_a^\prime\right)$ in (\ref{eq:v1_g_int_b})). When a virtual user is not added (when the Z channel terminates), an additional gap between the Gaussian outer bound and its linear deterministic equivalent results from the quantization of the channel gain. For example
\begin{align}
 \sum_{\ell=1}^{u_a^+} \Lambda_{a,\ell}\left(\widehat{H}_a^\prime\right)
	\leq {}& \max_{p({X_a^n})} \sum_{\ell=1}^{u_a^+} \Lambda_{a,\ell}\left(\widehat{H}_a^\prime\right)
	\leq  n u_a^+ + \log(3).
\end{align}

In summary, the outer bounds constructed for View~1 in the GIC and LDIC have a gap that increases by $\log(6)$ per interferer considered, and by $\log(3)$ at the end of the Z chain (component bounds that are non-terminating lack the $\log(3)$ gap). We find gaps between the component bounds (\ref{eq:view1_thm_1})--(\ref{eq:view1_thm_14}) and their Gaussian IC equivalents using this result, and detail them in Table~\ref{tab:v1_g}.

\begin{table}[ht]
\centering
\begin{tabular}{ | c | c | c | c | } 
	\hline
	\textbf{Component Outer Bound} & \textbf{Component} & \textbf{Gap} & \textbf{Maximum $\ell$}\\
	\hline 
	\hline 
	(\ref{eq:view1_proof_comp1}) & $r_a$ & $\log(3) + \log(6)$ & --- \\
	\hline 
	(\ref{eq:view1_proof_comp2}) & $r_b$ & $\log(3) + \log(6)$ & --- \\
	\hline 
	\hline 
	(\ref{eq:view1_proof_comp3}) & $\overline{r}_a^c$ &  $\log(3) + (2\ell+1)\log(6)$ 
		& $\left\lceil \frac{g_{bb}}{\delta} \right\rceil$ \\
	\hline 
	(\ref{eq:view1_proof_comp4}) & $\overline{r}_a^c$ &  $\log(3)$ & --- \\
	\hline 
	(\ref{eq:view1_proof_comp5}) & $\overline{r}_b^c$ &  $\log(3) + \log(6)$ & --- \\
	\hline 
	(\ref{eq:view1_proof_comp6}) & $\overline{r}_b^c$ &  $\log(3) + (2\ell+1)\log(6)$ 
		& $\left\lceil \frac{g_{ba}}{\delta} \right\rceil$ \\
	\hline 
	\hline
	(\ref{eq:view1_proof_comp7}) & $\overline{r}_a^c + r_b$ & $\log(3) + (2\ell+1)\log(6)$ 
		& $\left\lceil \frac{g_{bb}}{\delta} \right\rceil$ \\
	\hline 
	(\ref{eq:view1_proof_comp8}) & $\overline{r}_b^c + r_a$ & $\log(3) + \log(6)$ & --- \\
	\hline 
	(\ref{eq:view1_proof_comp9}) & $\overline{r}_b^c + r_b$ & $\log(3) + (2\ell)\log(6)$ 
		& $\left\lceil \frac{g_{ba}}{\delta} \right\rceil - 1 $ \\
	\hline 
	\hline 
	(\ref{eq:view1_proof_comp10}) & $\overline{r}_a^c + r_b - \overline{r}_b^c$ & $(2\ell+1)\log(6)$ 
		& $\left\lceil \frac{(g_{bb}-g_{ba})^+}{\delta} \right\rceil$ \\
	\hline 
	(\ref{eq:view1_proof_comp11}) & $\overline{r}_b^c + r_a - \overline{r}_a^c$ & $(2\ell+1)\log(6)$ 
		& $\left\lceil \frac{g_{ba}}{\delta} \right\rceil$\\
	\hline 
	\hline 
	(\ref{eq:view1_proof_comp12}) & $r_a - \overline{r}_a^c$ & $\log(3) + (2\ell+1)\log(6)$ 
		& $\left\lceil \frac{g_{bb}}{\delta} \right\rceil$\\
	\hline 
	(\ref{eq:view1_proof_comp13}) & $r_a - \overline{r}_a^c$ & $\log(3) + \log(6)$ & --- \\
	\hline 
	(\ref{eq:view1_proof_comp14}) & $r_b - \overline{r}_b^c$ & $\log(3) + (2\ell-1)\log(6)$ 
		& $\left\lceil \frac{(g_{bb}-g_{ba})^+}{\delta} \right\rceil$\\
	\hline 
	\end{tabular}
		\caption{Gap between component bounds (\ref{eq:view1_proof_comp1})--(\ref{eq:view1_proof_comp14}) and Gaussian counterparts.}
	\label{tab:v1_g}

\end{table}

From Table~\ref{tab:v1_g}, and referring to the method of combining expressions (\ref{eq:view1_proof_comp1})--(\ref{eq:view1_proof_comp14}), we also compute per-user gaps by combining the respective component bound gaps of Table~\ref{tab:v1_g}, include the potential gap in achievable policy rates, and arrive at the bound in Table~\ref{tab:gauss_result}.

\subsubsection{View 2}

Like View~1, by Lemma~\ref{lem:codegap} a Gaussian policy based on HK schemes that uses rates prescribed by the associated LDIC policy less 4 bits per user is achievable. To bound the gap between LDIC and GIC outerbounds, note that, as in the LDIC version of View~2, two interference cases are sufficient to define a set of outer bounds. WLOG, we consider Transmitter~$a$'s response and mimic the derivation of bounds (\ref{eq:view2_parreg_1})--(\ref{eq:view2_parreg_4}). 
Following the derivation of (\ref{eq:view2_parreg_1}) we have
\begin{align}
 nr_a\left(\widehat{G}_a\right) \leq{}&  \max I(X_a^n;Y_a^n)\\ 
	\leq{}& \max I(X_a^n;Y_a^n|X_b^n)\\ 
	\leq{}& \max \sum_{i=1}^{g_{aa}} \Lambda_a\left(\widehat{G}_a\right)\\ 
	\leq{}& n\max \log(1+|h_{aa}|^2)\\
	\leq{}& n\max \log(1+2^{g_{aa}+1})\\
	\leq{}& ng_{aa} + n\log(3). \label{eq:view2_gap_1}
\end{align}

From the derivation of (\ref{eq:view2_parreg_2}) we have
\begin{align}
nr_a\left(\widehat{G}_a\right) \leq{}&  I(X_a^n;Y_a^n)\\ 
	\leq{}& \left( h(Y_a^n|W_{aa,u_a^+},W_{ba,u_a^-}) - n\log(2\pi e) 
		- \sum_{\ell=1}^{g_{ba}} \Lambda_{b,\ell}\left(\widehat{G}_b\right) \right)
		+ \left(\sum_{\ell=1}^{u_a^+} \Lambda_{a,\ell}\left(\widehat{G}_a\right)
		+ \sum_{\ell=1}^{u_a^-} \Lambda_{b,\ell}\left(\widehat{G}_b\right)\right)\\
	\leq{}& \left( h(Y_a^n|W_{aa,u_a^+},W_{ba,u_a^-}) - n\log(2\pi e) 
		- nr_b\left(\widehat{G}_b\middle)\right|_{g_{bb}=g_{ba}} \right)
		+ \left(\sum_{\ell=1}^{u_a^+} \Lambda_{a,\ell}\left(\widehat{G}_a\right)
		+ \sum_{\ell=1}^{u_a^-} \Lambda_{b,\ell}\left(\widehat{G}_b\right)\right)\\
	\leq{}& \left( h(Y_a^n|W_{aa,u_a^+},W_{ba,u_a^-}) - n\log(2\pi e) 
		- ng_{ba}\tau_b\left(g_{ab},g_{ba}\right) \right)
		+ \left(\sum_{\ell=1}^{u_a^+} \Lambda_{a,\ell}\left(\widehat{G}_a\right)
		+ \sum_{\ell=1}^{u_a^-} \Lambda_{b,\ell}\left(\widehat{G}_b\right)\right)\\
	\leq{}& \left( n\log(6) + n\min\left\{g_{aa},g_{ba}\right\} 
		- ng_{ba}\tau_b\left(g_{ab},g_{ba}\right) \right)
		+ \left(nu_a^+ + nu_a^- +n\log(3)\right)\\
	\leq{}& n\left[ g_{ba}\tau_a\left(g_{ab},g_{ba}\right) + \left(g_{aa} - g_{ba}\right)^+ 
		+ \log(6) + \log(3)\right]. \label{eq:view2_gap_2}
\end{align}

If we consider Transmitter $a$'s potential impact on Link~$b$, we have
\begin{align}
 nr_b\left(\widehat{G}_b\middle)\right|_{g_{bb}=g_{ab}} 
	\leq{}&  I(X_b^n;Y_b^n)\\ 
	\leq{}& \left( h(Y_b^n|W_{bb,u_b^+},W_{ab,u_b^-}) - n\log(2\pi e) 
		- \sum_{\ell=1}^{g_{ab}} \Lambda_{a,\ell}\left(\widehat{G}_a\right) \right)
		+ \left(\sum_{\ell=1}^{u_b^+} \Lambda_{b,\ell}\left(\widehat{G}_b\right)
		+ \sum_{\ell=1}^{u_b^-} \Lambda_{a,\ell}\left(\widehat{G}_a\right)\right)\\
	\leq{}& ng_{ab} - \sum_{\ell=1}^{g_{ab}} \Lambda_{a,\ell}\left(\widehat{G}_a\right)  + n\log(6),
\end{align}
or
\begin{align}
\sum_{\ell=1}^{g_{ab}} \Lambda_{a,\ell}\left(\widehat{G}_a\right)
	\leq{}& ng_{ab} -nr_b\left(\widehat{G}_b\middle)\right|_{g_{bb}=g_{ab}}  + n\log(6)\\
	\leq{}& n\left[g_{ab}\tau_a\left(g_{ab},g_{ba}\right)  + \log(6)\right],
\end{align}
which gives us
\begin{align}
nr_a\left(\widehat{G}_a\right) \leq{}&  \max I(X_a^n;Y_a^n)\\ 
	\leq{}& \max I(X_a^n;Y_a^n|X_b^n)\\ 
	\leq{}& \max \sum_{i=1}^{g_{aa}} \Lambda_a\left(\widehat{G}_a\right)\\ 
	\leq{}& \max \sum_{i=1}^{g_{ab}} \Lambda_a\left(\widehat{G}_a\right) 
		+ \sum_{i=g_{ab}+1}^{g_{aa}} \Lambda_a\left(\widehat{G}_a\right) \\ 
	\leq{}& n\left[g_{ab}\tau_a\left(g_{ab},g_{ba}\right)  + \log(6)
		+ \left(g_{aa}-g_{ab}\right)^+ + \log(3)\right], \label{eq:view2_gap_3}
\end{align}
and 
\begin{align}
nr_a\left(\widehat{G}_a\right) \leq{}&  \max I(X_a^n;Y_a^n)\\ 
	\leq{}& \max I(X_a^n;Y_a^n|X_b^n)\\ 
	\leq{}& \max \sum_{i=1}^{g_{aa}} \Lambda_a\left(\widehat{G}_a\right)\\ 
	\leq{}& \max \sum_{i=1}^{g_{ab}} \Lambda_a\left(\widehat{G}_a\right) 
		+ \sum_{i=g_{ab}+1}^{g_{aa}} \Lambda_a\left(\widehat{G}_a\right) \\ 
	\leq{}& \max ng_{ab}\tau_a\left(g_{ab},g_{ba}\right)  + n\log(6)
		+ \sum_{i=g_{ab}+1}^{g_{aa}-g_{ba}} \Lambda_a\left(\widehat{G}_a\right) 
		+ \sum_{i=g_{aa}-g_{ba}+1}^{g_{aa}} \Lambda_a\left(\widehat{G}_a\right),\\
	\leq{}& n\left[ g_{ab}\tau_a\left(g_{ab},g_{ba}\right)  + \log(6)
		+ \left(g_{aa} - g_{ab} - g_{ba}\right)^+ + \log(3)
		+ g_{ba}\tau_a\left(g_{ab},g_{ba}\right) + \log(6)\right].\label{eq:view2_gap_4}
\end{align}

Consequently, we bound the Link~$a$ capacity gap (and Link~$b$ by parallel analysis) comparing (\ref{eq:view2_gap_1}), (\ref{eq:view2_gap_2}), (\ref{eq:view2_gap_3}), and (\ref{eq:view2_gap_4}) to (\ref{eq:view2_parreg_1})--(\ref{eq:view2_parreg_4}), and arrive at the stated bound. 

\subsubsection{Views 3 \& 5}

Proof of Views~3 and 5 relies on bounding disjoint sets of $g_{bb}$ consecutive levels of Transmitter~$a$'s input with the expression (\ref{eq:view3_proof6}). In the Gaussian IC version, by applying (\ref{eq:per_vir_user_gap}) we have
\begin{align}
\sum_{i=\kappa+1}^{\kappa+g_{bb}}\Lambda_{a,i}\left(\widehat{G}_a\right) 
	\leq{}& n\left(g_{bb}\tau_a(g_{aa},g_{bb}) + \log(6)\right).
\end{align}

To bound the total per-user gap to the Gaussian IC capacity region we track how the gap accumulates in the construction of the virtual Z-channel used in the linear deterministic proof. Let $\tilde{\Delta}_3[\ell]$ be the total gap after $\ell$ virtual Link~$a$s have been considered. When $\ell=1$, we find
\begin{align}
 \tilde{\Delta}_3[1] = \left(\frac{g_{aa}-\theta_0}{g_{bb}}\right)\log(6).
\end{align}
Recall that if $\theta_0 = 0$, then $g_{aa}$ was evenly divisible by $g_{bb}$ and the proof ends. If $\theta_0 \neq 0$ we can find through induction
\begin{align}
 \tilde{\Delta}_3[\ell] 
	={}& \left(\tilde{\Delta}_3[\ell-1] + 2 + \frac{g_{aa}-\theta_\ell - (g_{bb}-\theta_\ell-1)}{g_{bb}}\right)\log(6)\\
	={}& \left(\frac{\ell g_{aa} - \theta_{\ell-1}}{g_{bb}} +\ell - 1\right)\log(6).
\end{align}

When $\theta_{\ell-1} = 0$, the chain of substitutions in the proof of Theorem~\ref{thm:views35} ends, implying $\ell g_{aa}$ is evenly divisible by $g_{bb}$. Consequently, if we let $\ell^\star$ be the minimum value of $\ell$ where this occurs, $\ell^\star g_{aa}$ is the least common multiple of $g_{aa}$ and $g_{bb}$, and it becomes clear that
\begin{align}
 \Delta_3 ={}& \left(\frac{\ell^\star g_{aa}}{g_{bb}} +\ell^\star - 1\right)\log(6)\\
	={}& \left(\frac{\ell^\star g_{aa}}{g_{bb}} +\frac{\ell^\star g_{aa}}{g_{aa}} - 1\right)\log(6)\\
	={}& \left(\frac{\lcm(g_{aa},g_{bb})}{g_{bb}} +\frac{\lcm(g_{aa},g_{bb})}{g_{aa}} - 1\right)\log(6).
\end{align}

\subsubsection{Views 4, 6, \& 7}
For View~4, when Transmitter~$a$ considers the case $g_{ba} = g_{bb} = g_{aa}$
\begin{align}
nr_a\left(\widehat{G}_a\right) \leq{}&  I(X_a^n;Y_a^n)\\ 
	\leq{}& \left( h(Y_a^n|W_{aa,u_a^+},W_{ba,u_a^-}) - n\log(2\pi e) 
		- \sum_{\ell=1}^{g_{ba}} \Lambda_{b,\ell}\left(\widehat{G}_b\right) \right)
		+ \left(\sum_{\ell=1}^{u_a^+} \Lambda_{a,\ell}\left(\widehat{G}_a\right)
		+ \sum_{\ell=1}^{u_a^-} \Lambda_{b,\ell}\left(\widehat{G}_b\right)\right)\\
	\leq{}& \left(  n\log(6) + n\min\left\{g_{aa},g_{ba}\right\} 
		- nr_b\left(\widehat{G}_b\middle)\right|_{g_{bb}=g_{aa}} \right)\\
	\leq{}& \left( n\log(6) + ng_{aa} - ng_{aa}\tau_b \right)\\
	\leq{}& n\left[ g_{aa}\tau_a + \log(6) \right]. \label{eq:view4_gap}
\end{align}

For View~6, consider the case $g_{ab} = g_{bb} = g_{aa}$
\begin{align}
 nr_a\left(\widehat{G}_a\right) \leq{}&  I(X_a^n;Y_a^n)\\ 
	\leq{}& \sum_{\ell=1}^{g_{aa}} \Lambda_{a,\ell}\left(\widehat{G}_a\right)\\
	\leq{}& ng_{aa} - nr_b\left(\widehat{G}_b\middle)\right|_{g_{bb}=g_{aa}}  + n\log(6)\\
	\leq{}& ng_{aa} - ng_{aa}\tau_b  + n\log(6)\\
	\leq{}& n\left[g_{aa}\tau_a  + \log(6)\right].
\end{align}

The analysis for View~7 may follow that of either View~4 or View~6, and comparison of the resulting expression with the linear deterministic analogue confirms the stated claim.

\bibliographystyle{IEEEtran}
\bibliography{references}

\end{document}